\input epsf
\documentclass{emulateapj}
\usepackage[]{graphicx}

\shorttitle{VIVA (VLA Imaging of Virgo Spirals in Atomic gas)-I. Atlas}
\shortauthors{Chung et al.}
\begin{document}

\newcommand{\hi}{\mbox{H{\sc i}}}
\newcommand{\shi}{$S_{\tiny\mbox{H{\sc i}}}$}
\newcommand{\mhi}{$M_{\tiny\mbox{H{\sc i}}}$}

\title{VIVA, V{\small LA} I{\small maging of} V{\small irgo Spirals
in} A{\small tomic gas:}\\ {\small I. The Atlas \& the H{\sc I}
Properties}}

\author{Aeree Chung\altaffilmark{1,2}, J. H. van
Gorkom\altaffilmark{1}, Jeffrey D. P. Kenney\altaffilmark{3}, Hugh
Crowl\altaffilmark{3,4} and Bernd Vollmer\altaffilmark{5}}

\altaffiltext{1}{Department of Astronomy, Columbia University, 550
                 West 120th Street, New~York, NY~10027, U.S.A.;
                 jvangork@astro.columbia.edu}
                 \altaffiltext{2}{Currently NRAO Jansky Postdoctoral
                 Fellow at the National Radio Astronomy Observatory, 
                 P.O. Box O, Socorro, NM 87801,                 
                 U.S.A.; achung@aoc.nrao.edu}
                 \altaffiltext{3}{Department of Astronomy, Yale
                 University, P.O. Box 208101, New Haven, CT 06520,
                 U.S.A.; kenney@astro.yale.edu, hugh@astro.yale.edu}
                 \altaffiltext{4}{Current address: Department of Astronomy, 
                 University of Massachusetts, 710 North Pleasant Street,
                 Amherst, MA 01003-9305, USA; hugh@astro.umass.edu}
                 \altaffiltext{5}{Observatoire astronomique de
                 Strasbourg,  11 rue de l'universite, 67000
                 Strasbourg, France; bvollmer@astro.u-strasbg.fr}

\begin{abstract}
We present the result of a new VLA H{\sc i} Imaging survey of Virgo
galaxies, VIVA (the VLA Imaging survey of Virgo galaxies in Atomic
gas). The survey includes high resolution H{\sc i} data of 53
carefully selected late type galaxies (48 spirals and 5 irregular
systems). The goal is to study environmental effects on H{\sc i} gas
properties of cluster galaxies to understand which physical mechanisms
affect galaxy evolution in different density regions, and
to establish how far out the impact of the cluster reaches. As a
dynamically young cluster, Virgo contains examples of
galaxies experiencing a variety of environmental effects. Its nearness allows 
us to study each galaxy in
great detail.  We have selected Virgo galaxies with a range of star
formation properties in low to high density regions (at the projected
distance from M87, $d_{87}=$0.3-3.3~Mpc). Contrary to previous 
studies, more
than half of the galaxies in the sample ($\sim60\%$) are fainter than
12~mag in $B_T$. Overall, the selected galaxies  represent
the late type Virgo galaxies (S0/a to Sd/Irr) down to
$m_p\lesssim14.6$ fairly well in morphological type, systemic velocity, 
subcluster
membership, H{\sc i} mass and deficiency. The H{\sc i} observations
were done in CS configuration of the Very Large Array radio telescope,
with a typical spatial resolution of $15''$ and a column density
sensitivity of $\approx3-5\times10^{19}$~cm$^{-2}$ in $3\sigma$ per 10
km s$^{-1}$ channel.  The survey was supplemented with data of comparable
quality from the NRAO archive, taken in CS or C configuration. In this
paper (VIVA~I: the atlas and the H{\sc i} properties), we present
H{\sc i} channel maps, total intensity maps, velocity fields, velocity
dispersions, global/radial profiles, position-velocity diagrams and
overlays of H{\sc i}/1.4~GHz continuum maps on the optical images. We
also present the H{\sc i} properties such as total flux ($S_{\rm
HI}$), H{\sc i} mass ($M_{\rm HI}$), linewidths ($W_{\rm
20}$~\&~$W_{\rm 50}$), velocity ($V_{\rm HI}$), deficiency ($def_{\rm
HI}$), size ($D_{\rm HI}^{\rm eff}$~\&~$D_{\rm HI}^{\rm iso}$), and
describe the H{\sc i} morphology and kinematics of individual galaxies
in detail.  The survey has revealed details of H{\sc i} features that
were never seen before. In this paper we briefly discuss differences in 
typical H{\sc i} morphology for galaxies in regions of different
galaxy densities. 
We confirm that galaxies near the cluster core
($d_{87}\lesssim0.5~$Mpc) have small H{\sc i} disks compared to their
stellar disks ($D_{\rm HI}/D_{25}<0.5$).
Most of these galaxies in the core also show 
gas displaced from the disk, which is either currently
being stripped, or falling back after a stripping event.  At
intermediate distances ($d_{87}\sim$1~Mpc) from the center 
we find a remarkable number of
galaxies with long one sided H{\sc i} tails pointing away from M87. In
a previous letter we argue that these are galaxies recent arrivals, falling 
in for the first time into the Virgo core. In the outskirts we find many 
gas-rich galaxies, with gas disks extending far beyond the optical. 
Interestingly we also find some galaxies with H{\sc i} disks that 
are small compared to their stellar disk at large clustercentric
distances.
\end{abstract}
 

\section{Introduction}
It has long been known that cluster galaxies appear to be different
from  field galaxies in their morphological type and color
\citep[e.g.][]{hh31}. In the local universe $\sim90\%$ of the
population in the core regions of rich clusters consists of
ellipticals and S0's while spirals dominate in the field
\citep{dressler80}.  This could either mean that galaxies form
differently in dense environment or that galaxies are affected by
their surroundings.  Many mechanisms could drive environmental
evolution, for example ram-pressure stripping \citep{gg72},
turbulent/viscous stripping \citep{n82}, thermal evaporation
\citep{cs77}, starvation \citep{ltn80}, interaction with the cluster
potential \citep{bb99}, harassment, the cumulative effect of many fast
interactions \citep{mkldo96},  slow interactions between individual
galaxies \citep{mihosb04} and mergers.

Single dish 21~cm observations, such as \citet{dl73}, Chamaraux, Balkowski
\& Gerard (1980) and \citet{gasdef85} have found that spirals near the
cluster core regions are very deficient in their neutral atomic
hydrogen gas, H{\sc i}, compared to galaxies of the same morphological
type and size in the field.  \citet{gasdef85} first showed that not
only the gas content but also the size of the gas disks is affected.
Subsequent H{\sc i} imaging studies of nearby clusters like Virgo and
Coma \citep{warmels88,cvgbk90,bravo00}, showed that the H{\sc i} disks
of the highly H{\sc i} deficient galaxies are severely truncated to
within the stellar disk. These images of unperturbed stellar disks
with highly truncated gas disks strongly suggest that galaxies lose
their interstellar gas (ISM) through an interaction with the hot
intracluster medium (ICM).

However, there are still remaining questions. First,
\citet{dressler80} already pointed out that ram pressure stripping by the 
hot ICM  alone cannot be  responsible for the transformation of spirals into
 S0's. The morphology density relation changes
very smoothly and a significant fraction of S0's reside in low density
environment, and the bulge to disk ratios of S0's are systematically larger
in all density regimes. This cannot be caused by a simple ISM stripping.
More recently, in a
study of 18 nearby clusters, \citet{smgggh01}  show that the
H{\sc i} deficiency decreases  gradually with increasing
projected distance from the cluster center out to
 $\sim$2 Abell
radii ($\approx3h^{-1}$~Mpc). A similar trend is found in
star formation rate which begins to decrease at a clustercentric radius
of 3-4 virial radii or 1.5 Abell radii
\citep[e.g.][]{lewis02,gomez03}.  These results suggest that galaxies 
are already modified in much lower density environments, where ram pressure
is expected to be unimportant,
 and now the question has become: how exactly are 
galaxies affected in different density regions? That is, how far
out do galaxies feel the impact of the cluster, which mechanisms are
at work in lower  density environments, and what are the dominant  
environmental effects onto disks 
that galaxies experience as they come close to the cluster center?

\begin{figure*}
\plotone{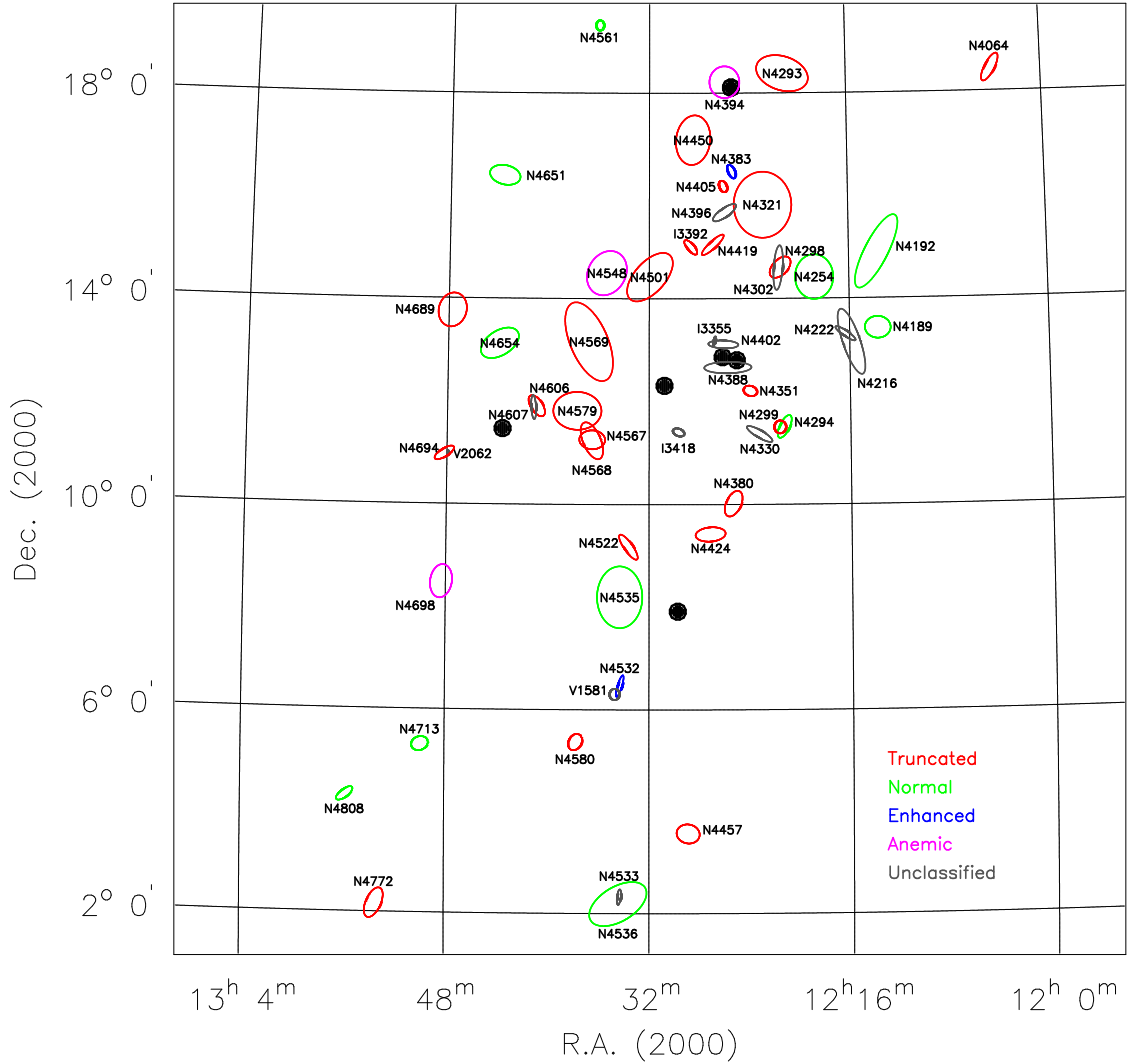}
\caption{The VIVA galaxies are shown at their proper positions with
NGC (N), IC (I) or VCC (V) names. Each galaxy is indicated by an ellipse
which represents $D_{25}\times10$ and is drawn using the position
angle and the inclination measured in the optical $B$-band. Galaxies
are color-coded based on the star formation properties classified in
the H$\alpha$ study by \citet[][]{kk04b}, except for light grey which 
indicates the ones not included in the sample of \citet[][b]{kk04a}. 
Six big ellipticals are shown in large black dots (M85, M86,
M84, M87, M60 and M49 in order of decreasing 
declination).\label{fig-viva}}
\end{figure*}

In order to answer these questions, we have probed the cluster
environment using high resolution H{\sc i} data on a sample of carefully 
selected
Virgo galaxies. The H{\sc i} gas is often a good tracer of different
physical processes as it gets affected both by the intracluster medium
(ICM) and gravitational interactions. Also, the outer gas disk is 
mostly in atomic form, where it is more vulnerable to the surroundings.  
In addition, it provides useful diagnostics for galaxy
evolution as it is the fuel for star formation.  The H{\sc i}
data were taken using the Very Large Array (VLA)\footnote{The VLA is
operated by the National Radio Astronomy Observatory, which is a
facility of the National Science Foundation (NSF), operated under
cooperative agreement by Associated Universities, Inc.}. 
The VLA has had significant improvements since the previous
Virgo survey. The L band (20~cm) receivers have been replaced
and the C array has been replaced by the C short (CS)
configuration.  Our high resolution, high sensitivity VLA H{\sc i}
data allow us to investigate not only the evolutionary history of
individual galaxies but also the overall impact of Virgo on its
members. 

Several results of the survey and a preceding pilot for the survey
have already been published. Kenney, van Gorkom \& Vollmer (2004), and
\citet{vbkvg04}
present an analysis of the data on NGC~4522, a galaxy far away
from M87, yet stripped to well within the optical disk and showing 
abundant evidence for currently ongoing stripping. They suggest that 
NGC~4522 possibly
shows evidence for enhanced ram pressure due to bulk motions or substructure 
of the ICM. \citet{ckvgv05} present  H{\sc i}, radio continuum
and high quality optical images of the edge-on galaxy NGC~4402,
which shows evidence for dense cloud ablation. One of the striking
results of the survey is a number of one sided long  H{\sc i} tails
pointing away from the cluster center and at intermediate distances from
M87. \citet{cvgkv07} argue that these are probably galaxies recently arrived
near the cluster and falling into the cluster for the first time. 
Some of our data have
already been used to constrain simulations of individual systems,
for example on NGC 4522 \citep{vollmer06} and NGC 4501 \citep{vsc08}.

In this work, we present the complete H{\sc i} atlas and describe the
H{\sc i} properties of individual galaxies in detail. In a second
paper we will present a statistical analysis of our results and
discuss the impact of the different environmental effects.

This paper is organized as follows. In \S~\ref{sec-sample} we describe
our selection criteria and present the general properties of the
sample. In \S~\ref{sec-obs} we present the observations and data
reduction. In \S~\ref{sec-atlas} we describe the H{\sc i}
atlas, which is appended at the end. In
\S~\ref{sec-hiprop} we measure the H{\sc i} quantities such as mass,
linewidth, velocity, deficiency, and size, and compare our H{\sc i}
fluxes with values found in the 
literature. We then present our main findings on the H{\sc i} morphology
in \S~\ref{sec-results}, followed
by a summary of the main results in \S~\ref{sec-summary}.
In an appendix we present the full H{\sc i} atlas and comment on 
individual galaxies.
Throughout this paper we assume that the distance to Virgo is 16 Mpc
 \citep{yfo97}.

\begin{table*}
\begin{center}
\caption[VIVA Sample and General Properties]{VIVA Sample and General Properties}
\label{tbl-sample}
\tiny
\begin{tabular}{crcclccccccc}
\hline\hline
\multicolumn{1}{c}{(1)}&
\multicolumn{1}{c}{(2)}&
\multicolumn{1}{c}{(3)}&
\multicolumn{1}{c}{(4)}&
\multicolumn{1}{c}{(5)}&
\multicolumn{1}{c}{(6)}&
\multicolumn{1}{c}{(7)}&
\multicolumn{1}{c}{(8)}&
\multicolumn{1}{c}{(9)}&
\multicolumn{1}{c}{(10)}&
\multicolumn{1}{c}{(11)}&
\multicolumn{1}{c}{(12)}\\
\multicolumn{1}{c}{Galaxy} &
\multicolumn{1}{r}{VCC}&
\multicolumn{1}{c}{$\alpha_{2000}$} &
\multicolumn{1}{c}{$\delta_{2000}$} &
\multicolumn{1}{c}{Type} &
\multicolumn{1}{c}{$D_{25}$} &
\multicolumn{1}{c}{$B_T$} &
\multicolumn{1}{c}{PA} &
\multicolumn{1}{c}{$i$} &
\multicolumn{1}{c}{$V$} &
\multicolumn{1}{c}{$d_{\rm M87}$} &
\multicolumn{1}{c}{SF} \\
\multicolumn{1}{l}{} &
\multicolumn{1}{c}{\hspace{-0.25cm}} &
\multicolumn{1}{c}{${hhmmss.s}$} &
\multicolumn{1}{c}{${ddmmss}$} &
\multicolumn{1}{l}{} &
\multicolumn{1}{c}{$'$} &
\multicolumn{1}{c}{mag} &
\multicolumn{1}{c}{$^\circ$} &
\multicolumn{1}{c}{$^\circ$} &
\multicolumn{1}{c}{km s$^{-1}$} &
\multicolumn{1}{c}{deg} &
\multicolumn{1}{c}{} \\
\hline
NGC 4064          &  ... &12 04 10.8 &+18 26 34 &SB(s)a: pec   & 4.37 &12.22 &150    &69  &1000& 9.0&T/C~~\\   
NGC 4189          &   89 &12 13 46.8 &+13 25 36 &SAB(rs)cd     & 2.40 &12.51 & 66$^c$&45  &1995& 4.4&N  ~~\\ 
NGC 4192          &   92 &12 13 48.2 &+14 53 43 &SAB(s)ab      & 9.77 &10.95 &155    &78  &-126& 4.9&N  ~~\\ 
NGC 4216          &  167 &12 15 53.1 &+13 08 58 &SAB(s)b       & 8.13 &10.99 & 19    &85  &  30& 3.8&...~~\\ 
NGC 4222$^*$      &  187 &12 16 22.7 &+13 18 31 &Sd            & 3.31 &13.86 & 56    &90  & 225& 3.7&...~~\\ 
NGC 4254          &  307 &12 18 49.4 &+14 25 07 &SA(s)c        & 5.37 &10.44 & 68$^c$&30  &2453& 3.6&N  ~~\\ 
NGC 4293          &  460 &12 21 13.0 &+18 22 58 &(R)SB(s)0     & 5.62 &11.26 & 72    &65  & 717& 6.5&T/A~~\\ 
NGC 4294          &  465 &12 21 17.4 &+11 30 40 &SB(s)cd       & 3.24 &12.53 &155    &70  & 421& 2.5&N  ~~\\ 
NGC 4298          &  483 &12 21 32.7 &+14 36 25 &SA(rs)c       & 3.24 &12.04 &140    &57  &1122& 3.2&T/N~~\\ 
NGC 4299          &  491 &12 21 40.6 &+11 30 15 &SAB(s)dm      & 1.74 &12.88 & 42$^c$&22  & 209& 2.5&T/E~~\\ 
NGC 4302          &  497 &12 21 42.5 &+14 36 05 &Sc            & 5.50 &12.50 &178    &90  &1111& 3.2&...~~\\ 
NGC 4321          &  596 &12 22 55.2 &+15 49 23 &SAB(s)bc      & 7.41 &10.05 & 30    &33  &1579& 4.0&T/N~~\\ 
NGC 4330          &  630 &12 23 16.5 &+11 22 06 &Scd?          & 4.47 &13.09 & 59    &90  &1567& 2.1&*  ~~\\ 
NGC 4351          &  692 &12 24 01.8 &+12 12 24 &SB(rs)ab: pec & 1.20 &13.03 & 61$^c$&49  &2388& 1.7&T/N~~\\ 
NGC 4380          &  792 &12 25 22.2 &+10 00 57 &SA(rs)b       & 3.47 &12.66 &153    &58  & 935& 2.7&T/A~~\\ 
NGC 4383          &  801 &12 25 25.6 &+16 28 12 &Sa: pec       & 1.95 &12.67 & 13$^c$&60  &1663& 4.3&E  ~~\\ 
NGC 4388          &  836 &12 25 47.0 &+12 39 42 &SA(s)b        & 5.62 &11.76 & 92    &83  &2538& 1.3&...~~\\ 
NGC 4394          &  857 &12 25 56.1 &+18 12 54 &(RS)B(r)b     & 3.63 &11.73 &113$^c$&28  & 772& 5.9&A  ~~\\ 
NGC 4396          &  865 &12 25 59.3 &+15 40 19 &SAd           & 3.31 &13.06 &125    &77  &-133& 3.5&...~~\\ 
NGC 4405          &  874 &12 26 07.5 &+16 10 50 &SA(rs)0       & 1.78 &13.03 & 20    &51  &1751& 4.0&T/N~~\\ 
NGC 4402          &  873 &12 26 07.9 &+13 06 46 &Sb            & 3.89 &12.55 & 90    &78  & 190& 1.4&...~~\\ 
IC 3355$^*$       &  945 &12 26 50.0 &+13 10 36 &Im            & 1.12 &15.18 &172    &68  & 127& 1.3&...~~\\ 
NGC 4419          &  958 &12 26 56.9 &+15 02 52 &SB(s)a        & 3.31 &12.08 &133    &74  &-224& 2.8&T/A~~\\ 
NGC 4424          &  979 &12 27 11.5 &+09 25 15 &SB(s)a        & 3.63 &12.34 & 95    &62  & 447& 3.1&T/C~~\\ 
NGC 4450          & 1110 &12 28 29.4 &+17 05 05 &SA(s)ab       & 5.25 &10.90 &175    &43  &2048& 4.7&T/A~~\\ 
IC 3392           & 1126 &12 28 43.7 &+15 00 05 &SAb           & 2.29 &12.99 & 40    &67  &1678& 2.7&T/N~~\\ 
NGC 4457          & 1145 &12 28 59.3 &+03 34 16 &(R)SAB(s)0    & 2.69 &11.76 & 82$^c$&33  & 738& 8.8&T/N~~\\ 
IC 3418           & 1217 &12 29 43.5 &+11 24 08 &IBm           & 1.48 &14.00 &...    &50  &  38& 1.0&*  ~~\\ 
NGC 4501          & 1401 &12 31 59.6 &+14 25 17 &SA(rs)b       & 6.92 &10.36 &140    &59  &2120& 2.1&T/N~~\\ 
NGC 4522          & 1516 &12 33 40.0 &+09 10 30 &SB(s)cd       & 3.72 &12.99 & 33    &79  &2332& 3.3&T/N~~\\ 
NGC 4532          & 1554 &12 34 19.4 &+06 28 12 &IBm           & 2.82 &12.30 &160    &70  &2154& 6.0&E  ~~\\ 
NGC 4535          & 1555 &12 34 20.3 &+08 11 53 &SAB(s)c       & 7.08 &10.59 &  0    &46  &1973& 4.3&N  ~~\\ 
NGC 4533$^*$      & 1557 &12 34 22.2 &+02 19 31 &SAd           & 2.09 &14.20 &161    &88  &1753&10.1&...~~\\ 
NGC 4536          & 1562 &12 34 26.9 &+02 11 19 &SAB(rs)bc     & 7.59 &11.16 &130    &67  &1894&10.2&N  ~~\\ 
HolmbergVII$^{*a}$& 1581 &12 34 44.8 &+06 18 10 &Im            & 1.29 &14.62 & 82$^c$&22  &2039& 6.2&...~~\\ 
NGC 4548          & 1615 &12 35 26.3 &+14 29 49 &SB(rs)b       & 5.37 &10.96 &150    &38  & 498& 2.4&A  ~~\\ 
NGC 4561          &  ... &12 36 08.6 &+19 19 26 &SB(rs)dm      & 1.51 &12.90 & 30    &33  &1441& 7.1&N  ~~\\ 
NGC 4567          & 1673 &12 36 32.8 &+11 15 31 &SA(rs)bc      & 2.95 &12.06 & 85    &49  &2213& 1.8&T/N~~\\ 
NGC 4568          & 1676 &12 36 34.7 &+11 14 15 &SA(rs)bc      & 4.57 &11.68 & 23    &66  &2260& 1.8&T/N~~\\ 
NGC 4569          & 1690 &12 36 50.1 &+13 09 48 &SAB(rs)ab     & 9.55 &10.26 & 23    &65  &-311& 1.7&T/N~~\\ 
NGC 4579          & 1727 &12 37 44.2 &+11 49 11 &SAB(rs)b      & 5.89 &10.48 & 95    &38  &1627& 1.8&T/N~~\\ 
NGC 4580          & 1730 &12 37 48.4 &+05 22 09 &SAB(rs)a: pec & 2.09 &11.83 &165    &40  &1227& 7.2&T/N~~\\ 
NGC 4606          & 1859 &12 40 57.8 &+11 54 41 &SB(s)a        & 3.24 &12.67 & 33    &62  &1653& 2.6&T/C~~\\ 
NGC 4607          & 1868 &12 41 12.2 &+11 53 09 &SBb           & 2.88 &13.75 &  2    &83  &2284& 2.6&...~~\\ 
NGC 4651          &  ... &12 43 42.6 &+16 23 40 &SA(rs)c       & 3.98 &11.39 & 80    &50  & 788& 5.1&N  ~~\\ 
NGC 4654          & 1987 &12 43 56.6 &+13 07 33 &SAB(rs)cd     & 4.90 &11.10 &128    &56  &1035& 3.4&N  ~~\\ 
NGC 4689          & 2058 &12 47 45.8 &+13 45 51 &SA(rs)bc      & 4.27 &11.60 &161$^c$&37  &1522& 4.5&T/N~~\\ 
VCC 2062$^{*b}$   & 2062 &12 47 59.9 &+10 58 33 &dE            & 0.69 &19.00 & 42$^c$&7   &1170& 4.5&...~~\\ 
NGC 4694          & 2066 &12 48 15.1 &+10 59 07 &SB0: pec      & 3.16 &12.06 &140    &63  &1211& 4.6&T/N~~\\ 
NGC 4698          & 2070 &12 48 23.5 &+08 29 16 &SA(s)ab       & 3.98 &11.46 &170    &53  &1032& 5.9&A  ~~\\ 
NGC 4713          &  ... &12 49 58.1 &+05 18 39 &SAB(rs)d      & 2.69 &12.19 &100    &52  & 631& 8.5&N  ~~\\ 
NGC 4772          &  ... &12 53 29.1 &+02 10 11 &SA(s)a        & 3.39 &11.96 &147    &62  &1042&11.7&T/N~~\\ 
NGC 4808          &  ... &12 55 49.6 &+04 18 14 &SA(s)cd       & 2.75 &12.35 &127    &68  & 738&10.2&N  ~~\\   
\hline
\end{tabular}
\end{center}
\noindent
{\footnotesize 
The data have been taken from {\it The Third reference catalogue of bright galaxies} (RC3; \prc3) 
unless noted.
(1) First names as they appear in RC3. $^*$Five bonus galaxies from the same field as the selected sample;
(2) Virgo Cluster Catalog (VCC) number \citep{bst85};(3) Right ascension in J2000;
(4) Declination in J2000; (5) Morphological type; 
(6) Optical size of major axis measured at 25~mag~$\Box''^{-1}$ in $B$-band;
(7) Position angle; 
(8) Inclination derived from the ratio of major to minor axis, using the Hubble formula for oblate
spheroids and an intrinsic axis ratio of 0.2, $i=cos^{-1} \sqrt{1.024~b^2/a^2-0.042}$ \citep{amh80};
(9) Optical velocity; (10) Projected distance from M87; 
(11) Star formation property classified based on H$\alpha$ surface profiles \citep{kk04b}: 
N-normal, T-truncated, C-compact, E-enhanced, and A-anemic. $^*$Galaxies not included in the sample of
\citet{kk04b} but selected for the VIVA survey due to the morphological pecularities in the $UV$ wavelength;
$^{a}$It will be refered with its VCC number (VCC~1581) hereafter;
$^{b}$The data were taken from \citet{bst85} since it is not available from RC3.
$^{c}$PA's determined by us using the H{\sc i} kinematics.}
\end{table*}

\section{Sample}
\label{sec-sample}
\subsection{The Virgo Cluster}
Virgo is the nearest rich galaxy cluster. Binggeli, Sandage \& Tammann
(1985) have
cataloged 2096 galaxies (VCC, Virgo Cluster Catalog) in
$\sim140~$deg$^2$ area centered on $\alpha,\delta_{1950}=12^{\rm
h}25^{\rm m},13^\circ$ ($\sim$1 degree northwest of M87). About 1300
galaxies have been identified as true members based on the
morphological appearance and the measured radial velocities. The X-ray
emission from the hot cluster gas \citep[][]{bohr94} shows plenty 
of substructures,
indicating that Virgo is far from being virialized but still growing
as several subclusters (the M86 and M49 group) merge into the main cluster
around M87.  The velocities and the surface brightness fluctuation
(SBF) distances of M87, M86 and M49, which are noted in black in
Figure~\ref{fig-viva}, are 1307, -244 and 997~km~s$^{-1}$
\citep{smith00}, and 16.1, 18.4 \citep{wb00} and 16.3~Mpc
\citep{fcj03}, respectively.  The M86 group is falling in from the back
and M49 is likely to be merging with the M87 group, falling in from the south 
in the plane of the sky \citep{ts84,sbb99}. For more detailed
discussions of the 3-dimensional structure of the Virgo cluster, see
for example \citet{gbspb99} and \citet{mei07}.

Since Virgo is nearby, it is an ideal target for H{\sc i} imaging
studies, and two major imaging surveys were done in the past.
 \citet{warmels88} and \citet{cvgbk90} have mapped 15 and
25 bright Virgo spirals with the Westerbork Synthesis Radio Telescope
(WSRT) and the VLA respectively. Those studies have shown 
that the H{\sc i} disks of the central galaxies are  truncated to well within 
the optical disks, making it likely that 
ICM-ISM interactions are  at least partly responsible for driving the
evolution of galaxies in the cluster inner region. Here we present the
results of a new survey that includes twice as many galaxies, covers
a much wider range in galaxy mass and probes the lower density outer
regions as well as the high density core.

\begin{figure}
\centering
\plotone{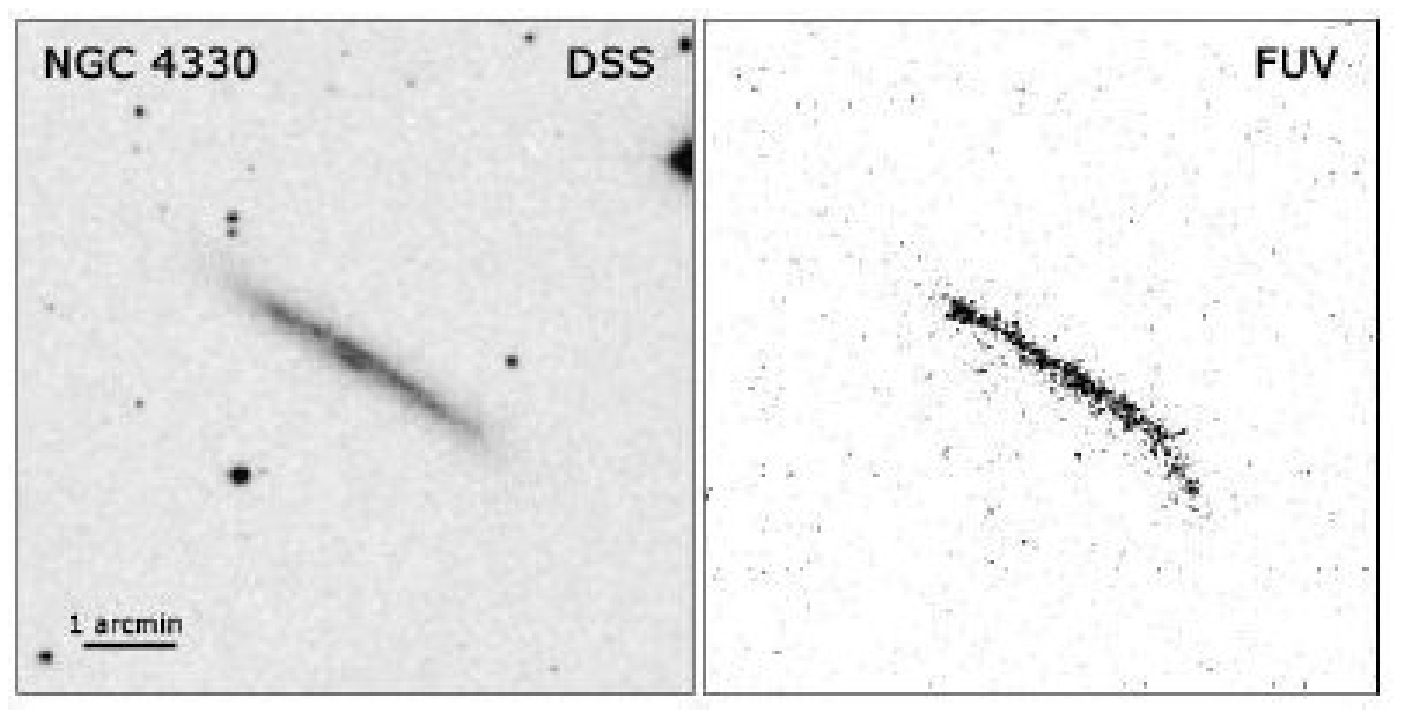}\\
\plotone{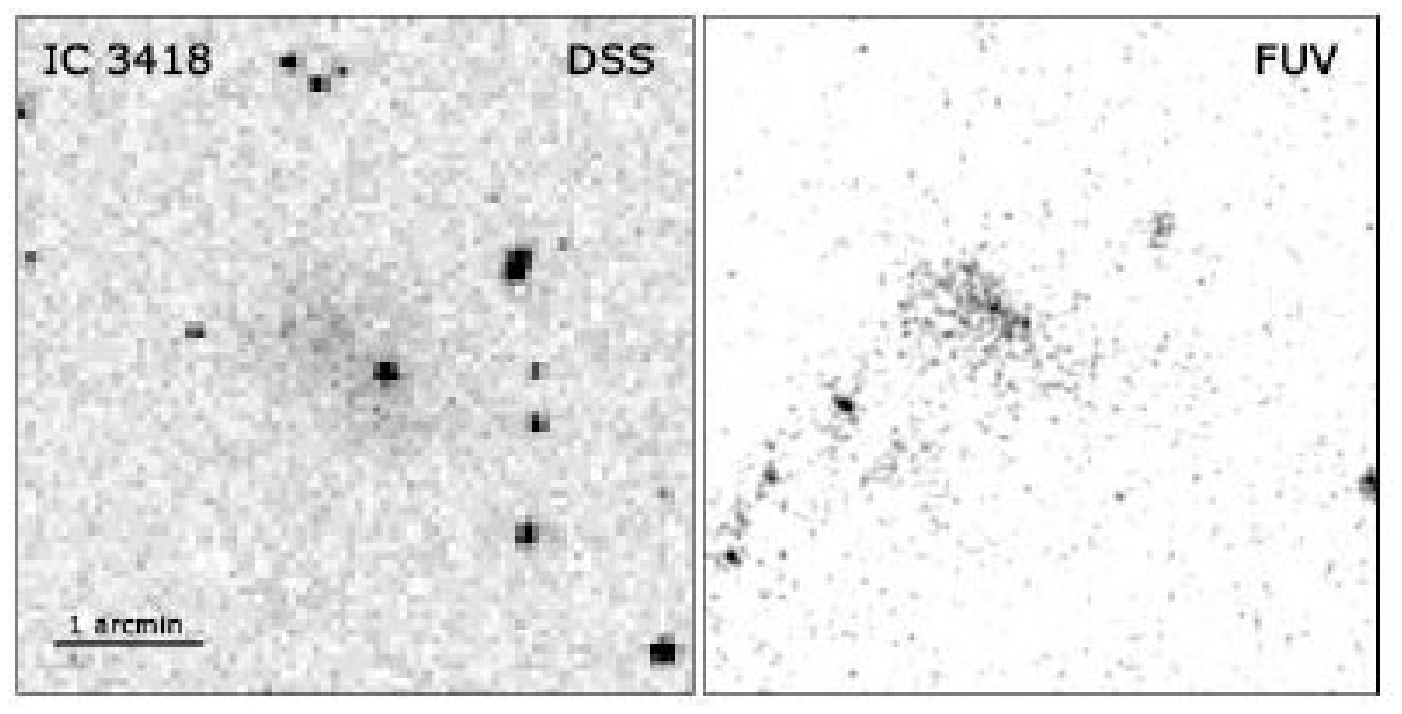}
\caption{Two galaxies, NGC~4330 and IC~3418, have been selected based
on the peculiarities in the $UV$. The Digitized Sky Survey (DSS) image
is shown on the left and the GALEX image at far ultra-violet
wavelength ($\lambda_{\rm center}=$1530~\AA) is shown on the right.  Note that NGC~4330 has
the $UV$ tail to the southwest where we do not find an optical
counterpart.  The $UV$ emission of IC~3418 is displaced from the
optical center and also shows a long $UV$ stream ($>2'$) to the
southeast. \label{fig-uvpec}}
\end{figure}

\begin{figure}
\centering
\includegraphics[width=9cm]{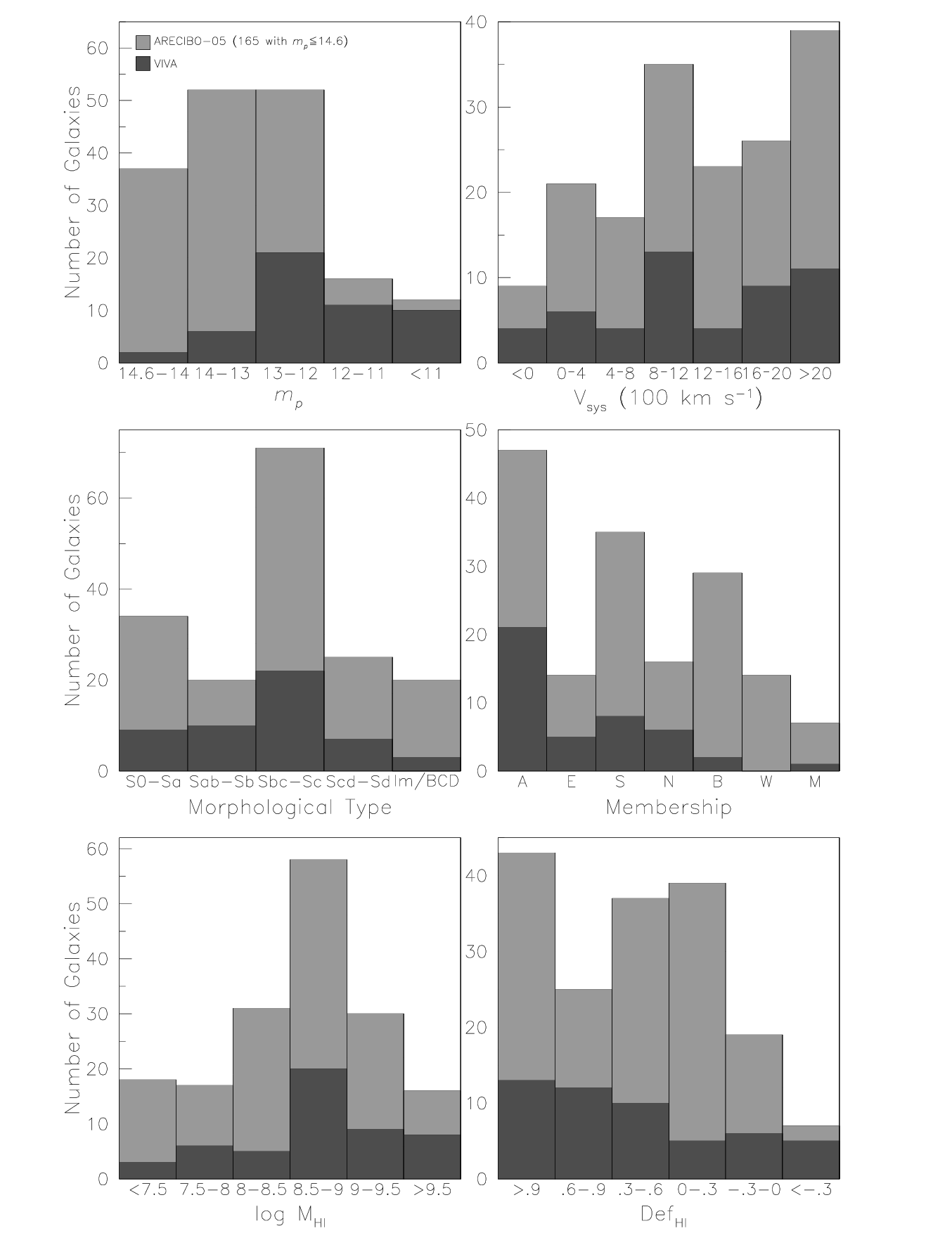}
\caption{The statistics of the global properties of the VIVA sample
compared to the ARECIBO-05 sample \citep{gbvdo05}. The VIVA sample of
53 galaxies is fairly representative for the late-type Virgo galaxies
down to $m_p\approx14.6$ in morphological type, velocity distribution
and H{\sc i} properties. Almost all of  the
selected galaxies are likely to be true members of the
cluster. \label{fig-stat}}
\end{figure}

\subsection{VIVA Sample}
  \citet[][]{kk04a} studied the H$\alpha$ morphology of 84
 Virgo galaxies, including fainter spirals which were never extensively
studied. Using  $R$-band and H$\alpha$ surface brightness profiles,
\citet[][]{kk04b} classified the star formation properties of these 52
Virgo galaxies into several categories; normal, anemic, enhanced, and
truncated.  They argue that these categories are likely to
reflect different evolutionary phases  and different types of interactions with
the cluster environment. Since we wanted to sample galaxies undergoing
different environmental effects 
 we have selected 46 Virgo
galaxies showing a range of star formation properties based on
\citet{kk04b}'s classification.

The survey also probes both the high and the low density regions,
covering angular distances of $\sim1^\circ-12^\circ$ from M87. 
At the distance of 16 Mpc this corresponds to 0.3-3.3~Mpc. 
Thus we probe galaxies out to a distance of  
$\sim$1.6 Abell radii \citep[$r_A\sim1.5~h^{-1}$~Mpc and
$H_0$=71~km~s$^{-1}$~Mpc$^{-1}$;][]{wmap03} or 4 virial radii
\citep[$r_{vir}\sim0.8$~Mpc for Virgo;][]{ts84}.

In addition we  selected two galaxies that are not in the 
\citet[][b]{kk04a} sample. These two galaxies,  
NGC~4330 and IC~3418, show
morphological peculiarities in the $UV$ (the GALEX nearby
galaxy survey). NGC~4330 is a highly inclined disk galaxy and has a
warped $UV$ tail extending beyond the optical disk on one side.
IC~3418 is optically a low surface brightness system, which shows in $UV$
a displacement from the optical disk with an extended broad
$UV$ tail on one side. 

Lastly we have included five galaxies that were found in the
H{\sc i} data cubes of our targets, i.e. they are spatially and in velocity
close to the target galaxies, bringing the total number of galaxies in
the VIVA sample to 53. 

In Figure \ref{fig-viva}, we show the locations of 53 selected
galaxies. The different colors represent different star formation
properties. In Figure~\ref{fig-uvpec}, the optical and the $UV$
images of the two galaxies that were not studied by 
\citet[][b]{kk04a} are shown. The general properties of the 53
galaxies are summarized in Table~\ref{tbl-sample}.
Note that more than half of the VIVA sample is fainter than 12~mag in
$B_T$. Fainter galaxies appear to be more disturbed in H$\alpha$
and may be more vulnerable to environmental effects than more massive
systems.

The VIVA sample probes the full range of the Virgo late type galaxy
population. In Figure \ref{fig-stat}, we compare the general
properties of the VIVA sample to those of a sample of 165 late-type
(Sa-Im-BCD) Virgo galaxies that is complete to $m_p\leq$14.6.  The
comparison sample is selected from a larger sample of 355 late-type
Virgo galaxies, which is complete to $m_p\leq$18.0
\citep{gbvdo05}. We have observed 32\% of the galaxies in the complete
sample of late type galaxies brighter than $m_p$=14.6
\citep[][ARECIBO-05 hereafter]{gbvdo05}, including 75\% of the
galaxies with $m_p\leq12$, 40\% of the galaxies with
$m_p=12-13$ and  eight galaxies which are fainter than $m_p>13$. 
We have good coverage over all velocity bins, and all
morphological types S0/a-Sd.
Based on the subcluster membership classification
by \citet{gbspb99}, all but 3 of the selected galaxies belong to the A, E,
S and N subclusters, and are likely to be true members of the cluster.  
Based on
an H$\alpha$ rotation curve and the H-band Tully Fisher relation
\citet{gbspb99} find that NGC~4380 might belong to the B cloud which
is located at 23~Mpc. The distances to NGC~4424 and NGC~4189 are
highly uncertain, but note that \citet{cortes08} using a stellar kinematics
based Tully-Fisher distance find that NGC~4424 does belong to the 
Virgo Cluster. It remains true that 
distances to some of the individual galaxies in Virgo are controversial
\citep{yfo97,sssgh02}.
The VIVA sample also covers a wide range of
H{\sc i} mass and deficiency.

\section{Observations \& Data Reduction}
\label{sec-obs}
\subsection{Observations}
Since the previous VLA Virgo survey by \citet{cvgbk90}, several
improvements have been made to the VLA.  In 1998 the C array was
replaced by the CS (C short) array by putting one antenna in the
center of the array. This means that CS and D array now have the same
shortest spacing. Compared to the C array the CS array has much better
short spacing baseline coverage while the longer spacings are
unchanged with spacings ranging from 0.035 to 3.4 km.  Hence the CS
array has better surface brightness sensitivity than the former C
array, and the same angular resolution.  In addition new L-band
receivers have been installed, which have a much lower system
temperature.

\begin{table*}
\begin{center}
\caption[VIVA Survey Observing Parameters]{VIVA Survey Observational Parameters}
\label{tbl-obsinfo}
\tiny
\begin{tabular}{cccccccccl}
\hline\hline
\multicolumn{1}{c}{(1)}&
\multicolumn{1}{c}{(2)}&
\multicolumn{1}{c}{(3)}&
\multicolumn{1}{c}{(4)}&
\multicolumn{1}{c}{(5)}&
\multicolumn{1}{c}{(6)}&
\multicolumn{1}{c}{(7)}&
\multicolumn{1}{c}{(8)}&
\multicolumn{1}{c}{(9)}&
\multicolumn{1}{l}{(10)} \\
\multicolumn{1}{c}{Galaxy} &
\multicolumn{1}{c}{Conf.} &
\multicolumn{1}{c}{$\alpha,\delta$$_{2000}$} &
\multicolumn{1}{c}{Obs. Date} &
\multicolumn{1}{c}{Int.} &
\multicolumn{1}{c}{$\Delta$B} &
\multicolumn{1}{c}{$v_{\rm obs}$} &
\multicolumn{1}{c}{Beam (PA)} &
\multicolumn{1}{c}{rms} &
\multicolumn{1}{l}{Ref.}\\
\multicolumn{1}{l}{} &
\multicolumn{1}{l}{} &
\multicolumn{1}{c}{$^{~~h~m~s}+^{\circ}~'~''$} &
\multicolumn{1}{c}{mon-yr} &
\multicolumn{1}{c}{hr} &
\multicolumn{1}{c}{MHz} &
\multicolumn{1}{c}{km~s$^{-1}$} &
\multicolumn{1}{c}{$''\times''$ (deg)} &
\multicolumn{1}{c}{mJy~$beam^{-1}$}&
\multicolumn{1}{c}{}\\
\noalign{\vspace{0.1cm}}
\hline
N4064&\hspace{-0.05cm}CS  &\hspace{-0.05cm}12 04 11.2+18 26 36 &\hspace{-0.05cm} Mar04             & 8   &3.125 &\hspace{-0.05cm}  930 &\hspace{-0.05cm} $16.28\times16.14$ ($-79$)&0.29&  \\
N4189&\hspace{-0.05cm}CS  &\hspace{-0.05cm}12 13 47.2+13 25 29 &\hspace{-0.05cm} Apr04             & 8   &3.125 &\hspace{-0.05cm} 2113 &\hspace{-0.05cm} $16.49\times15.55$ ($+33$)&0.32&  \\
N4192&\hspace{-0.05cm}C/D &\hspace{-0.05cm}12 13 48.1+14 53 46 &\hspace{-0.05cm} Jan91/May92       &3/2.5&3.125 &\hspace{-0.05cm} -160 &\hspace{-0.05cm} $27.99\times25.90$ ($+76$)&1.00&  \\
N4216&\hspace{-0.05cm}C   &\hspace{-0.05cm}12 15 53.7+13 08 42 &\hspace{-0.05cm} Jan91             &4.5  &3.125 &\hspace{-0.05cm}  120 &\hspace{-0.05cm} $16.44\times15.90$ ($-33$)&0.65&  \\
N4222&\hspace{-0.05cm}C   &\hspace{-0.05cm}12 16 22.5+13 18 25 &\hspace{-0.05cm} Jan91             &4.5  &3.125 &\hspace{-0.05cm}  120 &\hspace{-0.05cm} $16.44\times15.90$ ($-33$)&0.54&  \\
N4254&\hspace{-0.05cm}C/D &\hspace{-0.05cm}12 18 49.3+14 25 07 &\hspace{-0.05cm} Mar92/Apr91       & 8   &3.125 &\hspace{-0.05cm} 2408 &\hspace{-0.05cm} $26.78\times24.46$ ($+48$)&0.41&\citet{pvm93}\\
N4293&\hspace{-0.05cm}CS  &\hspace{-0.05cm}12 21 12.9+18 22 57 &\hspace{-0.05cm} Jul05             & 8   &3.125 &\hspace{-0.05cm}  893 &\hspace{-0.05cm} $16.58\times15.40$ ($-69$)&0.33&  \\
N4294&\hspace{-0.05cm}CS/D&\hspace{-0.05cm}12 21 17.8+11 30 40 &\hspace{-0.05cm} Apr04/Nov05,Jan06 &8/3.5&3.125 &\hspace{-0.05cm}  293 &\hspace{-0.05cm} $28.93\times26.74$ ($-34$)&0.29&  \\
N4298&\hspace{-0.05cm}CS  &\hspace{-0.05cm}12 21 32.8+14 36 22 &\hspace{-0.05cm} Jul05             & 8   &3.125 &\hspace{-0.05cm} 1142 &\hspace{-0.05cm} $16.85\times15.72$ ($-59$)&0.35&  \\
N4299&\hspace{-0.05cm}CS/D&\hspace{-0.05cm}12 21 40.5+11 30 11 &\hspace{-0.05cm} Apr04/Nov05,Jan06 &8/3.5&3.125 &\hspace{-0.05cm}  293 &\hspace{-0.05cm} $28.93\times26.74$ ($-34$)&0.29&  \\
N4302&\hspace{-0.05cm}CS  &\hspace{-0.05cm}12 21 42.5+14 35 52 &\hspace{-0.05cm} Jul05             & 8   &3.125 &\hspace{-0.05cm} 1142 &\hspace{-0.05cm} $16.85\times15.72$ ($-59$)&0.35&  \\
N4321&\hspace{-0.05cm}CS/D&\hspace{-0.05cm}12 22 54.8+15 49 21 &\hspace{-0.05cm} Mar04/Mar03       &8/2.3&2.629 &\hspace{-0.05cm} 1571 &\hspace{-0.05cm} $31.10\times28.11$ ($-26$)&0.37&  \\
N4330&\hspace{-0.05cm}CS/D&\hspace{-0.05cm}12 23 17.2+11 22 05 &\hspace{-0.05cm} Aug05/Dec05       & 8   &3.125 &\hspace{-0.05cm} 1565 &\hspace{-0.05cm} $26.36\times23.98$ ($-56$)&0.38&  \\
N4351&\hspace{-0.05cm}CS  &\hspace{-0.05cm}12 24 01.6+12 12 18 &\hspace{-0.05cm} Feb04             & 8   &3.125 &\hspace{-0.05cm} 2315 &\hspace{-0.05cm} $16.77\times16.31$ ($-38$)&0.30&  \\
N4380&\hspace{-0.05cm}CS  &\hspace{-0.05cm}12 25 22.1+10 01 01 &\hspace{-0.05cm} Aug05             & 8   &3.125 &\hspace{-0.05cm}  967 &\hspace{-0.05cm} $16.53\times15.51$ ($-47$)&0.37&  \\
N4383&\hspace{-0.05cm}CS/D&\hspace{-0.05cm}12 25 25.5+16 28 12 &\hspace{-0.05cm} Mar04/Dec05       & 8   &3.125 &\hspace{-0.05cm} 1710 &\hspace{-0.05cm} $44.58\times37.81$ ($-38$)&0.26&  \\
N4388&\hspace{-0.05cm}CS  &\hspace{-0.05cm}12 25 46.6+12 39 44 &\hspace{-0.05cm} Nov02             & 8   &3.125 &\hspace{-0.05cm} 2524 &\hspace{-0.05cm} $17.14\times15.12$ ($+01$)&0.36&  \\
N4394&\hspace{-0.05cm}CS  &\hspace{-0.05cm}12 25 55.6+18 12 50 &\hspace{-0.05cm} Jul05             & 8   &3.125 &\hspace{-0.05cm}  922 &\hspace{-0.05cm} $16.71\times15.17$ ($-59$)&0.32&  \\
N4396&\hspace{-0.05cm}CS/D&\hspace{-0.05cm}12 25 58.8+15 40 17 &\hspace{-0.05cm} Mar04             & 8   &3.125 &\hspace{-0.05cm} -128 &\hspace{-0.05cm} $27.39\times26.84$ ($-03$)&0.28&  \\
N4405&\hspace{-0.05cm}CS  &\hspace{-0.05cm}12 26 07.0+16 10 51 &\hspace{-0.05cm} Jul05             & 8   &3.125 &\hspace{-0.05cm} 1747 &\hspace{-0.05cm} $16.59\times15.36$ ($-61$)&0.36&  \\
N4402&\hspace{-0.05cm}CS  &\hspace{-0.05cm}12 26 07.8+13 06 43 &\hspace{-0.05cm} Jan03             & 8   &3.125 &\hspace{-0.05cm}  200 &\hspace{-0.05cm} $17.07\times15.27$ ($+07$)&0.33&\citet{ckvgv05}\\
I3355&\hspace{-0.05cm}CS  &\hspace{-0.05cm}12 26 51.1+13 10 33 &\hspace{-0.05cm} Jan03             & 8   &3.125 &\hspace{-0.05cm}  200 &\hspace{-0.05cm} $17.07\times15.27$ ($+07$)&0.36&  \\
N4419&\hspace{-0.05cm}CS  &\hspace{-0.05cm}12 26 56.4+15 02 50 &\hspace{-0.05cm} Oct02             & 8   &3.125 &\hspace{-0.05cm} -261 &\hspace{-0.05cm} $16.35\times15.26$ ($-09$)&0.32&  \\
N4424&\hspace{-0.05cm}CS  &\hspace{-0.05cm}12 27 11.5+09 25 14 &\hspace{-0.05cm} Apr04             & 8   &3.125 &\hspace{-0.05cm}  439 &\hspace{-0.05cm} $17.59\times15.53$ ($+36$)&0.39&  \\
N4450&\hspace{-0.05cm}CS  &\hspace{-0.05cm}12 28 29.5+17 05 06 &\hspace{-0.05cm} Jul05             & 8   &3.125 &\hspace{-0.05cm} 1954 &\hspace{-0.05cm} $16.45\times15.61$ ($-79$)&0.36&  \\
I3392&\hspace{-0.05cm}CS  &\hspace{-0.05cm}12 28 43.3+14 59 58 &\hspace{-0.05cm} Oct02             & 8   &3.125 &\hspace{-0.05cm} 1687 &\hspace{-0.05cm} $17.06\times15.06$ ($+15$)&0.28&  \\
N4457&\hspace{-0.05cm}CS  &\hspace{-0.05cm}12 28 58.9+03 34 14 &\hspace{-0.05cm} Jul05             & 8   &3.125 &\hspace{-0.05cm}  882 &\hspace{-0.05cm} $17.43\times16.28$ ($-36$)&0.47&  \\
I3418&\hspace{-0.05cm}CS  &\hspace{-0.05cm}12 29 43.8+11 24 09 &\hspace{-0.05cm} Aug05             & 8   &3.125 &\hspace{-0.05cm}38$^*$&\hspace{-0.05cm} $16.67\times15.77$ ($-64$)&0.43&  \\
N4501&\hspace{-0.05cm}C   &\hspace{-0.05cm}12 31 59.0+14 25 10 &\hspace{-0.05cm} Jan91             & 5   &3.125 &\hspace{-0.05cm} 2280 &\hspace{-0.05cm} $16.99\times16.56$ ($+51$)&0.57&  \\ 
N4522&\hspace{-0.05cm}CS  &\hspace{-0.05cm}12 33 39.7+09 10 31 &\hspace{-0.05cm} Mar00             & 8   &3.125 &\hspace{-0.05cm} 2330 &\hspace{-0.05cm} $18.88\times15.20$ ($-43$)&0.40&\citet{kvgv04}\\
N4532&\hspace{-0.05cm}C   &\hspace{-0.05cm}12 34 19.3+06 28 04 &\hspace{-0.05cm} Dec94             &5.5  &3.125 &\hspace{-0.05cm} 2000 &\hspace{-0.05cm} $17.36\times16.19$ ($+22$)&0.33&\citet{hoff99}\\
N4535&\hspace{-0.05cm}C/D &\hspace{-0.05cm}12 34 20.3+08 12 01 &\hspace{-0.05cm} Jan91/Jan94       & 5   &3.125 &\hspace{-0.05cm} 1950 &\hspace{-0.05cm} $24.98\times24.07$ ($+22$)&0.60&  \\
N4533&\hspace{-0.05cm}CS  &\hspace{-0.05cm}12 34 22.0+02 19 31 &\hspace{-0.05cm} Mar04             & 8   &3.125 &\hspace{-0.05cm} 1790 &\hspace{-0.05cm} $18.04\times16.18$ ($-12$)&0.33&  \\
N4536&\hspace{-0.05cm}CS  &\hspace{-0.05cm}12 34 27.0+02 11 17 &\hspace{-0.05cm} Mar04             & 8   &3.125 &\hspace{-0.05cm} 1790 &\hspace{-0.05cm} $18.04\times16.18$ ($-12$)&0.33&  \\
V1581&\hspace{-0.05cm}C   &\hspace{-0.05cm}12 34 45.3+06 18 02 &\hspace{-0.05cm} Dec94             &5.5  &3.125 &\hspace{-0.05cm} 2000 &\hspace{-0.05cm} $17.36\times16.19$ ($+22$)&0.33&\citet{hoff99}\\
N4548&\hspace{-0.05cm}CS  &\hspace{-0.05cm}12 35 26.4+14 29 47 &\hspace{-0.05cm} Mar04             & 5   &3.125 &\hspace{-0.05cm}  451 &\hspace{-0.05cm} $16.59\times15.81$ ($-37$)&0.30&  \\
N4561&\hspace{-0.05cm}C   &\hspace{-0.05cm}12 36 08.5+19 19 25 &\hspace{-0.05cm} Sep89             & 3   &3.125 &\hspace{-0.05cm} 1400 &\hspace{-0.05cm} $15.44\times14.00$ ($-39$)&1.40&  \\
N4567&\hspace{-0.05cm}CS  &\hspace{-0.05cm}12 36 32.7+11 15 28 &\hspace{-0.05cm} Jul05             & 8   &3.125 &\hspace{-0.05cm} 2265 &\hspace{-0.05cm} $17.12\times15.98$ ($-53$)&0.36&  \\
N4568&\hspace{-0.05cm}CS  &\hspace{-0.05cm}12 36 34.3+11 14 19 &\hspace{-0.05cm} Jul05             & 8   &3.125 &\hspace{-0.05cm} 2265 &\hspace{-0.05cm} $17.12\times15.98$ ($-53$)&0.36&  \\
N4569&\hspace{-0.05cm}CS  &\hspace{-0.05cm}12 36 49.8+13 09 46 &\hspace{-0.05cm} Apr04             & 8   &3.125 &\hspace{-0.05cm} -235 &\hspace{-0.05cm} $16.38\times16.27$ ($+10$)&0.33&  \\
N4579&\hspace{-0.05cm}CS/D&\hspace{-0.05cm}12 37 43.3+11 49 05 &\hspace{-0.05cm} Feb04/Mar03       &8/2.3&2.629 &\hspace{-0.05cm} 1519 &\hspace{-0.05cm} $42.42\times34.49$ ($+37$)&0.45&  \\
N4580&\hspace{-0.05cm}CS  &\hspace{-0.05cm}12 37 48.4+05 22 10 &\hspace{-0.05cm} May04             & 8   &3.125 &\hspace{-0.05cm} 1036 &\hspace{-0.05cm} $17.37\times16.34$ ($-03$)&0.31&  \\
N4606&\hspace{-0.05cm}CS  &\hspace{-0.05cm}12 40 57.6+11 54 40 &\hspace{-0.05cm} Aug05             & 8   &6.25  &\hspace{-0.05cm} 1961 &\hspace{-0.05cm} $16.68\times15.49$ ($-54$)&0.29&  \\
N4607&\hspace{-0.05cm}CS  &\hspace{-0.05cm}12 41 12.4+11 53 09 &\hspace{-0.05cm} Aug05             & 8   &6.25  &\hspace{-0.05cm} 1961 &\hspace{-0.05cm} $16.68\times15.49$ ($-54$)&0.29&  \\
N4651&\hspace{-0.05cm}CS  &\hspace{-0.05cm}12 43 42.6+16 23 36 &\hspace{-0.05cm} Mar04             & 8   &3.125 &\hspace{-0.05cm}  804 &\hspace{-0.05cm} $16.67\times16.25$ ($-69$)&0.40&  \\
N4654&\hspace{-0.05cm}C   &\hspace{-0.05cm}12 43 56.5+13 07 33 &\hspace{-0.05cm} Mar92             & 8   &3.125 &\hspace{-0.05cm} 1088 &\hspace{-0.05cm} $16.14\times15.52$ ($+35$)&0.45&\citet{pm95}\\
N4689&\hspace{-0.05cm}CS  &\hspace{-0.05cm}12 47 45.5+13 45 46 &\hspace{-0.05cm} Mar04             & 8   &3.125 &\hspace{-0.05cm} 1611 &\hspace{-0.05cm} $16.71\times15.85$ ($-37$)&0.27&  \\
V2062&\hspace{-0.05cm}CS  &\hspace{-0.05cm}12 47 59.9+10 58 33 &\hspace{-0.05cm} May04             & 8   &3.125 &\hspace{-0.05cm} 1117 &\hspace{-0.05cm} $16.35\times16.12$ ($+12$)&0.38&  \\
N4694&\hspace{-0.05cm}CS  &\hspace{-0.05cm}12 48 15.1+10 58 58 &\hspace{-0.05cm} May04             & 8   &3.125 &\hspace{-0.05cm} 1117 &\hspace{-0.05cm} $16.35\times16.12$ ($+12$)&0.38&  \\
N4698&\hspace{-0.05cm}CS  &\hspace{-0.05cm}12 48 22.9+08 29 14 &\hspace{-0.05cm} Apr04             & 8   &3.125 &\hspace{-0.05cm} 1000 &\hspace{-0.05cm} $16.96\times16.20$ ($-29$)&0.35&  \\
N4713&\hspace{-0.05cm}C   &\hspace{-0.05cm}12 49 58.0+05 18 38 &\hspace{-0.05cm} Sep89             & 2   &3.125 &\hspace{-0.05cm}  655 &\hspace{-0.05cm} $25.95\times22.13$ ($+67$)&1.96&  \\
N4772&\hspace{-0.05cm}CS  &\hspace{-0.05cm}12 53 29.1+02 10 06 &\hspace{-0.05cm} Jul05             & 8   &3.125 &\hspace{-0.05cm} 1040 &\hspace{-0.05cm} $17.80\times15.41$ ($-36$)&0.36&  \\
N4808&\hspace{-0.05cm}C/D &\hspace{-0.05cm}12 55 49.5+04 18 14 &\hspace{-0.05cm} Sep89/Nov05       & 2   &3.125 &\hspace{-0.05cm}  760 &\hspace{-0.05cm} $40.01\times35.53$ ($+08$)&0.59&  \\
\hline								  
\end{tabular}							  
\end{center}							  
\noindent								  
{\footnotesize {\it Footnote for Table 2:}
(1) NGC, IC or VCC names; (2) VLA configuration(s); (3) Field center; (4) Observation dates in month and year;
(5) Observation duration; (6) Total bandwidth. Note that NGC~4321 and NGC~4579 were observed in the way that
two 3.125-IF's were offset with each other with 14 channels overlapping around the velocity where the observations
were centered at, resulting in the total bandwidth of 2.629 MHz; (7) Heliocentric velocity of the central channel
using optical definition; (8) Synthesized beam FWHM (PA of the beam); (9) The rms per channel of the final
cube imaged with robust=1; (10) The literature where the same data have been presented.}
\end{table*}

All the new observations were done with the VLA in CS array, in a few
cases supplemented by D array.  Most galaxies were observed with 3.125
MHz bandwidth. The correlator was configured to produce 127 channels
and two polarizations. Online Hanning smoothing was applied after
which every other channel was discarded. This resulted in 63
independent channels with a velocity resolution of roughly
10~km~s$^{-1}$. Two galaxies, NGC 4606 and NGC 4607, were observed in
one pointing.  Since their velocities differ by 600 km~s$^{-1}$ a total
bandwidth of 6.25 MHz was used, two polarizations and no online
Hanning smoothing, resulting in 63 channels with 21 km~s$^{-1}$ width.  In
addition to these new observations we reprocessed archival data on
Virgo galaxies that were of comparable quality, taken in C or CS
array, including nine galaxies that were observed by us earlier.  We
have reached a column density sensitivity of
3-5$\times10^{19}$~cm$^{-2}$ in 3$\sigma$ per channel with a typical
spatial resolution of 15-16$''$ ($\lesssim1.1~$kpc at the Virgo
distance). This is a factor 3 and 4 better in  spatial and spectral
resolution respectively than the data of the previous 
VLA survey by \citet{cvgbk90}.

For some galaxies we suspected that we were missing extended diffuse
emission based on the images or based on a  comparison with single dish
measurements. For those galaxies we either obtained D array data 
ourselves or we used archival data of comparable quality.
For eleven galaxies in total we use the combined C/CS and D array data.
Observing parameters are summarized in Table~\ref{tbl-obsinfo}.

\subsection{Data Reduction}
Both the new and the archival data were reduced in the same way using
the Astronomical Imaging Processing System (AIPS).  After flux, phase and 
bandpass calibration, the continuum was subtracted by making a linear fit
to the $uv$ data for a range of line-free channels at both sides of
the band. High $uv$ points caused by interference were flagged after
continuum subtraction.  Two galaxies (NGC~4321 and NGC~4579) were
observed with overlapping IFs, which were offset by $\sim120~$km~s$^{-1}$. 
For those two galaxies we
converted the two IFs into one long spectrum by averaging the overlapping
channels, using {\tt UJOIN}. Several channels at both edges of the
IFs, where the spectral frequency response drops rather steeply,
were not included.

First we  made low resolution cubes covering a large
field of view ($1.4\times1.4$~deg$^2$) to search for H{\sc i} emission
of sources far away from the field center. We found five galaxies that
were fully covered in velocity and we added those to the VIVA survey (see
\S\ref{sec-sample}.2 and Table~\ref{tbl-sample}).

The final image cubes were made using {\tt ROBUST=1} \citep{bri95}  
to maximize  sensitivity while keeping good spatial resolution.
The cubes were cleaned to remove the sidelobes.
The final cubes were about 40 arcmin in size, slightly larger than the FWHP
(30 arcmin) of the primary beam of the VLA. 

All but IC~3418 were previously detected with single-dish. In our
survey IC~3418 is also the only target that was not detected in H{\sc
i}, down to $\sim8\times10^6 ~M_\odot$ per beam in 3 $\sigma$,
assuming a profile width of 100~km~s$^{-1}$. However, our sensitivity
would be less to a huge/diffuse H{\sc i} disk, which would be smooth
over 10 arcminutes in a single velocity channel. 

The total H{\sc i} image, the intensity weighted velocity field and the
velocity dispersion image were also produced using AIPS by taking
moments along the frequency axis (0th, 1st and 2nd moments).  The AIPS
task {\tt MOMNT} allows you to create a mask to blank the images at a given
cutoff level.  In creating a mask, we applied Gaussian and Hanning
smoothing in spatial and in velocity, respectively, to maximize the
signal-to-noise. We normally used $1\sim2\times$rms of the
cube as the cutoff. We applied those masks to the full resolution
cubes and  calculated moments on the full
resolution blanked cubes. Once image cubes and moment maps were
obtained, we performed further analysis using the Groningen Image
Processing SYstem (GIPSY).

We also made 1.4~GHz continuum images by averaging the line free
channels. In order to reduce the effects of interfering sources, which
may cause substantial sidelobes especially at low frequencies, we have
used the AIPS task {\tt PEELR}. It iteratively attempts to calibrate
on multi fields around bright continuum sources (self-calibration),
subtract the sources in those fields from the self-calibrated data, undo the
field-specific calibration from the residual data, and it finally
restores all fields to the residual data.  We used the same weighting
scheme ({\tt ROBUST=1}) as for the H{\sc i} images.  The quality of
our continuum data varies depending on the number of line free
channels of individual target galaxies.

\section{H{\sc i} Atlas: Descriptions}
\label{sec-atlas}
\begin{figure}
\plotone{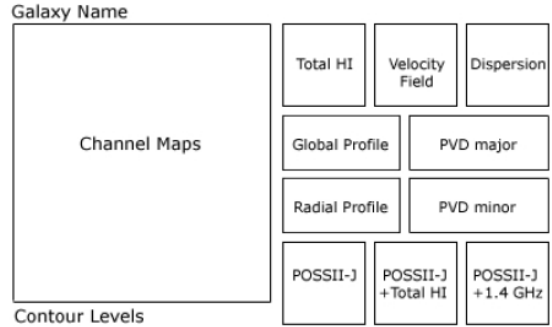}
\caption{This diagram illustrates how the figures are arranged in the
atlas, which is appended at the end.\label{fig-atlas}}
\end{figure}

In this section, we describe the atlas which is appended at the
end. Individual galaxies are presented in separate pages, except
IC~3418 which we did not detect in H{\sc i} and thus is not included
in the atlas. In Figure~\ref{fig-atlas} we show the outline of each
page. The contour levels of the H{\sc i} emission in the channel maps,
the H{\sc i} surface density in the total H {\sc i} image, the
velocities of the velocity field and velocity dispersion images, and 1.4~GHz 
continuum emission are shown at the left-bottom of
each page.

\bigskip
\centerline{\it Channel maps} 
We present the cubes of $\Delta
v\approx10.4$~km~s$^{-1}$ for all galaxies, except NGC~4606 and NGC~4607,
which have channels that are $\Delta
v\approx 21$~km~s$^{-1}$.
The lowest contours 
represent $\pm2\sigma$, where $\sigma$ is the rms per beam per
channel. The synthesized beam is shown at the left-bottom corner of
the first panel on the top-left, i.e. in the channel with the largest
velocity. In the same channel, we indicate the optical size, 
$D_{25}$,
the optical position angle and inclination with an
ellipse. In every channel, the optical center is shown with a cross
and, the velocity (in km~s$^{-1}$) is shown in the top-right corner.

\bigskip
\centerline{\it H{\small I} Distribution and Velocities} 
On the top of
the right half, the H{\sc i} surface density distribution (left), the
intensity weighted velocity field (middle) and the velocity dispersion (right)
are presented with contours overlaid on their own grayscale. The
synthesized beam is shown in the bottom-left of the H{\sc i} surface
density image. In the velocity field, the thick-white line represents
the H{\sc i} systemic velocity, $V_{\rm HI}$ measured using the linewidths
(\S~\ref{sec-hiprop}). In the total  H{\sc i} image  and the
velocity field, the optical major and minor axes ($D_{25}$) are shown
as dotted lines. For 11 systems the optically derived position angle is 
uncertain because they are either close to face-on with $i\lesssim45$ or highly
warped. For those galaxies we determined the PA kinematically using a 
tilted-ring model fit on the inner regions of the  H{\sc i} velocity field,
where the kinematics is fairly regular. 
Those galaxies
are NGC~4189, NGC~4254, NGC~4299, NGC~4321, NGC~4351, NGC~4394,
NGC~4457, VCC~1581, NGC~4689, VCC~2062. The kinematically derived PA's 
are used throughout the atlas for these galaxies.

\bigskip
\centerline{\it Global \& Radial profiles} 
The H{\sc i} flux density
profile and the azimuthally averaged radial H{\sc i} surface density
distribution are shown  below the H{\sc i} surface density
distribution. To make the global profiles we measured in each channel   
the flux density ($F_{\rm HI}$) in a
tight box around the H{\sc i} area where emission is seen. Throughout the
cube we used the
areas outside of the H{\sc i} emission
to measure the rms as shown with the error bars.  The H{\sc i} systemic
velocity, measured from the H{\sc i} linewidth (\S~\ref{sec-hiprop})
is indicated with an upward-pointing arrow.

The azimuthally averaged H{\sc i} profiles have been derived by fitting tilted
ring models, adopting the optically defined center, position angle, and
inclination.  For galaxies where we derived the
position angle based on the H{\sc i} velocity field, the kinematic PA's were
used. The surface density profiles are corrected to face-on and are given 
in $M_\odot~$pc$^{-2}$. The galactocentric radius is given
in kpc. The dashed line is the fit using the entire disk, 
while  open circles and  solid triangles
are fits to the east and the west side separately.
  For comparison, the optical size of the disk, $R_{25}$, is
indicated with an upper arrow. Of course azimuthally averaged profiles
can be misleading, and especially for some of the highly inclined galaxies with 
known extraplanar gas, such as NGC~4522, NGC~4569, NGC~4330 and NGC~4402,
the profiles are of limited use.

\bigskip
\centerline{\it PVDs} 
On the right side of the global and radial
profiles, the position-velocity diagrams (PVDs) along the major axis
(upper) and the minor axis (lower) are presented. Again, we adopted
the optical center and the position angle for most of the sample to
make slices as shown on the upper-right corner of the figures, but we
used the kinematically derived PA's for the 11 galaxies mentioned above.  
The  optical center of the cut and
the H{\sc i} systemic velocity as derived from the linewidths
(\S~\ref{sec-hiprop}) are indicated with dashed lines.

\bigskip
\centerline{\it Miscellaneous} 
On the bottom of the right side of each
page, we present the POSSII$-J$ (optical $B$) image (left) and
overlays of the H{\sc i} (middle) and 1.4~GHz continuum contours
(right) on the optical image. The optical image itself is shown in high
contrast to bring out better  the structure of the inner stellar disk,
while lower contrast images are used  for the overlays to bring out the extent
of the stellar disk in comparison to the extent of the H{\sc i}
 and the radio continuum emission.

Fully reduced H{\sc i} data (cubes, moments, XV slices) from the VIVA
survey are available
in the following URL: {\tt http://www.astro.yale.edu/viva}.
\begin{table*}
\begin{center}
\caption[VIVA Sample and H{\sc i} Properties]{VIVA Sample and H{\sc i} Properties}
\label{tbl-hiprop}
\tiny
\begin{tabular}{crrrrrrrrrcccr}
\hline\hline
\multicolumn{1}{c}{(1)} &
\multicolumn{1}{c}{(2)} &
\multicolumn{1}{c}{(3)} &
\multicolumn{1}{c}{(4)} &
\multicolumn{1}{c}{(5)} &
\multicolumn{1}{c}{(6)} &
\multicolumn{1}{c}{(7)} &
\multicolumn{1}{c}{(8)} &
\multicolumn{1}{c}{(9)} &
\multicolumn{1}{c}{(10)} &
\multicolumn{1}{c}{(11)} &
\multicolumn{1}{c}{(12)} &
\multicolumn{1}{c}{(13)} &
\multicolumn{1}{c}{(14)} \\
\multicolumn{1}{c}{Galaxy} &
\multicolumn{1}{c}{$S_{\rm HI}$} &
\multicolumn{1}{c}{$M_{\rm HI}$} &
\multicolumn{1}{c}{$W_{\rm HI}^{20\%}$} &
\multicolumn{1}{c}{$W_{\rm HI}^{50\%}$} &
\multicolumn{1}{c}{$V_{\rm HI}$} &
\multicolumn{1}{c}{$D_{\rm HI}^{\rm iso}{^{\Delta E}_{\Delta W}}$} &
\multicolumn{1}{c}{$D_{\rm HI}^{\rm eff}{^{\Delta E}_{\Delta W}}$} &
\multicolumn{1}{c}{$def_{\rm HI}$} &
\multicolumn{1}{c}{log$\frac{M_{\rm HI}}{L_{B}}$} &
\multicolumn{1}{c}{log$\frac{M_{\rm HI}}{L_{K}}$} &
\multicolumn{1}{c}{$\frac{D_{\rm HI}^{\rm iso}}{D_{B}}$} &
\multicolumn{1}{c}{$\frac{D_{\rm HI}^{\rm iso}}{D_{K}}$} &
\multicolumn{1}{c}{$F_{1.4~\rm GHz}$} \\
\multicolumn{1}{l}{} &
\multicolumn{1}{c}{Jy km s$^{-1}$} &
\multicolumn{1}{c}{$10^8~M_\odot$} &
\multicolumn{3}{c}{--- km s$^{-1}$ ---} &
\multicolumn{2}{c}{--- arcmin ---} &
\multicolumn{1}{c}{} &
\multicolumn{1}{c}{} &
\multicolumn{1}{c}{} &
\multicolumn{1}{c}{} &
\multicolumn{1}{c}{} &
\multicolumn{1}{c}{mJy}\\
\hline
NGC 4064 &  0.66$\pm$0.39 &  0.40$\pm$ 0.24 &\hspace{-0.35cm} 175 & 110~ &  934~~ &\hspace{-0.45cm} 0.91$^{+0.02}_{-0.03}$ &  $0.86^{-0.06}_{+0.01}$ &  1.79$\pm$0.20 &\hspace{-0.45cm}  -2.15 & -2.61 & 0.21 & 0.38 &  10.1$\pm$~1.2 \\
NGC 4189 &  9.40$\pm$1.23 &  5.67$\pm$ 0.74 &\hspace{-0.35cm} 245 & 232~ & 2131~~ &\hspace{-0.45cm} 2.86$^{+0.04}_{-0.07}$ &  $1.51^{+0.04}_{-0.03}$ &  0.25$\pm$0.04 &\hspace{-0.45cm}  -0.90 & -1.26 & 1.19 & 1.47 &  17.1$\pm$~0.8 \\
NGC 4192 & 70.50$\pm$6.08 & 42.52$\pm$ 3.67 &\hspace{-0.35cm} 476 & 448~ & -156~~ &\hspace{-0.45cm} 9.92$^{+0.45}_{-0.82}$ &  $9.47^{-0.44}_{-0.27}$ &  0.51$\pm$0.20 &\hspace{-0.45cm}  -0.65 & -1.29 & 1.01 & 1.72 &  78.5$\pm$~4.2 \\
NGC 4216 & 29.34$\pm$6.17 & 17.70$\pm$ 3.72 &\hspace{-0.35cm} 538 & 518~ &  137~~ &\hspace{-0.45cm} 6.17$^{+0.28}_{-5.97}$ &  $4.45^{+0.24}_{-0.23}$ &  0.76$\pm$0.20 &\hspace{-0.45cm}  -1.01 & -1.84 & 0.76 & 0.85 &  14.1$\pm$~1.1 \\
NGC 4222 & 10.59$\pm$2.45 &  6.39$\pm$ 1.48 &\hspace{-0.35cm} 246 & 229~ &  228~~ &\hspace{-0.45cm} 3.71$^{-0.04}_{+0.04}$ &  $2.36^{+0.04}_{+0.00}$ &  0.32$\pm$0.04 &\hspace{-0.45cm}  -0.31 & -0.79 & 1.12 & 1.84 &     $<$0.3~~~  \\
NGC 4254 & 73.42$\pm$7.00 & 44.28$\pm$ 4.22 &\hspace{-0.35cm} 250 & 218~ & 2395~~ &\hspace{-0.45cm} 0.15$^{+0.05}_{-1.89}$ &  $4.49^{-0.04}_{-0.41}$ & -0.10$\pm$0.02 &\hspace{-0.45cm}  -0.85 & -1.23 & 1.88 & 2.98 & 449.5$\pm$~8.8 \\
NGC 4293 &  0.44$\pm$0.50 &  0.27$\pm$ 0.30 &\hspace{-0.35cm} 252 & 242~ &  929~~ &\hspace{-0.45cm}         ....~~~~~      &  $1.02^{-0.24}_{+0.07}$ &  2.25$\pm$0.20 &\hspace{-0.45cm}  -2.75 & -3.24 & ...~ & ...$^\dag$~ &  17.7$\pm$~2.7 \\
NGC 4294 & 27.08$\pm$2.03 & 16.33$\pm$ 1.22 &\hspace{-0.35cm} 221 & 192~ &  363~~ &\hspace{-0.45cm} 4.38$^{-0.05}_{+0.21}$ & $15.59^{-0.04}_{-8.17}$ & -0.11$\pm$0.02 &\hspace{-0.45cm}  -0.44 & -0.48 & 1.37 & 2.44 &  26.7$\pm$~1.1 \\
NGC 4298 &  8.21$\pm$1.46 &  4.95$\pm$ 0.88 &\hspace{-0.35cm} 240 & 225~ & 1136~~ &\hspace{-0.45cm} 3.03$^{-0.01}_{+0.08}$ &  $1.72^{+0.04}_{-0.03}$ &  0.41$\pm$0.02 &\hspace{-0.45cm}  -1.15 & -1.64 & 0.95 & 1.21 &  16.8$\pm$~0.8 \\
NGC 4299 & 18.20$\pm$0.84 & 10.98$\pm$ 0.51 &\hspace{-0.35cm} 137 &  93~ &  227~~ &\hspace{-0.45cm} 3.45$^{-0.31}_{+0.49}$ &  $5.19^{-3.24}_{-0.48}$ & -0.43$\pm$0.02 &\hspace{-0.45cm}  -0.47 & -0.42$^\ast$ & 2.03 & 0.10 &  18.7$\pm$~0.9 \\
NGC 4302 & 24.60$\pm$3.95 & 14.84$\pm$ 2.38 &\hspace{-0.35cm} 383 & 362~ & 1146~~ &\hspace{-0.45cm} 5.21$^{+0.16}_{-0.09}$ &  $3.55^{+0.14}_{-0.14}$ &  0.39$\pm$0.02 &\hspace{-0.45cm}  -0.49 & -1.39 & 0.95 & 1.11 &  31.6$\pm$~1.7 \\
NGC 4321 & 47.71$\pm$2.67 & 28.78$\pm$ 1.61 &\hspace{-0.35cm} 268 & 244~ & 1571~~ &\hspace{-0.45cm} 7.66$^{+0.19}_{-0.31}$ &  $4.88^{+0.14}_{-0.20}$ &  0.35$\pm$0.12 &\hspace{-0.45cm}  -1.17 & -1.54 & 1.03 & 1.52 & 284.7$\pm$~8.0 \\
NGC 4330 &  7.37$\pm$1.73 &  4.45$\pm$ 1.04 &\hspace{-0.35cm} 275 & 247~ & 1566~~ &\hspace{-0.45cm} 2.72$^{-0.05}_{+0.07}$ &  $5.54^{-1.54}_{+0.36}$ &  0.80$\pm$0.04 &\hspace{-0.45cm}  -0.76 & -1.17 & 0.60 & 0.89 &  18.7$\pm$~1.0 \\
NGC 4351 &  4.96$\pm$0.67 &  2.99$\pm$ 0.40 &\hspace{-0.35cm} 124 &  99~ & 2319~~ &\hspace{-0.45cm} 2.44$^{-0.31}_{+0.08}$ &  $1.46^{-0.44}_{+0.06}$ &  0.23$\pm$0.20 &\hspace{-0.45cm}  -0.96 & -0.95 & 1.22 & 1.90 &   2.0$\pm$~0.2 \\
NGC 4380 &  2.10$\pm$0.92 &  1.27$\pm$ 0.55 &\hspace{-0.35cm} 291 & 274~ &  969~~ &\hspace{-0.45cm} 2.21$^{+0.01}_{-0.01}$ &  $1.63^{+0.04}_{-0.11}$ &  1.13$\pm$0.20 &\hspace{-0.45cm}  -1.48 & -2.15 & 0.63 & 0.93 &     $<$0.6~~~  \\
NGC 4383 & 48.38$\pm$5.15 & 29.18$\pm$ 3.10 &\hspace{-0.35cm} 233 & 213~ & 1708~~ &\hspace{-0.45cm} 8.39$^{-0.06}_{+0.24}$ &  $7.80^{-1.53}_{+0.73}$ & -0.81$\pm$0.20 &\hspace{-0.45cm}  -0.11 & -0.42 & 4.19 & 7.19 &  44.3$\pm$~4.1 \\
NGC 4388 &  6.10$\pm$3.66 &  3.68$\pm$ 2.21 &\hspace{-0.35cm} 396 & 368~ & 2519~~ &\hspace{-0.45cm} 3.10$^{+0.02}_{+0.00}$ &  $2.05^{-0.14}_{+0.21}$ &  1.16$\pm$0.12 &\hspace{-0.45cm}  -1.39 & -1.89 & 0.55 & 1.00 & 169.0$\pm$17.1 \\
NGC 4394 &  7.27$\pm$0.51 &  4.38$\pm$ 0.31 &\hspace{-0.35cm} 173 & 162~ &  914~~ &\hspace{-0.45cm} 3.74$^{+0.01}_{-0.02}$ &  $2.57^{-0.04}_{+0.06}$ &  0.62$\pm$0.20 &\hspace{-0.45cm}  -1.32 & -1.76 & 1.04 & 1.15 &     $<$0.2~~~  \\
NGC 4396 & 14.31$\pm$2.81 &  8.63$\pm$ 1.69 &\hspace{-0.35cm} 213 & 199~ & -121~~ &\hspace{-0.45cm} 3.94$^{-0.09}_{+0.09}$ &  $3.16^{-0.74}_{+0.19}$ &  0.30$\pm$0.04 &\hspace{-0.45cm}  -0.49 & -0.44 & 1.20 & 3.27 &  21.0$\pm$~0.8 \\
NGC 4405 &  0.75$\pm$0.44 &  0.45$\pm$ 0.27 &\hspace{-0.35cm} 169 & 155~ & 1740~~ &\hspace{-0.45cm} 0.92$^{+0.02}_{-0.02}$ &  $0.38^{+0.04}_{-0.01}$ &  0.95$\pm$0.20 &\hspace{-0.45cm}  -1.78 & -2.27 & 0.51 & 0.70 &   5.6$\pm$~0.4 \\
NGC 4402 &  6.13$\pm$2.60 &  3.70$\pm$ 1.57 &\hspace{-0.35cm} 288 & 249~ &  236~~ &\hspace{-0.45cm} 2.92$^{-0.29}_{+0.07}$ &  $2.11^{-0.14}_{-0.03}$ &  0.74$\pm$0.12 &\hspace{-0.45cm}  -1.06 & -1.76 & 0.75 & 0.80 &  68.3$\pm$~3.4 \\
 IC 3355 &  3.20$\pm$0.19 &  1.93$\pm$ 0.11 &\hspace{-0.35cm}  61 &  38~ &  -10~~ &\hspace{-0.45cm} 1.82$^{-0.07}_{+0.05}$ &  $9.42^{-0.24}_{+0.17}$ &  0.09$\pm$0.06 &\hspace{-0.45cm}  -0.29 & -0.06$^\ast$ & 1.65 & 0.05 &     $<$0.4~~~  \\
NGC 4419 &  0.96$\pm$1.37 &  0.58$\pm$ 0.83 &\hspace{-0.35cm} 382 & 367~ & -200~~ &\hspace{-0.45cm} 1.17$^{+0.03}_{-0.04}$ &  $1.11^{-0.14}_{+0.12}$ &  1.37$\pm$0.20 &\hspace{-0.45cm}  -2.07 & -2.85 & 0.35 & 0.41 &  50.7$\pm$~7.4 \\
NGC 4424 &  3.19$\pm$0.52 &  1.92$\pm$ 0.31 &\hspace{-0.35cm} 108 &  56~ &  434~~ &\hspace{-0.45cm} 1.42$^{-0.11}_{+0.11}$ &  $7.61^{-0.04}_{-0.78}$ &  0.97$\pm$0.20 &\hspace{-0.45cm}  -1.42 & -1.73 & 0.39 & 0.65 &   6.5$\pm$~0.7 \\
NGC 4450 &  4.72$\pm$0.88 &  2.85$\pm$ 0.53 &\hspace{-0.35cm} 322 & 304~ & 1955~~ &\hspace{-0.45cm} 2.98$^{-0.65}_{+0.14}$ &  $2.23^{+0.14}_{-0.08}$ &  1.17$\pm$0.20 &\hspace{-0.45cm}  -1.84 & -2.39 & 0.57 & 0.80 &   7.1$\pm$~1.1 \\
 IC 3392 &  0.72$\pm$0.57 &  0.43$\pm$ 0.34 &\hspace{-0.35cm} 197 & 163~ & 1683~~ &\hspace{-0.45cm} 1.03$^{+0.08}_{-0.08}$ &  $0.73^{-0.04}_{-0.04}$ &  1.15$\pm$0.12 &\hspace{-0.45cm}  -1.83 & -2.33 & 0.45 & 0.56 &   3.3$\pm$~0.2 \\
NGC 4457 &  3.21$\pm$0.81 &  1.94$\pm$ 0.49 &\hspace{-0.35cm} 161 & 143~ &  889~~ &\hspace{-0.45cm} 1.76$^{+0.08}_{-0.08}$ &  $0.90^{+0.04}_{-0.06}$ &  0.92$\pm$0.20 &\hspace{-0.45cm}  -1.65 & -2.29 & 0.65 & 0.85 &  33.6$\pm$~2.6 \\
 IC 3418 &     $<$0.13~~~ &  $<$0.08~~~     &\hspace{-0.35cm}...~ &...~~ &....~~~ &\hspace{-0.45cm}       ....~~~~~        &        ...~~~~~      &  ...~~~~~      &\hspace{-0.45cm}$<$-2.16&  ...$^\diamond$~ & ...~ & ...~ &  $<$0.8~~~  \\
NGC 4501 & 27.46$\pm$4.10 & 16.56$\pm$ 2.47 &\hspace{-0.35cm} 532 & 508~ & 2278~~ &\hspace{-0.45cm} 6.32$^{+0.72}_{-1.27}$ &  $3.60^{+0.34}_{-0.28}$ &  0.58$\pm$0.12 &\hspace{-0.45cm}  -1.30 & -1.97 & 0.92 & 1.22 & 306.0$\pm$~7.3 \\
NGC 4522 &  5.63$\pm$1.67 &  3.40$\pm$ 1.01 &\hspace{-0.35cm} 240 & 214~ & 2331~~ &\hspace{-0.45cm} 2.90$^{-0.01}_{+0.06}$ &  $5.01^{-1.14}_{+0.47}$ &  0.86$\pm$0.02 &\hspace{-0.45cm}  -0.91 & -1.22 & 0.78 & 1.48 &  22.6$\pm$~1.3 \\
NGC 4532 & 32.45$\pm$2.33 & 19.57$\pm$ 1.41 &\hspace{-0.35cm} 208 & 160~ & 2016~~ &\hspace{-0.45cm} 5.31$^{-0.11}_{+1.45}$ & 1$3.00^{-1.64}_{-4.52}$ & -0.06$\pm$0.06 &\hspace{-0.45cm}  -0.43 & -0.59 & 1.90 & 2.54 &  86.7$\pm$~5.4 \\
NGC 4535 & 54.34$\pm$2.23 & 32.77$\pm$ 1.34 &\hspace{-0.35cm} 292 & 272~ & 1971~~ &\hspace{-0.45cm} 8.86$^{+0.02}_{-0.02}$ &  $5.85^{-0.04}_{-0.11}$ &  0.41$\pm$0.12 &\hspace{-0.45cm}  -0.88 & -1.17 & 1.25 & 2.42 &  67.4$\pm$~2.9 \\
NGC 4533 &  4.44$\pm$1.15 &  2.68$\pm$ 0.69 &\hspace{-0.35cm} 192 & 182~ & 1742~~ &\hspace{-0.45cm} 2.17$^{+0.00}_{-0.01}$ &  $3.53^{+0.24}_{-0.86}$ &  0.51$\pm$0.04 &\hspace{-0.45cm}  -0.53 &... $^\diamond$~ & 1.03 & 0.07 &     $<$0.5~~~  \\
NGC 4536 & 78.45$\pm$3.85 & 47.32$\pm$ 2.32 &\hspace{-0.35cm} 348 & 325~ & 1802~~ &\hspace{-0.45cm} 8.74$^{+0.07}_{-0.10}$ &  $5.33^{+0.04}_{-0.08}$ &  0.16$\pm$0.12 &\hspace{-0.45cm}  -0.49 & -0.97 & 1.15 & 2.31 & 191.2$\pm$25.4 \\
VCC 1581 &  5.20$\pm$0.52 &  3.14$\pm$ 0.31 &\hspace{-0.35cm} 117 & 107~ & 2045~~ &\hspace{-0.45cm} 2.18$^{+0.06}_{-0.14}$ & 1$2.19^{+2.34}_{-0.40}$ & -0.06$\pm$0.06 &\hspace{-0.45cm}  -0.29 & -0.14$^\ast$ & 1.67 & 0.07 &     $<$0.2~~~  \\
NGC 4548 & 10.65$\pm$0.72 &  6.42$\pm$ 0.43 &\hspace{-0.35cm} 249 & 233~ &  480~~ &\hspace{-0.45cm} 4.71$^{+0.02}_{-0.04}$ &  $3.49^{+0.04}_{-0.04}$ &  0.82$\pm$0.12 &\hspace{-0.45cm}  -1.48 & -1.97 & 0.87 & 1.34 &   3.3$\pm$~0.2 \\
NGC 4561 & 23.21$\pm$1.86 & 14.00$\pm$ 1.12 &\hspace{-0.35cm} 171 & 132~ & 1404~~ &\hspace{-0.45cm} 5.50$^{-0.14}_{+0.05}$ &  $2.82^{+0.31}_{-0.44}$ & -0.71$\pm$0.02 &\hspace{-0.45cm}  -0.34 & -0.16 & 3.66 & 5.77 &   1.6$\pm$~0.3 \\
NGC 4567 & 15.64$\pm$1.16 &  9.43$\pm$ 0.70 &\hspace{-0.35cm} 204 & 197~ & 2275~~ &\hspace{-0.45cm} 9.57$^{-0.91}_{+0.17}$ &  $5.94^{-0.64}_{+0.21}$ &  0.13$\pm$0.12 &\hspace{-0.45cm}  -0.86 & -1.36 & 3.19 & 3.69 &  14.7$\pm$~0.7 \\
NGC 4568 & 25.11$\pm$2.83 & 15.14$\pm$ 1.71 &\hspace{-0.35cm} 337 & 314~ & 2249~~ &\hspace{-0.45cm} 4.66$^{+3.90}_{-0.60}$ &  $4.45^{+0.04}_{-2.06}$ &  0.38$\pm$0.12 &\hspace{-0.45cm}  -0.81 & -1.49 & 1.01 & 1.33 & 143.1$\pm$~7.4 \\
NGC 4569 & 10.29$\pm$2.38 &  6.21$\pm$ 1.44 &\hspace{-0.35cm} 406 & 387~ & -212~~ &\hspace{-0.45cm} 4.11$^{+0.40}_{-1.07}$ &  $2.29^{+0.14}_{-0.17}$ &  1.47$\pm$0.20 &\hspace{-0.45cm}  -1.79 & -2.23 & 0.43 & 0.75 & 105.6$\pm$~5.0 \\
NGC 4579 &  9.34$\pm$2.49 &  5.63$\pm$ 1.50 &\hspace{-0.35cm} 371 & 358~ & 1516~~ &\hspace{-0.45cm} 4.11$^{+0.19}_{-0.16}$ &  $2.78^{+0.14}_{-0.14}$ &  0.95$\pm$0.20 &\hspace{-0.45cm}  -1.73 & -2.33 & 0.70 & 1.03 & 163.2$\pm$12.2 \\
NGC 4580 &  0.46$\pm$0.37 &  0.28$\pm$ 0.22 &\hspace{-0.35cm} 179 & 168~ & 1035~~ &\hspace{-0.45cm} 0.87$^{+0.06}_{-0.10}$ &  $0.52^{+0.04}_{+0.00}$ &  1.53$\pm$0.20 &\hspace{-0.45cm}  -2.47 & -2.71 & 0.42 & 0.47 &   3.5$\pm$~0.1 \\
NGC 4606 &  0.41$\pm$0.22 &  0.25$\pm$ 0.13 &\hspace{-0.35cm} 158 & 142~ & 1647~~ &\hspace{-0.45cm} 0.65$^{+0.07}_{-0.13}$ &  $0.54^{-0.04}_{+0.10}$ &  1.64$\pm$0.20 &\hspace{-0.45cm}  -2.19 & -2.58 & 0.20 & 0.35 &   1.1$\pm$~0.1 \\
NGC 4607 &  3.63$\pm$1.34 &  2.19$\pm$ 0.81 &\hspace{-0.35cm} 247 & 221~ & 2253~~ &\hspace{-0.45cm} 2.04$^{+0.06}_{-0.07}$ &  $1.93^{-0.14}_{+0.09}$ &  0.82$\pm$0.12 &\hspace{-0.45cm}  -0.82 & -1.55 & 0.70 & 0.75 &  19.6$\pm$~1.5 \\
NGC 4651 & 67.08$\pm$4.00 & 40.46$\pm$ 2.41 &\hspace{-0.35cm} 386 & 363~ &  801~~ &\hspace{-0.45cm} 9.28$^{-1.82}_{+0.01}$ &  $4.78^{-0.74}_{-0.44}$ & -0.30$\pm$0.02 &\hspace{-0.45cm}  -0.48 & -0.88 & 2.32 & 4.03 &  50.1$\pm$~2.0 \\
NGC 4654 & 49.19$\pm$3.17 & 29.67$\pm$ 1.91 &\hspace{-0.35cm} 310 & 288~ & 1031~~ &\hspace{-0.45cm} 7.47$^{+0.00}_{-2.38}$ &  $4.42^{+0.14}_{-1.74}$ &  0.12$\pm$0.02 &\hspace{-0.45cm}  -0.73 & -1.08 & 1.53 & 2.33 & 115.7$\pm$~3.7 \\
NGC 4689 &  7.81$\pm$0.84 &  4.71$\pm$ 0.51 &\hspace{-0.35cm} 197 & 180~ & 1615~~ &\hspace{-0.45cm} 2.99$^{+0.06}_{-0.06}$ &  $1.76^{+0.04}_{-0.07}$ &  0.68$\pm$0.12 &\hspace{-0.45cm}  -1.33 & -1.65 & 0.70 & 1.17 &   1.6$\pm$~0.0 \\
VCC 2062 &  5.32$\pm$0.22 &  3.21$\pm$ 0.13 &\hspace{-0.35cm}  73 &  45~ & 1139~~ &\hspace{-0.45cm} 4.87$^{-3.07}_{+0.00}$ &  $3.89^{-2.74}_{-0.17}$ &  ...~~~~~      &\hspace{-0.45cm}   1.44 & ...$^\diamond$ & 6.95 & 0.15 &     $<$0.4~~~  \\
NGC 4694 &  4.19$\pm$0.25 &  2.53$\pm$ 0.15 &\hspace{-0.35cm} 116 &  84~ & 1176~~ &\hspace{-0.45cm} 1.47$^{-0.88}_{+0.44}$ &  $9.82^{-3.14}_{+0.26}$ &  0.83$\pm$0.20 &\hspace{-0.45cm}  -1.44 & -1.68 & 0.46 & 0.83 &   3.1$\pm$~0.4 \\
NGC 4698 & 27.15$\pm$2.00 & 16.38$\pm$ 1.21 &\hspace{-0.35cm} 432 & 413~ & 1009~~ &\hspace{-0.45cm} 8.84$^{-1.16}_{+0.58}$ &  $6.18^{-0.04}_{-0.55}$ &  0.02$\pm$0.20 &\hspace{-0.45cm}  -0.85 & -1.44 & 2.21 & 3.10 &     $<$1.0~~~  \\
NGC 4713 & 48.01$\pm$3.30 & 28.90$\pm$ 1.98 &\hspace{-0.35cm} 185 & 167~ &  654~~ &\hspace{-0.45cm} 8.49$^{-1.65}_{+0.21}$ &  $3.69^{-0.54}_{+0.56}$ & -0.31$\pm$0.04 &\hspace{-0.45cm}  -0.31 & -0.27 & 3.15 & 7.77 &  10.2$\pm$~0.4 \\
NGC 4772 & 13.86$\pm$1.94 &  8.36$\pm$ 1.17 &\hspace{-0.35cm} 463 & 437~ & 1044~~ &\hspace{-0.45cm} 2.83$^{-0.02}_{+0.02}$ &  $2.51^{-0.24}_{+0.44}$ &  0.15$\pm$0.20 &\hspace{-0.45cm}  -0.94 & -1.42 & 0.83 & 1.36 &   2.5$\pm$~0.3 \\
NGC 4808 & 59.15$\pm$4.20 & 35.68$\pm$ 2.53 &\hspace{-0.35cm} 280 & 260~ &  760~~ &\hspace{-0.45cm} 7.95$^{-0.61}_{+0.76}$ &  $6.05^{+0.13}_{-0.34}$ & -0.58$\pm$0.04 &\hspace{-0.45cm}  -0.17 & -0.51 & 2.84 & 4.37 &  45.0$\pm$~3.9 \\
\hline
\end{tabular}
\end{center}
\noindent
{\footnotesize {\it Footnote for Table 3:}
(1) NGC, IC or VCC names; (2) Integrated H{\sc i} flux$\pm\sigma$; (3) Total H{\sc i} mass$\pm\sigma$; (4) Linewidth measured at 20\% of the peak flux;
(5) Linewidth measured at 50\% of the peak flux; (6) H{\sc i} velocity determined using $W_{20}$ and $W_{50}$; (7) Isophotal diameter determined at 
1~$M_\odot~$pc$^{-2}$ ($^{\Delta\rm east}_{\Delta\rm west}$); (8) Effective diameter measured at $4\pi \int_{r} \Sigma_{\rm HI}(r)\cdot r^2~dr$=0.5$S_{\rm HI}$ 
($^{\Delta\rm east}_{\Delta\rm west}$); (9) Type independent H{\sc i} deficiency$\pm$uncertainty from the morphological classification; (10) Log of H{\sc i}
mass-to-light ratio in $B$ ($M_\odot/L_\odot$); (11) Log of H{\sc i} mass-to-light ratio in $K$ ($M_\odot/L_\odot$)
except for the ones marked with an asterisk where GOLDMINE $K$ is used or with a diamond which are not available
in either database. Note that there is a systematic offset between the 2MASS $K$ mag (DENIS, Cohen et al. 2003, AJ, 
126, 1090) and the GOLDMINE $K$ mag (Johnson) due to the different bandwidths. The mean offset
for 47 VIVA galaxies available in both databases, $<K_{\rm 2MASS}-K_{\rm GOLDMINE}>$ is 0.217$\pm0.140$.; 
(12) The ratio of H{\sc i} isophotal 
diameter-to-optical $B$ diameter at 25~mag~$\Box''^{-1}$; (13) The ratio of H{\sc i} isophotal diameter-to-optical $K$ diameter at 20~mag~$\Box''^{-1}$. $^\dag$Not available since the H{\sc i} surface brightness is too low 
($<1~M_\odot$) to define $D^{\rm iso}_{\rm HI}$ for this galaxy.
(14) The 1.4 GHz continuum flux$\pm\sigma$.}
\end{table*}

\section{H{\scriptsize I} Properties}
\label{sec-hiprop}
\subsection{H{\small I} Quantities} 
In this section, we describe how the H{\sc i} properties have been 
determined. The result is presented in Table~\ref{tbl-hiprop}.

\bigskip
\centerline{\it Flux {\rm (}$S_{\rm HI}{\rm )}$ \& Mass {\rm (}$M_{\rm HI}${\rm )}}
\noindent
Column 2 \& 3: We have measured the total flux by integrating the
global profile along the velocity axis,
\begin{equation}
S_{\rm HI}=\sum F_{\rm HI} \cdot \Delta v \pm ({\sum \sigma_{\rm
HI}^2})^{\frac{1}{2}} \cdot \Delta v
\end{equation}
in Jy~km~s$^{-1}$, where $F_{\rm HI}$ and $\sigma_{\rm HI}$ are the H{\sc i} flux and the
rms in Jy at each channel, and $\Delta v$ is the channel width
($\approx10.4$ km~s$^{-1}$). The H{\sc i} mass in $M_\odot$ can be
then determined by,
\begin{equation}
M_{\rm HI}=2.356\times10^5 S_{\rm HI} D_{\rm Mpc}^2
\end{equation}
in $M_\odot$, where $S_{\rm HI}$ is the total flux in Jy~km~s$^{-1}$ and $D$ is the
distance to the galaxy in Mpc (assumed to be 16 Mpc for all galaxies).

\bigskip
\centerline{\it Linewidths {\rm (}$W_{20}$, $W_{50}${\rm )} \& H{\small I} velocity {\rm (}$V_{\rm HI}${\rm )}}
\noindent
Column 4, 5 \& 6:
The linewidths have been measured at 20\% and 50\% level of the peak fluxes on both the receding 
and the approaching sides of the profile,
\begin{equation}
\begin{array}{c}
W_{20}=V_{20}^{R}-V_{20}^{A}\\
{}\\
W_{50}=V_{50}^{R}-V_{50}^{A}\\
\end{array}
\end{equation}
where $V_{20}^{R}$, $V_{50}^{R}$ and $V_{20}^{A}$, $V_{50}^{A}$ are the velocities with 20\%, 50\%
of the peak flux on the receding and the approaching sides, respectively. The H{\sc i} velocity
has been determined with $V_{20}^{R}$, $V_{50}^{R}$ and $V_{20}^{A}$, $V_{50}^{A}$ using the
following definition,
\begin{equation}
V_{\rm HI}=0.25~(V_{20}^{A}+V_{50}^{A}+V_{50}^{R}+V_{20}^{R}).
\end{equation}
The uncertainties in $W_{20}$, $W_{50}$ and $V_{\rm HI}$ are 
approximately 10.4~km~s$^{-1}$.

\bigskip
\centerline{\it Diameters {\rm (}$D_{\rm HI}^{\rm iso}~\&~$$D_{\rm HI}^{\rm eff}${\rm)}}
\noindent
Column 7 \& 8: To determine the isophotal diameter we use the
radius where the azimuthally averaged H{\sc i} surface density
($\Sigma_{\rm HI}$) drops to 1~$M_\odot$~pc$^{-2}$. If there is
more than one radius with $\Sigma_{\rm HI}$=1~$M_\odot$~pc$^{-2}$
(e.g. in case an H{\sc i} hole is present in the central area on the
disk), we take the outermost position to derive the isophotal
diameter.  NGC~4293 is
the only galaxy where $D_{\rm HI}^{\rm iso}$ is not defined in this way since
$\Sigma_{\rm HI}$ is always below 1~$M_\odot$~pc$^{-2}$. For this galaxy 
we use as  effective diameter
the region  that contains 50\% of the
total flux. We also determined  isophotal and the effective
diameters for the east and the west sides of the disk, separately. The
difference between these and the diameters measured over the entire disk,
i.e. $\Delta {\rm E}=D_{\rm east}-\bar{D}$ and $\Delta {\rm W}=D_{\rm
west}-\bar{D}$, is a useful measure of the morphological
asymmetry.

\bigskip
\centerline{\it Deficiency {\rm (}$def_{\rm HI}${\rm )}}
\noindent
Column 9: The H{\sc i} deficiency is an indicator of how H{\sc i}
deficient individual galaxies are compared to field galaxies of the same
size and morphological type.
 \citet{hg84} have defined  $def_{\rm HI}$ as
follows,

\begin{equation}
def_{\rm HI}=<{\rm log}~\bar{\Sigma}{\rm_{HI}} {\rm (T)}>-{\rm
log}~\bar{\Sigma}_{\rm HI},
\end{equation}
where $\bar{\Sigma}_{\rm HI}\equiv S_{\rm HI}/D_{opt}^2$, is the mean
H{\sc i} surface density within the optical disk. This mean surface
density  varies only slightly with
 Hubble type, T, for types Sab-Sm, but varies more for types Sa and earlier. 
 For isolated galaxies, \citet{hg84} empirically
determined $<{\rm log}~\bar{\Sigma}{\rm_{HI}} (T)>=$0.24, 0.38, 0.40,
0.34, and 0.42 for Sa/Sab, Sb, Sbc, Sc, and  later types than
Sc respectively. However, \citet{kkt98} have shown that Hubble classification
does not work for many cluster spiral galaxies in Virgo, due to environmental
processes which remove gas and greatly reduce star formation rates.
 We therefore prefer to use the type independent H{\sc i} deficiency 
parameter, which compares all morphological types to a  mean
H{\sc i} surface density \citep[$<{\rm
log}~\bar{\Sigma}{\rm_{HI}}>=0.37$;][]{hg84}. We use as the uncertainty
in the deficiency the difference between the type independent deficiency 
and the type dependent deficiency using the morphological types from the 
RC3 catalog  (Table~\ref{tbl-sample}), $\Delta_{def_{\rm HI}}=|def_{\rm
HI}(T)-def_{\rm HI}|$.

\bigskip
\centerline{\it H{\small I} mass-to-light ratio {\rm (}$M_{\rm
HI}/L${\rm ) in $B$ \& $K$}}
\noindent
Column 10 \& 11: The H{\sc i} mass to light ratio in $B$ and $K$-band in solar unit ($M_\odot/L_\odot$)
has been measured using the following equations,
\begin{equation}
\begin{array}{l}
\frac{M_{\rm HI}}{L_B}=1.51\times10^{-7}S_{\rm HI}10^{0.4(m_B-A_B)} \frac{M_\odot}{L_{\odot,B}},\\
\frac{M_{\rm HI}}{L_K}=1.15\times10^{-6}S_{\rm HI}10^{0.4(m_K-A_K)} \frac{M_\odot}{L_{\odot,K}}
\end{array}
\end{equation}
where $S_{\rm HI}$ is in Jy~km~s$^{-1}$, and $A_B$ and $A_K$ are the Galactic extinction in $B$ and $K$-band,
taken from the Lyon-Meudon Extragalactic Database (LEDA; Paturel et
al. 1997) and NED (NASA/IPAC Extragalactic Database). In
Table~\ref{tbl-hiprop}, the values are given in logarithmic scale. The
$K$-band magnitude (Kron magnitude measured at 20~mag~arcsec$^{-2}$) 
has been obtained from the 2MASS \citep[The Two Micron
All Sky Survey;][]{2mass} database.

\bigskip
\centerline{\it H{\small I}-to-optical size {\rm (}$D_{\rm HI}^{\rm iso}/D_{\rm opt}${\rm )} in $B$ \& $K$}
\noindent
Column 12 \& 13: The ratio of the H{\sc i} isophotal diameter to the $B$ and $K$-band optical diameters
are presented. The $B$ band diameters (listed in Table 1) are D$_{25}$ from RC3. 
The $K$-band diameters have been obtained from the 2MASS database (Kron
isophotal diameters at 20~mag~arcsec$^{-2}$).
No photometric measurements are available  for several systems 
that are faint in $K$.


\subsection{Comparison of total H{\sc i} flux with values in the literature}

In this subsection we compare the VIVA fluxes (VLA C or CS array) with
either the fluxes measured in the Arecibo Legacy Fast ALFA 
(ALFALFA) survey of the Virgo region
\citep{kent08} or for the few galaxies that have not yet been
observed with ALFALFA with the most reliable fluxes listed in
ARECIBO-05. We also compare our flux values with the earlier imaging
surveys by Cayatte et al. (1990; VLA D array) and Warmels (1988a and
b; WSRT imaging and one-dimensional strip scans, respectively). The
latter surveys have much lower signal to noise than the VIVA data, but
have different UV coverage.  Arecibo has a filled aperture and is less
likely to miss flux. However, its beam is quite large ($\sim3.5'$ at
21~cm) and the total flux within one beam can be confused by other
systems. It can also miss some flux in case a galaxy with H{\sc i}
extent larger than the beam is observed with a single pointing.
However the new seven element Arecibo L-band Feed Array (ALFA)
receiver system makes a complete image of the area and we consider the
ALFALFA fluxes the best measure of the total amount of H{\sc i}.
Meanwhile interferometers cannot measure structures on angular scales
larger than the fringe spacing formed by the shortest spacing
\citep{tup04}. As a result they can miss some flux in extended
features. The VLA CS array has the same shortest spacing as the VLA D
array, and it should in principle be able to measure extended features
equally well (the maximum extended structure that can be imaged is
15~arcmin at 20~cm).  However since it has fewer short spacings than D
array it still somewhat less sensitive to faint extended
structure. With the VLA C array, the maximum extent visible is
6~arcmin and we could possibly have missed some flux from very
extended structures in the galaxies observed with the C
configuration. Our conclusion is that in general there is very good
agreement with the ALFALFA fluxes.
\begin{figure}
\plotone{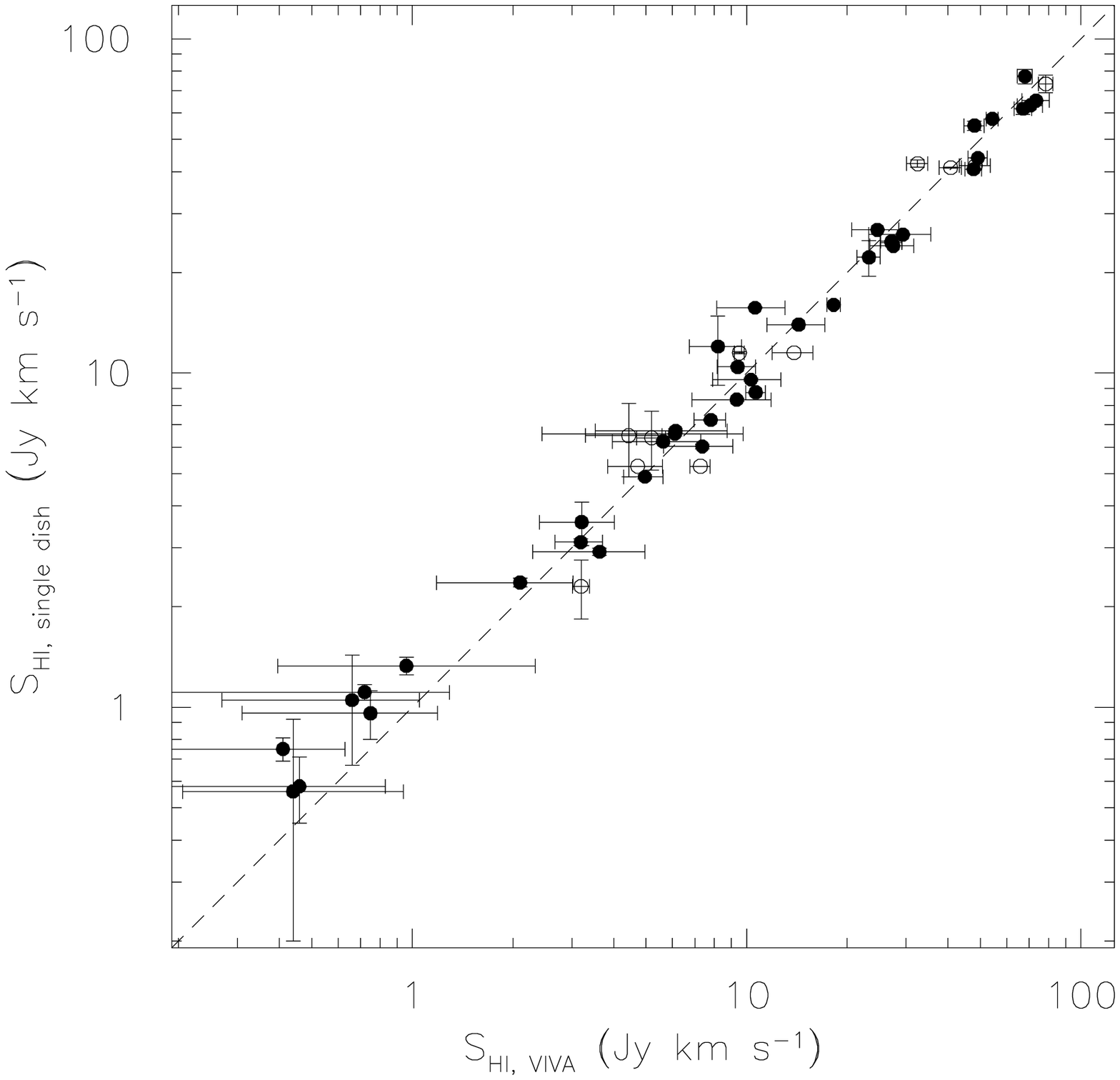}\\
\plotone{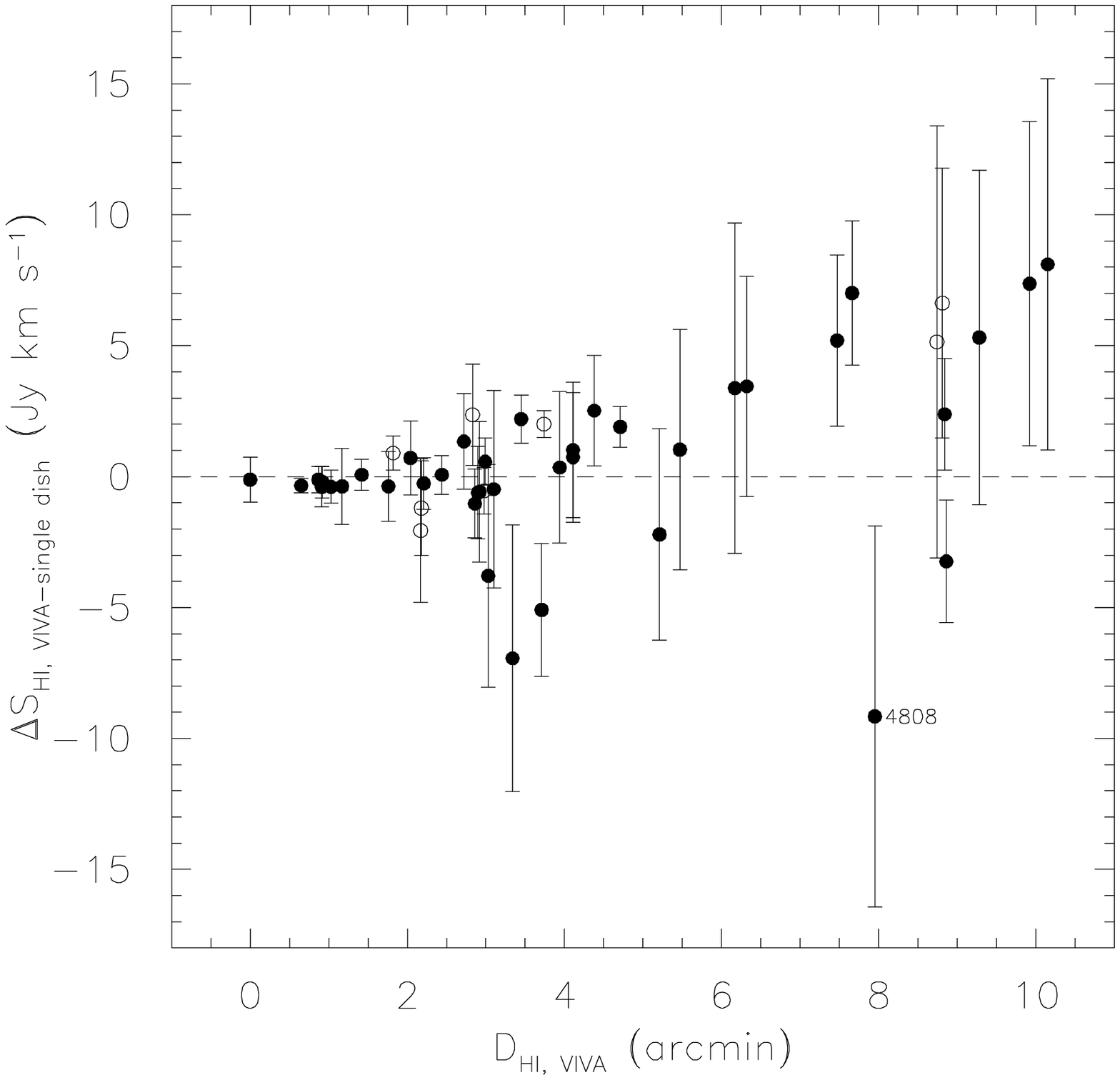}
\caption{Top) A comparison of the single dish fluxes and the
VIVA fluxes. The ALFALFA and the ARECIBO-05 fluxes are shown in
filled and open circles, respectively. Bottom) The difference
between the VIVA and the single dish flux as function of H{\sc i}
extent.\label{fig-compub1}}
\end{figure}

\begin{figure}
\plotone{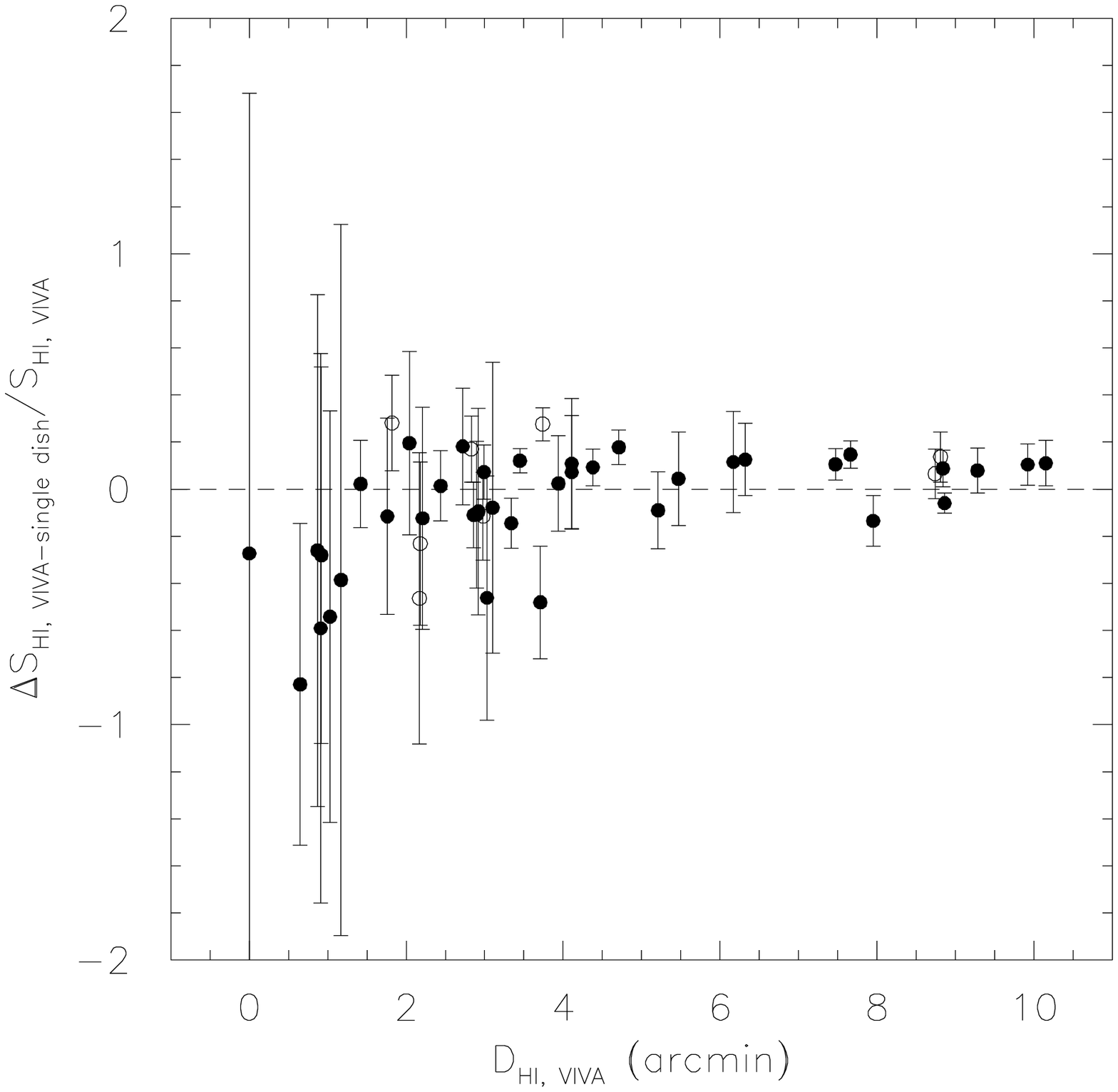}\\
\plotone{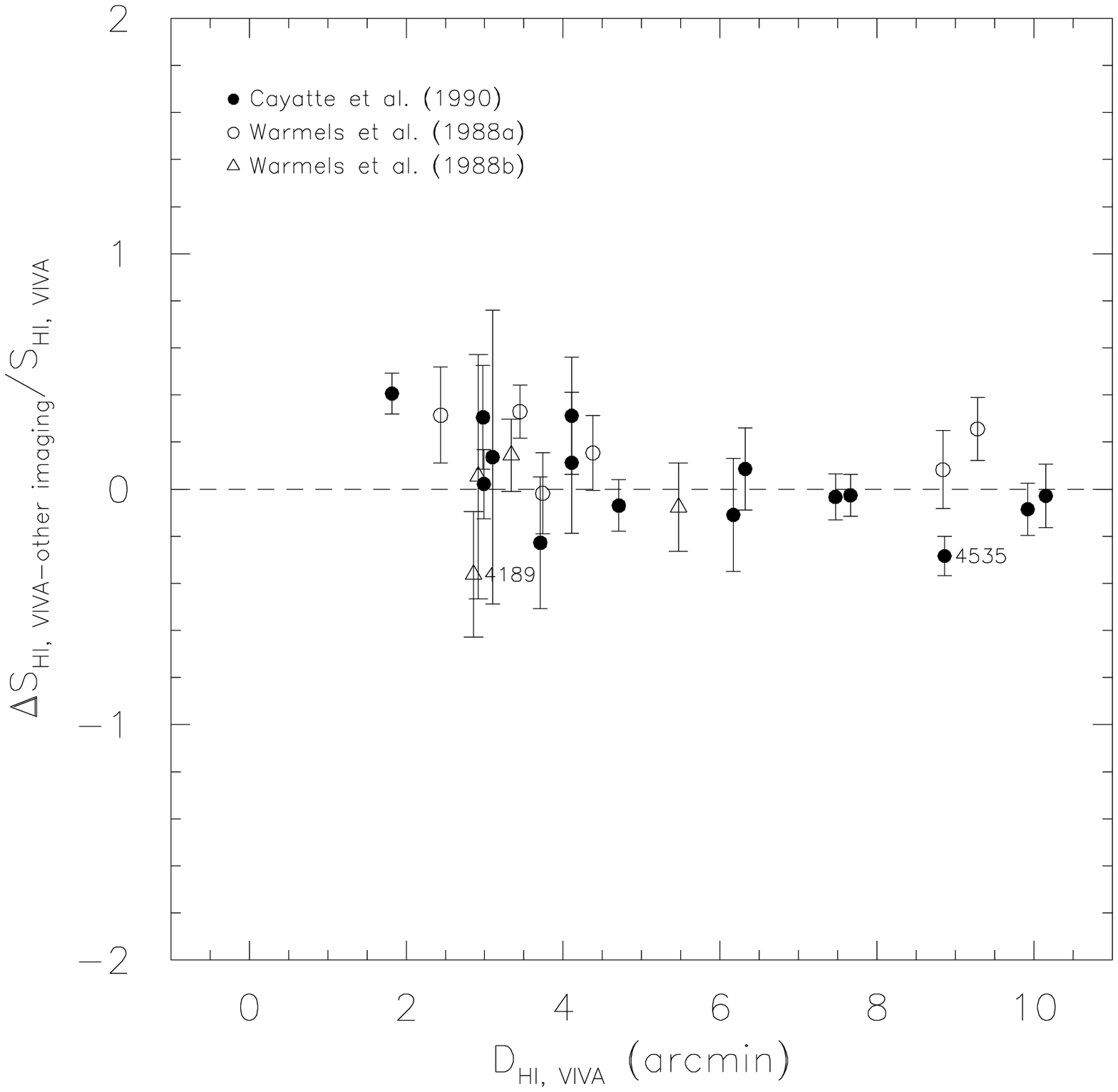}
\caption{Top) The difference between the VIVA and the single dish
flux normalized by the VIVA flux as function of H{\sc i} extent.
The same symbols are used as Figure~\ref{fig-compub1}.
Bottom) Same as above but a comparison with the 
previous VLA imaging study (Cayatte et al. 1990; filled circle) 
or WSRT (Warmels 1988a; open circle) imaging study and 1-dimensional
observations (Warmels 1988b; open triangle).
\label{fig-compub2}}
\end{figure}

In the upper plot in Figure~\ref{fig-compub1} we show a comparison 
of the ALFALFA
fluxes with the VIVA fluxes (filled symbols). Open symbols are single
dish measurements taken from ARECIBO-05. There is good agreement
between the interferometer and the single dish values.  
Since there is a distinct possibility that the interferometer
resolves out some of the most extended flux, we show in the lower plot
the difference between the single dish values and the VIVA flux as function of 
H{\sc i} extent. As expected the scatter in the total flux goes up in
absolute value for large sources, but the fractional error goes down.
There is no evidence that VIVA fluxes are less than ALFALFA fluxes for
the large diameter sources. Rather, there is a marginally significant
suggestion that VIVA fluxes are on average greater than ALFALFA fluxes
for the largest sources. 
To estimate how significant the uncertainties
in the flux values are we show in Figure~\ref{fig-compub2} (upper plot) 
the same differences, but  normalized
by the VIVA fluxes. Clearly there is good agreement for the large
size sources. 
Interestingly there
appears to be a very small systematic bias for the smaller size
sources. The VIVA fluxes are all below the single dish value although
this is a small effect compared to the size of the error bars.
We believe this to be the result of the way we make the total H{\sc i}
images, by using a cutoff in the smoothed images. 

\begin{figure*}
\plotone{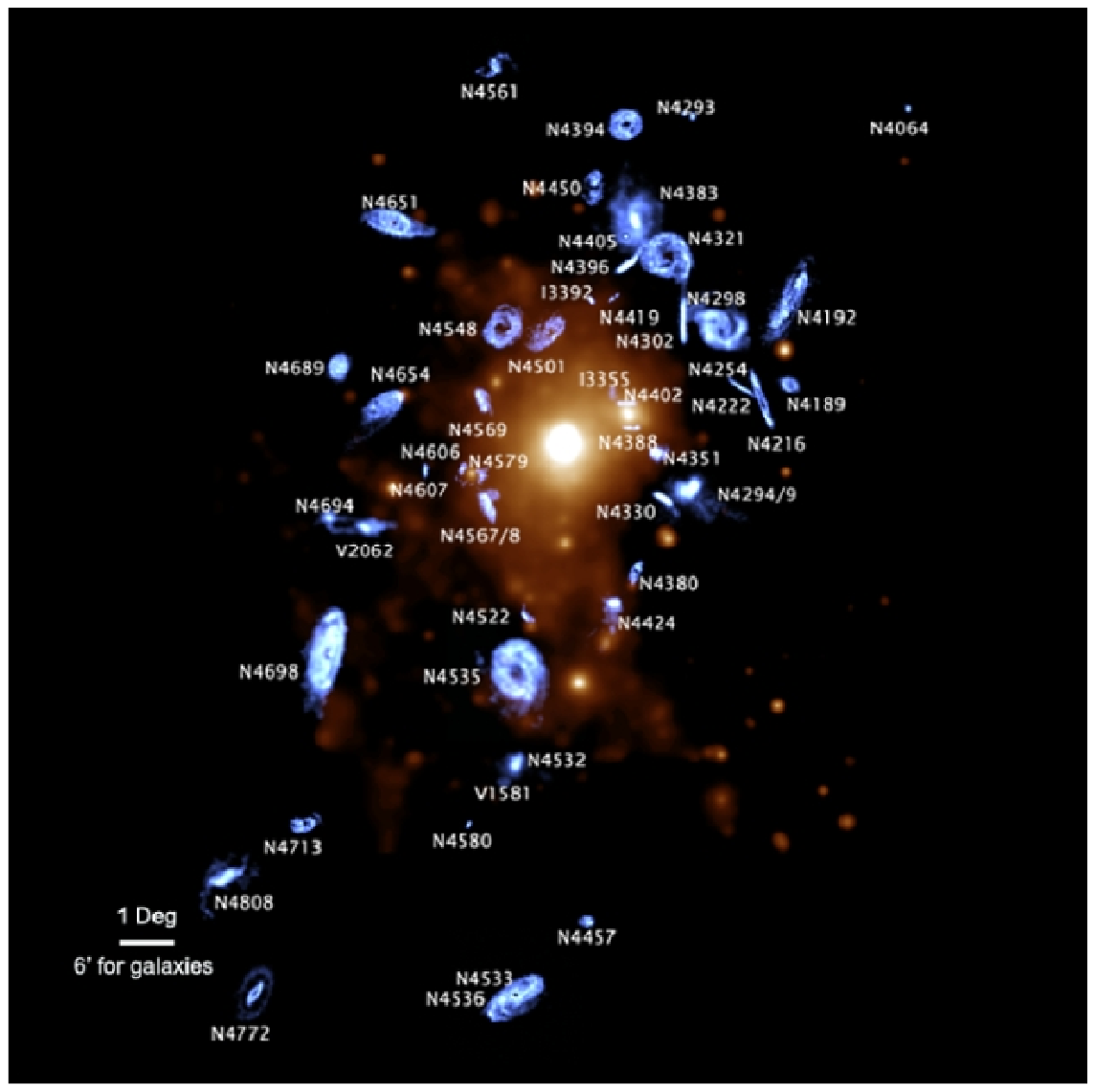}
\caption{A composite image of  the total H{\sc i} images of the individual
galaxies (in blue) overlaid on the ROSAT X-ray image (orange)  
by \citet{bohr94}.
The galaxies are located at the proper position in
the cluster but each H{\sc i} image is magnified by a factor 10 to show
the details of the H{\sc i} distribution.  The picture shows clearly how non uniform the
mass distribution in Virgo is, with enhanced X-ray emission from the three
subclusters centered at the ellipticals, M87, M86 and M49.
\label{fig-poster}}
\end{figure*}  

Finally, mostly for historical interest, we show in the bottom plot of
Figure 6 the difference between the VIVA flux and the values measured
by Cayatte et al. (1990) and/or Warmels (1988a, b), normalized by the VIVA flux as function of
H{\sc i} diameter. There is excellent agreement, except for NGC~4535.
Interestingly both Cayatte et al. (1990) and Warmels (1988b) find 
a 20\% larger flux for this galaxy,
on the other hand there is excellent agreement between the ALFALFA and
VIVA flux. We have no explanation for this discrepancy. 

In conclusion we find that there is excellent agreement between the
VIVA and ALFALFA fluxes and there is no indication that the interferometer
has missed any very extended flux.

\section{The H{\sc i} morphology in different environments}
\label{sec-results}

\begin{figure*}
\plotone{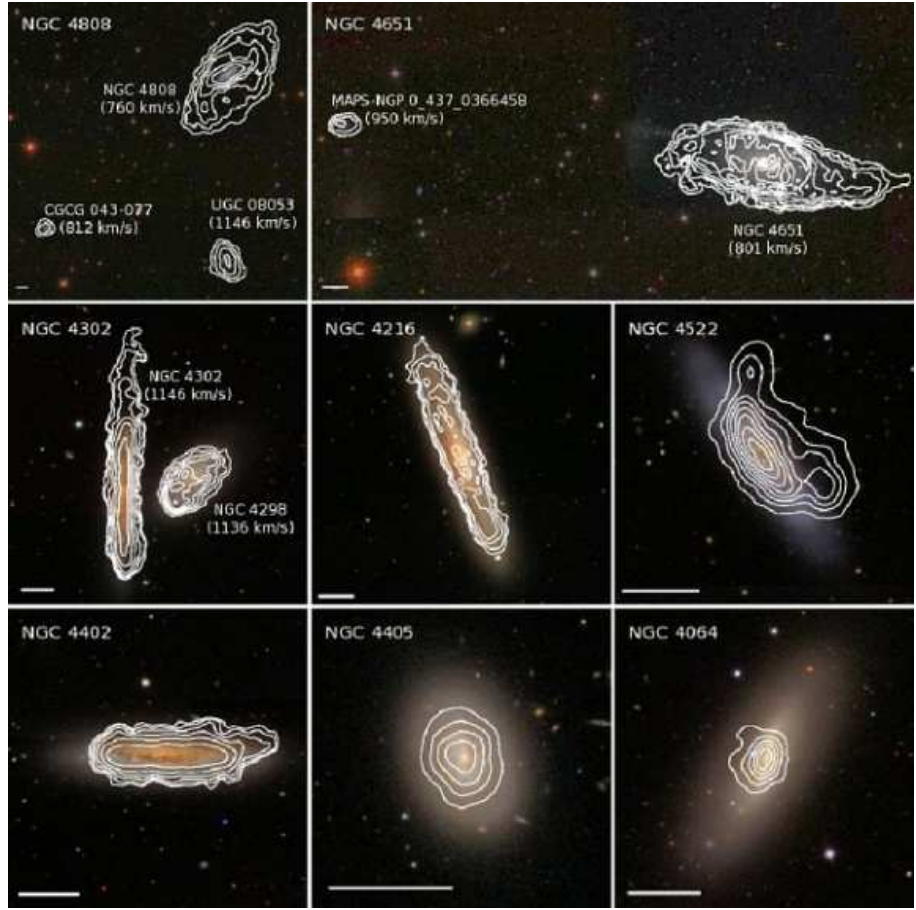}
\caption{Examples of the different H{\sc i} morphologies found in
the survey. Total H{\sc i} images are shown in white contours 
overlaid on the Sloan Digital Sky Survey (SDSS)
images. The thick white bar in the bottom left 
corner indicates 1 arcmin in each panel. The top row shows examples
of gas rich galaxies in gas rich environments in the outskirts,
the middle row shows galaxies at intermediate distances,
while the bottom row shows examples of severely truncated
H{\sc i} disks at a range of projected distances from M87.
\label{fig-examples}}
\end{figure*}

In this section we describe the range of H{\sc i} morphologies found
in the different locations in Virgo. We present results for individual
galaxies in an appendix.  In Figure~\ref{fig-poster} we show a
composite image of the total H{\sc i} images of the individual
galaxies (in blue) overlaid on the ROSAT X-ray image (orange) by
\citet{bohr94}. The galaxies are located at the proper position in the
cluster but each H{\sc i} image is magnified by a factor 10 to show
the details of the H{\sc i} distribution.  The picture shows how
non-uniform the mass distribution in Virgo is, with enhanced X-ray
emission from the cluster and subclusters centered at the giant
ellipticals, M87, M86 and M49 respectively.  There is a huge range in
H{\sc i} size of the galaxies.  In general the galaxies at larger
projected distances have larger H{\sc i} sizes, while galaxies in the
core have small H{\sc i} sizes, but there are exceptions. In
Figure~\ref{fig-examples} we show typical examples for the range of
morphologies that we see.

{\bf H{\sc i} rich galaxies in the outskirts of the cluster} H{\sc i}
rich galaxies are exclusively found in the lower density regions of
the cluster outskirts at the projected distance from M87 
$d_{87}\gtrsim1~$Mpc.  These
galaxies usually have H{\sc i} extending well beyond the stellar disk
in all directions.  Small, kinematically distinct H{\sc i} features
with or without optical counterparts are quite common around these
systems. A typical example is NGC~4808 shown in
Figure~\ref{fig-examples}.  Many of the galaxies in the outskirts look
morphologically peculiar showing tails or rings in H{\sc i}, in the
stellar distribution or both. Some show a kinematical decoupling
between inner and outer gaseous disks. These galaxies seem to be
experiencing gravitational interactions and possibly continuing infall
of gas from the halo.  For example NGC~4651
(Figure~\ref{fig-examples}) shows in H{\sc i} an extension to the
west, while deep optical images show a stellar tail to the opposite
side of the H{\sc i}, which ends with a low surface brightness
arc. Kinematically the H{\sc i} disk shows a discontinuity in position
angle between inner and outer disk.

{\bf Long one-sided H{\sc i} tails pointing away from M87} At intermediate 
distances from M87 ($0.6\lesssim d_{\rm M87}\lesssim1~$Mpc), we find
seven galaxies with long one-sided H{\sc i} tails pointing away from
M87. An example is NGC~4302 (Figure~\ref{fig-examples}). The H{\sc i} is
mildly truncated within the stellar disk in the south, and the gas tail
is extended to the north, with no optical counterpart. Although there 
is a nearby companion, NGC~4298, NGC~4302  looks optically
undisturbed. In \citet{cvgkv07} we argue that these galaxies
have only recently arrived in the cluster and are falling in to the
center, likely on highly radial orbits as hinted by the direction of
the tails. A simple estimate suggests that all but two of
the tails could have been formed by ram pressure stripping of the
gas in the very outer parts of the disk. Some of these galaxies
have close neighbours, suggesting that tidal interactions
may have moved gas outward, making it more susceptible to 
ram pressure stripping. Apparently galaxies begin to lose
their gas already at intermediate distances from the cluster
center through ram pressure stripping and tidal interactions 
or a combination of both.

{\bf Symmetric H{\sc i} disks with $\bf{D_{\rm HI}/D_{opt}\approx1}$ at
  intermediate distances} At similar distances from M87 as the H{\sc
  i} tails we find galaxies with fairly symmetric H{\sc i} disks that
are comparable in size to their stellar disk, e.g. NGC~4216 shown in
Figure~\ref{fig-examples}.  Some are quite H{\sc i} deficient, despite
the fact that the H{\sc i} extent is comparable to the optical
extent. Their H{\sc i} surface density is down by up to a factor two.
These systems might be under the influence of a process that slowly
affects the entire face of the galaxy, such as turbulent viscous
stripping or thermal evaporation \citep{n82,caya94}.

{\bf Small H{\sc i} disks near the cluster center} Near the 
cluster core ($d_{\rm M87}<0.5~$Mpc) galaxies always have
gas disks that are truncated to within the optical disk.
These galaxies often show highly asymmetric H{\sc i} distributions
as they are currently undergoing strong ram pressure stripping.
An example is NGC~4402 (Figure~\ref{fig-examples}), which has been
studied in detail by \citet{ckvgv05}. Galaxies appear
to lose most of their H{\sc i} gas ($>70\%$) in these
regions through a strong interaction with the ICM.

{\bf Severely stripped H{\sc i} disks beyond the cluster core} 
Interestingly, we also find a number of galaxies that are stripped to well within
the stellar disk at large projected distances from the cluster
center ($\gtrsim1~$Mpc). Examples are NGC~4522, NGC~4405,
and NGC~4064 (Figure~\ref{fig-examples}). Some of these
may have been stripped while crossing the cluster center. 
As the galaxies move out from the high ICM density region
stripped and disturbed gas that is still bound to the galaxy
may resettle onto the disk and form a small symmetric 
gas disk in the center (NGC~4405). However, a detailed
study of the mean stellar age at the truncation radius
by \citet{ck08} shows that some of these
galaxies have been forming stars until recently ($<0.5~$Gyr).
This leaves not enough time for these galaxies to have
been stripped in the center and then travelled to their 
current location. NGC~4064 is a good example of this. It must
have lost its gas at large distances from the cluster center.

A particularly interesting case is NGC~4522 \citep{kvgv04},
which shows abundant evidence for current ongoing strong 
ram pressure stripping despite its large projected distance
from M87 ($\approx1$~Mpc).  An estimate of the mean 
stellar age at the stripping radius \citep{ck06}
also suggests that stripping is ongoing, yet an
estimate of the ram pressure at that location based
on a smooth distribution of the ICM would indicate
that pressure is too low by a factor 10. \citet{kvgv04}
argue that merging of the sub cluster M49 with Virgo 
could locally enhance the ram pressure, due to bulk motions,
clumpy density distributions and variations in the temperature of the 
ICM gas. A temperature map of the
X-ray emission \citep{smy01} does show that
NGC~4522 is located near strong variations in 
the X-ray temperature. The results on galaxies such as NGC~4064 
and NGC~4522 fit in nicely with recent work by Tonnesen, Bryan \& van Gorkom (2007)
and \citet{tb08}, who find that ram pressure  can
vary by more than a factor 10 at a given distance from the cluster
center due to structure in the ICM. This makes it
possible for some galaxies to get stripped in the outskirts,
without ever making it to the center of the cluster,
something we may be witnessing in Virgo.

\section{Summary}
\label{sec-summary}

We present the results of a new H{\sc i} imaging survey of 53 galaxies
in the Virgo cluster. The goal is to study the impact of different
environmental effects on the H{\sc i} disks of the galaxies. Virgo
is ideal for this type of study as it is dynamically young and potentially
contains galaxies that are affected by a wide range of environmental
effects. Its nearness allows us to study individual galaxies in
great detail.

We have selected 48 galaxies and obtained data on 5 additional
galaxies that were in the same field and velocity range as the target
galaxies. The galaxies were selected to cover a wide range of
star formation properties from anemic to starburst, and to be located 
in a wide range of local galaxy densities, from the dense core
to the outskirts of the cluster.
The target galaxies are at projected distances from 0.3 to 3.3 Mpc 
from the cluster center and as such the survey covers a region that
is 2 to 3 times larger than the area explored in the previous VLA
survey by \citet{cvgbk90}. Many of our galaxies had never been 
imaged in H{\sc i}.
The new survey was done with the VLA CS configuration. Its spatial
and spectral resolution are a factor 3 and 4 better than that of 
the previous survey. 
The VIVA survey has not only confirmed results from previous 
H{\sc i} imaging studies, it has found many features that were never seen
before in Virgo, or any other cluster.  To summarize our main results:

\begin{enumerate}
\item We confirm that galaxies near the cluster center have H{\sc i} disks
that are much smaller than the optical disk. We see however extraplanar
gas near some of the galaxies, providing the first direct evidence
for ongoing ram pressure stripping and fall back of stripped gas.
\item At intermediate distances from the center (0.6-1 Mpc) we
find galaxies with long one-sided H{\sc i} tails pointing away from
M87. \citet{cvgkv07} argue that these are most likely galaxies falling into
the cluster on highly radial orbits. The tails are due to ram pressure 
stripping and in a few cases to the combined effect of gravitational
interactions and rampressure stripping. Thus the impact of ram pressure 
begins to affect galaxies already at intermediate distances from the center.
\item We found several galaxies in the outskirts of Virgo ($d_{87}>$1.5 Mpc)
that have also H{\sc i} disks that are much smaller than the stellar disks. 
Some of these were already known to be strongly H{\sc i} deficient
\citep{sanchis02}. Although these galaxies are as H{\sc i} deficient
as the galaxies in the core, none of them shows signs of ongoing/recent
stripping. Some of these galaxies may have been stripped earlier when passing 
through the center of Virgo, but at least some of them have been
forming stars in the stripped part of the disks until quite recently
\citep{ck08}. The latter galaxies almost certainly have been
stripped of their gas in the outskirts of the cluster.
\item In the outskirts we find several extended H{\sc i} bridges and
optical disturbances, which indicate that the systems are gravitationally
interacting.
\end{enumerate}

In Paper~{\sc ii} we will do a statistical analysis of our H{\sc i} 
imaging results and discuss the importance of various environmental
effects on the evolution of cluster galaxies. 

\acknowledgments

The VIVA collaboration has been growing over time. We are grateful 
for many useful discussions with our colleagues who have joined
more recently, David Schiminovich, Eric Murphy, Tomer Tal, Anne Abramson,
Ivy Wong, Tom Oosterloo. We thank the ALFALFA consortium for making
their data so promptly available to the scientific community. We thank
the anonymous referee for comparing the VIVA data to single dish data
from GOLDMINE and providing us with plots of the excellent
agreement. This work has been supported by NSF grants 00-98294 and
06-07643 to Columbia University and by grant 00-71251 to Yale University.
This work has been supported by NASA grant 1321094.
This research has made use of the NASA/IPAC Extragalactic Database (NED) 
which is operated by the Jet Propulsion Laboratory, California Institute 
of Technology, under contract with the National Aeronautics and Space 
Administration. The Digitized Sky Survey was produced at the Space Telescope
Science Institute under US goverment grant NAG W-2166. 
This publication makes use of data products from the Two Micron All 
Sky Survey, which is a joint project of the University of
Massachusetts and the Infrared Processing and Analysis 
Center/California Institute of Technology, funded by the 
National Aeronautics and Space Administration and the National 
Science Foundation.
Funding for the SDSS and SDSS-II has been provided by 
the Alfred P. Sloan Foundation, the Participating Institutions, the 
National Science Foundation, the U.S. Department of Energy, the National 
Aeronautics and Space Administration, the Japanese Monbukagakusho, the 
Max Planck Society, and the Higher Education Funding Council for England. 
The SDSS Web Site is {\tt http://www.sdss.org/}.

\appendix

\section{Comments on Individual Galaxies}
\label{sec-indiv}

In this section, we describe the H{\sc i} morphology and kinematics of
individual galaxies in detail and compare them with data at other wavelengths.
 Unless otherwise mentioned, the optical $R$-band and H$\alpha$
morphology and surface brightness profiles are from \citet{kky01} and
\citet[][b]{kk04a}. For the radio continuum emission we refer to our own 
1.4~GHz continuum data, otherwise references are given.
\bigskip

\noindent{\bf NGC~4064} 
The H{\sc i} extends to only about one fifth
of the stellar disk ($<4~$kpc) and might be slightly extended to the NE.
Optically, NGC~4064 has a relatively undisturbed outer stellar disk, with
a strong central bar that smoothly connects with open spiral arms in the
outer disk. It has strong star formation in the central 1 kpc but virtually no
H$\alpha$ emission beyond.  Strong radio continuum emission from the
central kpc is roughly coincident with the circumnuclear string of luminous HII
regions.
A detailed morphological and kinematical study of the central regions
of NGC~4064 is presented by C$\acute{\rm o}$rtes, Kenney \& Hardy (2006). 
Along the bar,  the stellar, molecular (CO) and ionized (H$\alpha$) gas
velocity fields show strong non-circular motions indicative of radial streaming
out to a radius of at least 1.5 kpc.
The H{\sc i} velocity field shows no evidence of non-circular motions, but this may be
because the bar is not resolved at the resolution of the H{\sc i} data.
It is somewhat of a puzzle why this galaxy has such a severely
stripped H{\sc i} disk.  The galaxy is located in the outskirts of the
cluster ($d_{87}=$2.5~Mpc), which makes ongoing
ram-pressure stripping due to the ICM seem unlikely. \citet{ck08}
estimate that star formation in the stripped part of the disk
got quenched only 425 Myr ago, while it would take about 2 Gyr for the
galaxy to travel from the core to its current location. Thus the gas
has not all been stripped in the cluster core.
Although some galaxies appear to be stripped by ICM-ISM ram pressure
at surprisingly large cluster distances, perhaps due to a dynamic lumpy ICM
\citep{kvgv04,ck08},
NGC~4064 does not seem fully consistent with this scenario.
While the outer stellar disk looks undisturbed,
the large radial gas motions and circumnuclear starburst suggest
a recent gravitational interaction.
NGC~4064 may have experienced the combined effects of a gravitational
interaction and gas stripping in the cluster outskirts \citep{tbvg07}
although the details remain uncertain.

\begin{figure}
\plotone{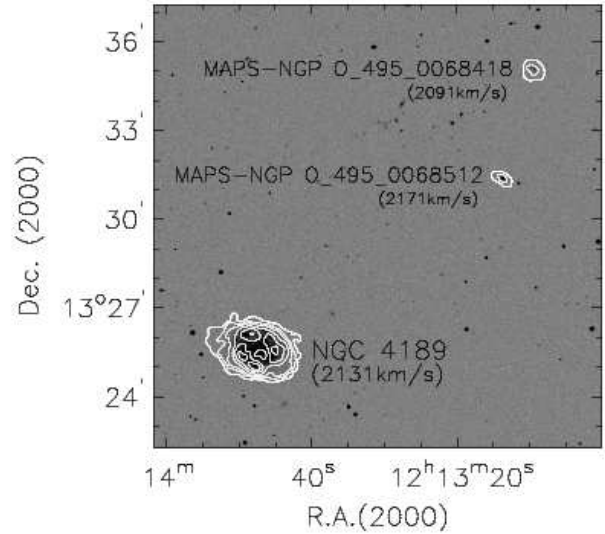}
\caption{The H{\sc i} distributions of NGC~4189 and two dwarf neighbors are shown in 
white contours overlaid on the Digital Sky Survey (DSS) image. \label{fig-4189}}
\end{figure}

\noindent{\bf NGC~4189} The H{\sc i} disk is slightly more extended than 
the optical disk. The velocity field and position velocity slices show that
the disk has a symmetric warp.
Enhanced H{\sc i} emission is found to the south east,
where a clumpy H$\alpha$ ridge is present.  The radio continuum 
shows enhanced emission along that same ridge in the south east. 
Its Tully Fisher distance estimate puts it significantly
further than the Virgo mean distance \citep{gbspb99, sssgh02}, 
and \citet{bst85} argue that it belongs to the M cloud. 
H{\sc  i} emission has been also detected from two dwarf galaxies at
similar velocities within $<50~$kpc distances (Fig.~\ref{fig-4189}).
This makes it even more likely that NGC~4189 is in the background.
If close to the cluster, the dwarfs would have been stripped of their H{\sc i}. 

\noindent{\bf NGC~4192} The H{\sc i} is more extended to the south east
which is clearly seen in the PVD and the flux density profile. The
velocity field shows distortion in the center which might be due to
the presence of a bar \citep{warmels88,bosma81}. The outer H{\sc i}
disk shows a warp. \citet{warmels88}
reported an extended disk emission in continuum at 1.4~GHz, which we
also have detected in spite of a small number of line-free
channels. It has been classified as normal in H$\alpha$. See also
\citet{cvgbk90}.

\begin{figure}
\plotone{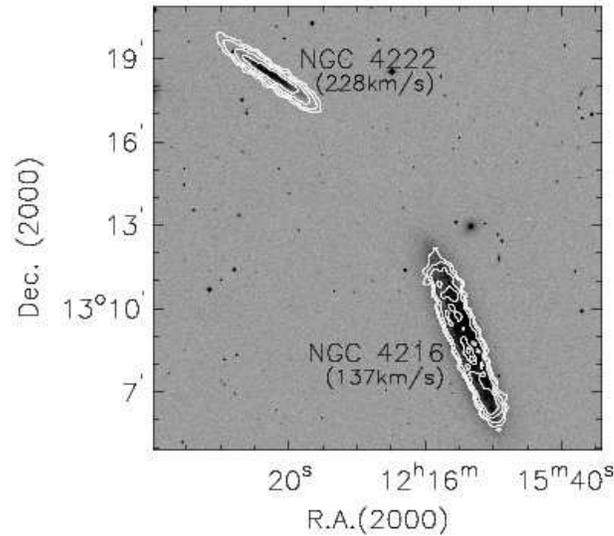}
\caption{The H{\sc i} distributions of NGC~4216 and NGC~4222 are overlaid on the DSS image. 
Both galaxies appear to be warped in the outer H{\sc i} disk. 
\label{fig-4216}}
\end{figure}

\noindent{\bf NGC~4216} It is fairly regular in the H{\sc i}
distribution but has a depression in the center as shown by the radial
surface profile. Its H{\sc i}
extent is slightly less than the optical extent. 
The overall H{\sc i} surface density, however, is low for a spiral 
galaxy of this size. It is its H{\sc i} surface density,
that makes this galaxy H{\sc i} deficient. 
The velocity field looks regular but shows non-circular motions 
in the outer parts (see the velocity field).
No radio continuum has been
detected.  A prominent dust lane is present. Perhaps this galaxy has been 
affected by a process that affects the entire surface of the disk,
such as viscous turbulent stripping or thermal evaporation.

\noindent{\bf NGC~4222} The H{\sc i} extent is larger than the optical extent. 
The gaseous disk appears to be warped in the southwest as is apparent from
the H{\sc i} velocity field and the PVD. This galaxy was
found in the same field as NGC~4216 ($d\approx56$~kpc in projection,
$\Delta v=280$~km~s$^{-1}$, see Figure~\ref{fig-4216}).  We do not
find any clear signatures of interactions between the two, although it
is interesting to see mild distortions at the edge of the disk in both
systems. Unlike NGC~4216, NGC~4222 is not H{\sc i} deficient and does
not have an unusually low H{\sc i} surface density, yet there is one more 
similarity, it is also not detected in radio continuum. 

\noindent{\bf NGC~4254} As a prototypical one armed spiral galaxy,
NGC~4254 has been the subject of many studies. 
The H{\sc i} morphology is highly asymmetric
with a low surface brightness extension to the north. 
Optically, the one outstanding spiral arm winds
around from the south to the southwest. The H{\sc i} follows this
stellar arm. However just south of this arm, faint H{\sc i} emission
can be seen in the total H{\sc i} image. These are the peaks of 
a giant H{\sc i} tail imaged with Arecibo by Haynes, Giovanelli
and Kent (2007). The tail  wraps around NGC~4254 in the west, 
and then extends  north for a total length of 250 kpc.
Most of its velocity is around 2000 km~s$^{-1}$, just outside the velocity
ranged probed by the VLA. In that sense it is reminiscent of the
long tail found near NGC~4388 \citep{ovg05}. The fact that both tails
cover a velocity range well outside that of the associated galaxy is perhaps
good evidence that these galaxies really are inside the cluster.
The tail feels the additional cluster potential. There are however
important differences between the tails, NGC~4388 is close
to the center of Virgo, and its tail is almost certainly due to
ram pressure stripping \citep{ovg05}, while NGC 4254 is further from the center
of Virgo. \citet{hgk07} argue that the tail of NGC~4254
is probably the result of galaxy harassment, but simulations of
\citet{db08} show that the morphology can be reproduced by one rapid 
close flyby.


\begin{figure*}
\plotone{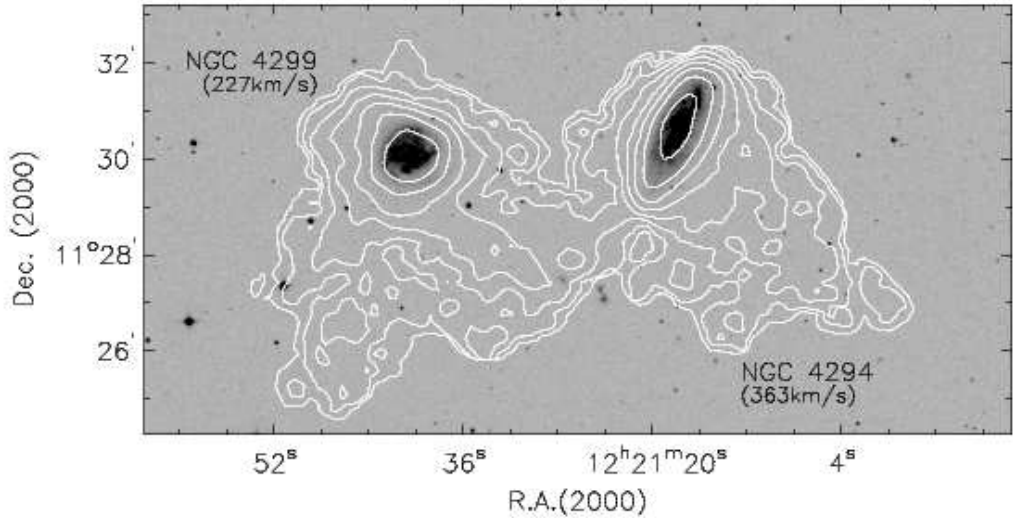}
\caption{The H{\sc i} distributions (C+D) of NGC~4294 and NGC~4299 are overlaid on the DSS image. 
Note that both tails are pointing to the same direction to the southwest. The C+D image has revealed
a longer tail of NGC~4294 and the second tail of NGC~4299 to the southeast more clearly. This pair
of galaxies are located at $\sim0.7~$deg distance in projection to the southwest of M87.
\label{fig-tango}}
\end{figure*}

\noindent{\bf NGC~4293} It is one of the two most H{\sc i} deficient
galaxies ($def_{\rm HI}>2$) in the sample. The azimuthally averaged
H{\sc i} surface density is everywhere below $1~M_\odot$~pc$^{-2}$ and the
H{\sc i} isophotal diameter ($D_{\rm HI}^{\rm iso}$) is not
defined. There is no  H{\sc i} emission in the center but its central
radio continuum is pretty strong and the emission may be hidden by absorption.
This means that in reality we have a lower limit to the total H{\sc i} mass and
an upper limit to the H{\sc i} deficiency.  
In fact weak redshifted H{\sc i} absorption is seen
in the center which indicates the presence of non-circular
motion.  It is truncated/anemic in H$\alpha$ morphology. The
misalignment between the outer stellar envelope and the inner stellar
disk suggests a gravitational disturbance (Cort{$\rm\acute{e}$}s 2005).
This may be responsible for both the truncation in H{\sc i} and
non-circular motion in the center.

\noindent{\bf NGC~4294} ({\it Tango~I}) Within the stellar disk, the
H{\sc i} morphology and kinematics are quite regular.  The H{\sc i} is
slightly more extended to the southeast but the asymmetry is not
significant along the major axis. Along the minor axis however, a long
H{\sc i} tail is found on one side to the southwest, which had been
completely missed by \citet{warmels88}.  The length of the tail is
about 23~kpc. The full data set, including both C and D-array data, 
show that the length of the tail is about 27 kpc.
The H{\sc i} tail does not have a stellar counterpart
 down to a limiting magnitude
of $r=26$ mag~arcsec$^{-2}$ in the Sloan Digitized Sky Survey. 
The stellar disk looks more diffuse in the southeast
while a strong spiral arm can be seen the northwest side of the disk. The
star formation property has been classified as normal \citep{kk04b}.
However the H$\alpha$ morphology is quite asymmetric in the same way as
the radio continuum with more emission to the northwest
\citep{kky01}. 
The lack of a stellar counterpart makes it less likely that a tidal interaction
is the only cause of the tail. However, NGC~4299 is only 27~kpc away,
and the two galaxies have almost the same velocity, 
with $\Delta v\approx120~$km~s$^{-1}$.  A  gravitational interaction
between the two galaxies cannot be ruled out (Fig.~\ref{fig-tango}).
See also the comments on NGC~4299 and \citet{cvgkv07}.

\begin{figure}
\plotone{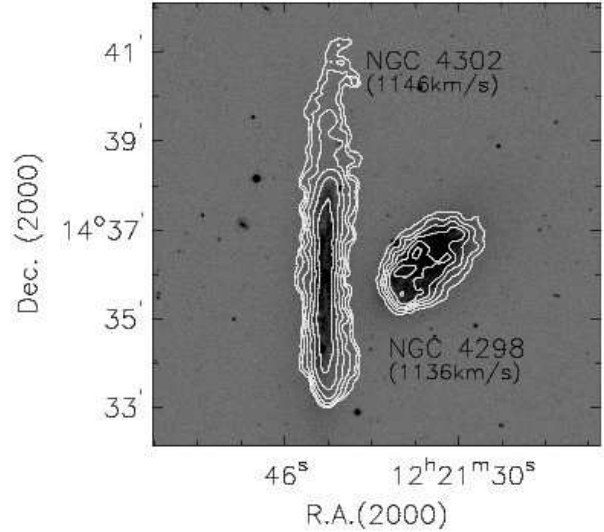}
\caption{The H{\sc i} distributions of NGC~4298 and NGC~4302 are overlaid on the DSS image. 
They are only 11~kpc and $\approx$30~km~s$^{-1}$ away from each other. This pair of galaxies
are located at 0.9~deg distance to the northwest of M87 with velocities close to the cluster
mean motion.\label{fig-4298+02}}
\end{figure}

\noindent{\bf NGC~4298} The H{\sc i} is more extended and diffuse to
the northwest while the other side of the disk shows a sharp cutoff
(see Fig.~\ref{fig-4298+02}). 
The H{\sc i} rotation
curve is also slightly asymmetric, rising more steeply in the north west.
The radio continuum emission shows a central point source and extended
emission to the south east, coinciding with the compressed H{\sc i}.
The stellar disk on the other hand is more extended to the north west
at the opposite side of the H{\sc i} compression. 
NGC~4302 is only 11~kpc
away in projection with $\Delta v\approx30$~km~s$^{-1}$. The two
galaxies are likely to be a physical pair, and an interaction might
well have caused the lobsidedness of NGC~4298. 
A tidal
interaction is less likely to be solely responsible for the H{\sc i}
morphology of NGC~4302.  See also \citet{cvgkv07}.

\noindent{\bf NGC~4299} ({\it Tango~II}) H{\sc i} compression is
seen to the southwest and a long H{\sc i} tail is found to the
southwest. This tail is pointing to the same direction as NGC~4294's
H{\sc i} tail without a stellar counterpart down to the SDSS
limit. The fact that the tails are parallel makes it less likely that
it is only a tidal interaction which has caused the tail.  However, NGC~4299
shows some hints for tidal interactions as well. Optically it has a
very small/weak bulge and weak spiral arms which are highly
asymmetric. Also as shown in Figure~\ref{fig-tango}, a much broader
tail H{\sc i} is found to the southeast, which looks similar to 
tidal debris \citep{me03}. As we argue in \citet{cvgkv07}, the pair,
NGC~4294 and 99 are likely to be under the influences of both ICM
pressure and as tidal interaction.  The H$\alpha$ emission is
enhanced to the south.

\noindent{\bf NGC~4302} The H{\sc i} is mildly truncated within the
optical disk to the south while a long tail is present to the
opposite side \citep[Fig.~\ref{fig-4298+02}, see also][]{cvgkv07}.
Its H{\sc i} PVD presented here and in \citet{cvgkv07} shows a
{\lq\lq}figure-of-eight{\rq\rq} feature. NGC~4302 appears to be
box-shaped in the optical. These features suggest an edge-on view of a
thickened bar \citep{bf99}. The radio continuum is found along the
prominent dust lane. The mild truncation of the H{\sc i} to within
the optical disk to the south suggests that this galaxy is undergoing
ram pressure stripping, possibly also the cause of the H{\sc i} tail
seen to the north \citep{cvgkv07}. 

\noindent{\bf NGC~4321} This galaxy has an H{\sc i} disk which is
slightly larger than the optical disk. It has a low surface brightness
extension to the south west, coinciding with a very faint optical
arm. The velocity field shows that the disk is slightly warped in 
the south west. An H{\sc i} depression is present in the
center. NGC~4321 has a nuclear stellar bar and a prominent ring of 
circumnuclear star forming regions \citep{wdh98}. The radio continuum
emission peaks in the center possibly contributing to the central
depression in H{\sc i}. A faint optical bridge is present to the
northeast which is connected to a dwarf companion, NGC~4323, at only
24~kpc distance away in projection,  with a velocity difference of  $\Delta
v\approx300~$km~s$^{-1}$.

\noindent{\bf NGC~4330} The H{\sc i} is much more extended to the southern 
side of the disk than the northern side. The northeast side of the H{\sc i}
disk ends with a sharp cutoff. 
On the other side an extended H{\sc i} tail is seen, starting well within the 
optical disk, but curving south-southwest. Along the western edge
of the tail, radio continuum emission is seen. There is no optical
counterpart down to the SDSS surface brightness limit \citep{cvgkv07}, but  
the western side of the H{\sc i} tail is also detected by
GALEX (Fig.~\ref{fig-uvpec}), indicative of recent star formation. 
\citet{cvgkv07} have argued that the
galaxy is undergoing ram-pressure stripping as it enters the cluster
for the first time. More extensive discussion based on
multi-wavelength data on this particular galaxy will appear in
Abramson et al. (2009, in prep).

\noindent{\bf NGC~4351} ({\it Stubby tail}) 
The H{\sc i} shows a short
and broad extension in the southwest and a sharp cutoff to the
northeast.
It has a modest H{\sc i} deficiency, and the H{\sc i} and optical extents are similar.
Its PVD is also asymmetric, showing a flat velocity
gradient to the northeast but a decreasing velocity gradient to the
southwest.  The radio continuum is quite strong to the northeast, the
same side where the H{\sc i} compression is found. Its location
($d_{87}\lesssim0.5~$Mpc) and extreme velocity w.r.t.  Virgo ($\Delta
v>1000~$km~s$^{-1}$) suggest that the galaxy may be radially falling
into the cluster, and possibly experiencing strong ICM ram pressure.
However, 
the nucleus and the optically brightest parts of the inner galaxy are significantly
offset from the centroid of the outer galaxy isophotes,
and the outer stellar disk shows suggestions of shell-like structure.
Thus some of the galaxy's peculiarities are
likely caused by a gravitational disturbance.

\noindent{\bf NGC~4380} The H{\sc i} distribution shows a mild
asymmetry in a sense that the northwest disk is overall more dense
compared to the southeast disk which shows a slightly low H{\sc i}
surface density but extended. In the middle, little H{\sc i} is
found and the radio continuum is very weak. Optically, the galaxy has
a very weak bulge with a low surface brightness stellar disk without
clear spiral features. Its H$\alpha$ morphology has been classified as
truncated/anemic. There is a hint of a stellar ring to the northwest
where the highest H{\sc i} column density is found. The galaxy is
somewhat H{\sc i} deficient without any obvious signatures of ongoing
or recent ICM-ISM interaction.

\noindent{\bf NGC~4383} ({\it Crazy})
This is one of the most H{\sc i} rich
galaxies in the sample with $def_{\rm HI}\lesssim-0.8$. The H{\sc i}
extent is enormous compared to the optical disk ($D_{\rm
 HI}^{\rm iso}/D_{25}>4$).
Within the stellar disk, its H{\sc i}
kinematics and morphology seem fairly regular, although
even within the central 2$'$
the major axis PVD shows some low velocity gas and
the minor axis PVD shows some  non-circular motions.
Beyond this radius, the H{\sc i} distribution and kinematics are irregular.
There is a clear kinematical distinction between the inner
(within the optical disk) and the outer H{\sc i}, with
different kinematic major axes suggestive of an irregular warp.
Along the entire eastern side there is a sharp kinematical discontinuity
$\sim$2-3$'$ from the nucleus, whereas in the west the
transition is more gradual.
The outer H{\sc i} shows weak m=2 spiral structure, with a weak arm to the
south-southeast,
and a somewhat stronger one to the north-northwest.
The outer H{\sc i} in the east-northeast forms a single irregular arm 
unrelated to the
m=2 pattern.
Along the same eastern side as this irregular arm, but
beyond the main body of H{\sc i} there are two distinct gas clouds:
one to the east just beyond the main body of H{\sc i}, but with a velocity
which is $\sim$20 km~s$^{-1}$  offset, and the other 7$'$  to the
southeast. It is not impossible that both clouds are high surface
brightness peaks in a more extended very low surface brightness structure.  
H$\alpha$ and UV emission are observed from roughly the same area as
the stellar disk, with very little from the region of extended H{\sc i},
except for a $UV$ counterpart to the inner H{\sc i} spiral arm in the north.
NGC~4383 is a starburst galaxy, with strong H$\alpha$ and UV emission from the
central $\sim$1$'$, and
a biconical H$\alpha$ nebulosity along the minor axis
suggestive of an outflow.
The galaxy is likely influenced by a combination of
gas accretion and tidal interaction.
The small galaxy 2.5 arcmin to the southwest of NGC~4383 is UGC~07504
(VCC~0794), a Virgo cluster galaxy with a velocity of 918 km~s$^{-1}$.  This
velocity is offset by 800 km~s$^{-1}$ from NGC~4383, thus they are not
gravitationally bound, and probably not physically associated.  The
velocity range of UGC~07504 is entirely outside the range of our VLA
H{\sc i} observations, however the galaxy is undetected in radio continuum,
H$\alpha$, and single dish H{\sc i} observations (ARECIBO-05).

\noindent{\bf NGC~4388} Our VLA data show that the H{\sc i} is very
deficient, truncated and asymmetric within the stellar disk, with a
much greater extent to the east. There is an H{\sc i} "upturn"
extending vertically upwards (north) from the outer edge of the H{\sc
  i} disk in the west, suggestive of ongoing ram pressure from the
southwest.  WSRT observations show a $\sim100~$kpc long plume of H{\sc
  i} extending toward the northwest of NGC~4388 \citep{ovg05}.  The
H{\sc i} mass in this plume is similar to the H{\sc i} mass remaining
in NGC~4388.  The plume is smoothly connected with NGC~4388 both
spatially and in velocity and has no optical counterpart, suggesting
that the H{\sc i} plume is gas that was ram pressure stripped from
NGC~4388 within the last few hundred Myrs.  This plume was missed in the
VIVA survey due to the limited bandwidth and reduced sensitivity at
large distances from the field center.  The highest H{\sc i} surface
densities of the plume gas occur beyond the halfwidth of the primary
beam of the VLA and at that point the velocities of the gas are
outside the velocity window of our observations. We note that the VIVA
integrated profile is very asymmetric with an excess of H{\sc i} at
the high velocity side, where the plume connects with the
disk. \citet{yoshida02} and \citet{kenney08} find very extended
extraplanar H$\alpha$ emission which might have originated from
stripped cool ISM shocked by the hot ICM and/or the central AGN. This
ionized gas is almost certainly part of the HI tail. 
Both
the major and minor axis PVD's show redshifted absorption, indicative
of non-circular motions.  NGC~4388 was the first Seyfert galaxy
discovered in Virgo \citep{phm82}, and its nuclear activity has been
detected at many wavelengths
\citep[e.g.][]{vbctm99,yoshida02,iwfy03}. It was also one of the first
spiral galaxies in which anomalous radio continuum was detected with
an elongated component crossing the nucleus and perpendicular to the
optical disk \citep{hvgk83}.  Here we detect strong radio continuum
emission from the AGN, as well as emission from the star-forming disk
with almost the same extent and asymmetry as the H{\sc i} disk.

\noindent{\bf NGC~4394} ({\it Fruit Loop or Life Saver}) The H{\sc i}
is quite deficient and is mildly truncated within the stellar disk.
The H{\sc i} morphology and kinematics are remarkably regular and symmetric.
An H{\sc i} hole is found in the middle which is quite common for
strongly barred galaxies like NGC~4394.  Optically the galaxy is
slightly asymmetric in the sense that the northeast spiral arm is more
prominent than the one in the southwest.  Its H$\alpha$ classification
is anemic, and its radio continuum emission is correspondingly very
weak.  It has a large apparent neighbor, the S0 galaxy NGC~4382 (M85)
at $\approx35$~kpc projected distance with a velocity difference of
$\Delta v\approx200~$km~s$^{-1}$. Although NGC~4382 has stellar shells
suggesting that it might be a merger remnant \citep{ss92} we find
neither in H{\sc i} nor at other wavelengths any signatures of a tidal
interaction between the two galaxies.

\noindent{\bf NGC~4396} ({\it Crocodile}) 
This galaxy has a prominent H{\sc i} tail to the northwest
\citep{cvgkv07} but unlike other
Virgo galaxies with H{\sc i} tails clearly caused by ram pressure,
the origin of NGC~4396's tail is unclear.
H{\sc i} contours are compressed to the south, but not at the southeast end
of the major axis, as would be expected if this were the leading
edge of a strong ICM-ISM interaction (as is seen in NGC~4330, NGC~4388,
NGC~4402).
H$\alpha$ and broadband images show that the distribution of star formation is
asymmetric with only one prominent spiral arm found to the southeast.
Radio continuum emission is strong in the center, and within the disk it is
stronger in the southeast where the one prominent H$\alpha$ spiral arm is
located.
There is no radio continuum or UV emission associated with the 
H{\sc i} tail,
contrary to what is seen in NGC~4330, NGC~4402, and NGC~4522.
Deep optical imaging reveals that the outer NW stellar disk is
gas-free and the H{\sc i} tail leaves the stellar disk, consistent
with a ram pressure stripping origin. However, there is no ``radio
deficit'' observed at the SE outer edge \citep{mkhch09}, as expected
for strong active ram pressure, thus the origin of the features in 
NGC~4396 are not completely clear.

\begin{figure*}
\plotone{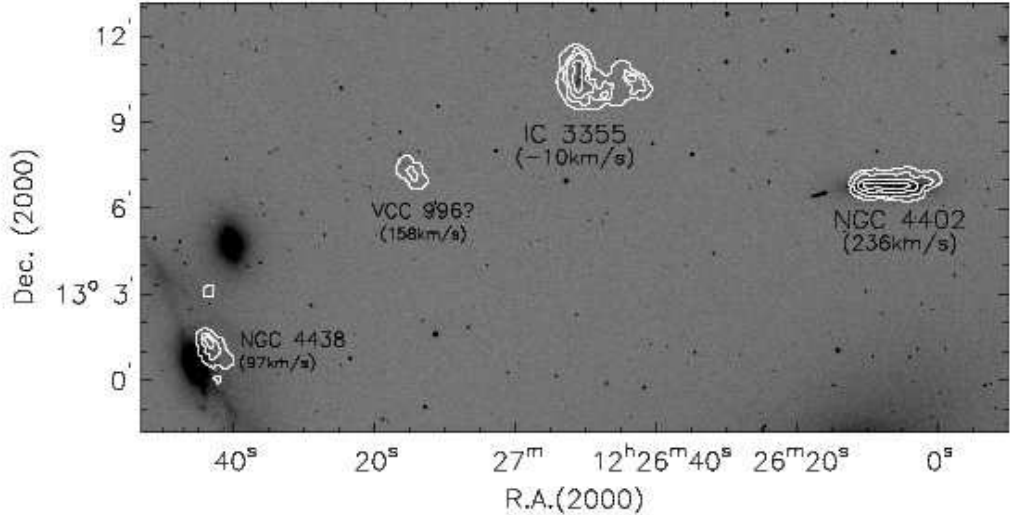}
\caption{The H{\sc i} distributions of NGC~4402 and its neighbors 
are overlaid on the DSS image. The projected distance between NGC~4402 
and NGC~4438 is $\approx122~$kpc at the Virgo distance.
\citet{mhfm05} have presented a deeper optical image of this region.
\label{fig-4402}}
\end{figure*}

\noindent{\bf NGC~4405} 
The H{\sc i} is highly deficient and strongly truncated, and the stellar disk
appears normal. The H$\alpha$ image shows relatively normal star formation in
the
central 30\% of the stellar disk and is sharply truncated beyond that.
There are no compressed H{\sc i} contours or significantly disturbed kinematics,
so there is no evidence for ongoing ICM pressure.
Its radio continuum is quite strong and slightly
asymmetric with a possible extension to the southwest.
Its properties are consistent with a strong ram pressure stripping event
at least a few hundreds Myr ago \citep{ck08}.

\noindent{\bf NGC~4402} The H{\sc i} is moderately deficient,
moderately truncated within the undisturbed stellar disk, and
asymmetric.  Within the disk, the H{\sc i} extends further west than
east.  The H{\sc i} contours are compressed in the SE and a modest
H{\sc i} tail exists to the NW, suggesting active ram pressure from
the SE.  The radio continuum has an extended tail to the NW, extending
further from the disk than the H{\sc i}. The position angle of this
tail matches those of elongated dust filaments which \citet{ckvgv05}
have interpreted as dense clouds being ablated by ram pressure.
Further evidence of strong ram pressure acting from the SE comes from
the enhanced radio continuum polarization \citep{vsb07} and the radio
continuum deficit region in this area \citep{mkhch09}.
As shown in Figure~\ref{fig-4402}, several
neighbors have been detected in the same field: IC~3355, which is in the
VIVA sample, and NGC~4438, which is known for its optical
peculiarities with a small H{\sc i} disk. We also found a small H{\sc i}
cloud somewhere between IC~3355 and NGC~4438 at the velocity of
$\approx160~$km~s$^{-1}$.  It seems very possible that VCC~996
($v=-28$~km~s$^{-1}$), whose optical center is only only 55$''$ from
the H{\sc i} peak of the cloud, is its optical counterpart.

\noindent{\bf IC~3355} ({\it Casper}) This is a peculiar low-mass,
H{\sc i}-rich object whose origin is unclear.  Optical images show an
elongated stellar body with no clear nucleus, a high surface
brightness ridge with a sharp cutoff in the east, and lower surface
brightness emission extending to the west.  H{\sc i} emission covers
the entire stellar body, with two peaks that straddle the galaxy
center. There are no compressed contours and no clear evidence of
active ram pressure.  The H{\sc i} velocity field shows a very modest
gradient across the main body, and the overall pattern is not one
dominated by rotation.  A small H{\sc i} cloud with no known optical
counterpart is present 2$'$ to the west, connected by an H{\sc i}
bridge which extends from the southern part of the stellar body.  The
small H{\sc i} cloud exhibits a larger velocity gradient ($\sim$50
km~s$^{-1}$) than the main body of the galaxy, but there is no clear
rotation pattern.  The line-of-sight velocity of IC~3355 is near zero,
and its membership to Virgo is controversial \citep{hhs88,cvgbk90}.
However, it is unlikely to be in the Local Group, since the stellar
light distribution is too smooth for such a nearby system.  If it is
in the Virgo cluster, its velocity and proximity on the sky to M86
make it very likely that this galaxy is associated with the M86 group.
This galaxy is located 11.5$'$ ($\sim55~$kpc projected) from the
spiral NGC~4402, $\sim$16.9$'$ ($\sim79~$kpc projected) from M86, and
$\sim$16.8$'$ ($\sim78~$kpc projected) from the disturbed spiral
NGC~4438.  However, its Virgo membership might be questioned because
1) more massive galaxies in this neighborhood are severely stripped in
H{\sc i} (e.g. NGC~4438; Cayatte et al. 1990 and NGC~4402; Crowl et al.
2005),
while this galaxy is H{\sc i} rich and 2) in a recent deep optical
image of this region \citep{mhfm05}, this galaxy appears to be located
outside of a huge common envelope of intracluster light (ICL) around
the M86 group.  Nonetheless, it is possible that IC~3355 is a member
of the M86 group and has not yet been gas stripped.  Its H{\sc i}
properties suggest it may have been disturbed by a gravitational
interaction.

\noindent{\bf NGC~4419} The H{\sc i} is highly deficient and severely
stripped within the stellar disk ($D_{\rm HI}^{\rm iso}/D_{25}<0.5$).
It has very strong radio continuum emission from a nuclear source,
likely an AGN \citep{decarli07}, plus weaker extended emission
co-spatial with the H$\alpha$ emission that traces the anemic star
formation in the disk.
Some H{\sc i} is observed in absorption against the bright nuclear
source, as shown in the PVD diagrams. This absorbing gas is redshifted
by $\sim$100 km~s$^{-1}$ with respect to the nucleus, providing evidence
for non-circular motions in the central region.  The presence of
absorption has reduced the amount of H{\sc i} seen in emission, and
accounts for the relative paucity of H{\sc i} emission observed near
the nucleus.  Thus the plots of the integrated H{\sc i} emission and
the radial distribution underestimate the amount of H{\sc i} present
near the nucleus.  This is the only galaxy for which the otherwise
excellent agreement with linewidths measured by ALFALFA breaks down,
the width measured with VIVA is almost 200~km~s$^{-1}$ larger. However
the profile shapes are in very good agreement and we suspect that
ALFALFA must have been misled by the absorption and not included the
component at -100~km~s$^{-1}$. Optical images show an undisturbed
stellar disk. There is neither extraplanar H{\sc i} emission detected
in this highly
inclined spiral, nor compressed H{\sc i} contours, so there is no
direct evidence of ongoing ram pressure.  There may however be some
hints of rather recent stripping.  The outer H{\sc i} extent is
somewhat asymmetric with more emission in the SE part of the disk, and
optical images show disturbed dust lanes. CO emission is also strongly
asymmetric in the disk of NGC~4419, but in the opposite sense from the
H{\sc i}, since there is more CO to the NW than to the SE
\citep{kenney90}. The stellar population
study of \citet{ck08} shows that star formation in the gas-free outer
disk stopped $\sim500~$Myr ago, and the galaxy could be experiencing
fall-back after peak ram pressure.

\noindent{\bf NGC~4424} ({\it Jellyfish}) This is a very peculiar
H{\sc i}-deficient galaxy. It is one of the galaxies with long one-sided
H{\sc i} tails pointing away from M87 \citep{cvgkv07}.  The stellar
disk is strongly disturbed, with shells and banana-shaped isophotes.
There is strong star formation in a bar-like string of HII complexes
in the central 1 kpc, and no star formation beyond \citep{kkry96}.
The radio continuum is quite strong in the circumnuclear region, and
shows two distinct off-nuclear peaks coincident with the brightest
star forming complexes.  The central region also contains disturbed
dust lanes and disturbed ionized gas kinematics \citep{cortes06}.
This all clearly indicates a strong gravitational interaction, either
a merger or close collision.  \citet{cvgkv07} find that the H{\sc i}
extent is much smaller than the optical disk along the major axis
while it has a remarkably long H{\sc i} tail to the south
($\approx$18~kpc at least). 
One end of the tail is
curved to the southeast, pointing almost directly to M49, the giant
elliptical at the center of the M49 sub-cluster, which is
$\approx0.44~$Mpc away.  In a recent follow-up study with the WSRT
(Oosterloo, van Gorkom \& Chung, in preparation) the H{\sc i} tail
appears to extend over $>40~$kpc. Interestingly, the H{\sc i} cloud
that \citet{ste87} found near M49 is at the same velocity as
NGC~4424's H{\sc i} tail.  This suggests a collision between NGC~4424
and M49 as a possible explanation for the peculiarities of NGC~4424.
Further study is required to distinguish between this and other
scenarios.

\noindent{\bf NGC~4450} The H{\sc i} is overall moderately truncated
with low surface density. Less H{\sc i} is detected along the minor
axis but this could be because the low surface density gas is just
below our detection threshold on the minor axis, where gas is spread
out more in velocity.  An H{\sc i} depression is present in the center
where there is a weak stellar bar. Strong radio continuum is detected
from the nucleus, probably from an AGN.  Optically the galaxy shows
tightly wound but weak spiral structure and anemic star formation. No
obvious evidence for tidal or ongoing ICM-ISM interactions are found.

\noindent{\bf IC~3392} The H{\sc i} is severly truncated within the
undisturbed stellar disk with $D_{\rm HI}^{\rm iso}/D_{25}<$0.5.  In
the channel maps there is a hint of an H{\sc i} extension along the
minor axis, to the northwest.  The H{\sc i} distribution is fairly
symmetric, there are no compressed H{\sc i} contours, and the H{\sc i}
velocity field is regular, thus we do not find any signatures of
ongoing pressure.  The stellar population study of \citet{ck08} shows
that star formation in the gas-free outer disk stopped $\sim500~$Myr
ago, thus the galaxy seems to have been gas-stripped a while ago and
is no longer in an active phase of stripping.  It is located only
$\approx125~$kpc away in projection from NGC~4419 but the systemic
velocities differ by ($\Delta v\sim 1900~$km~s$^{-1}$), and the
galaxies are unlikely to be physically related.

\noindent{\bf NGC~4457} The H{\sc i} extent is smaller than the
optical disk and quite asymmetric with the H{\sc i} peak offset from
the optical center toward the southwest.  This H{\sc i} peak coincides
with the galaxy's one strong and peculiar spiral arm, which is very
prominent in H$\alpha$.  Quite strong radio continuum is present with
almost the same extent as the H{\sc i}.  There are no compressed H{\sc i}
contours, and therefore there is no evidence for active ICM pressure.
The velocity field suggests that the H{\sc i} disk is slightly warped,
but there are no obvious signature of a tidal interaction in the H{\sc
  i} kinematics or optically.

\noindent{\bf IC~3418} ({\it Ghost}) This is a peculiar low surface
brightness system (IBm) found $\sim1^\circ$ to the southwest of M87,
with a tail of UV emission to the SE.  We have not detected H{\sc i}
in this galaxy over the velocity range we searched (-250 to 250 km~s$^{-1}$).
Until very recently, the galaxy's velocity and even its Virgo 
membership was not
well-known, and H{\sc i} could exist outside our search window.  Within
our velocity window, the H{\sc i} upper limit is $\sim8 \times
10^6~M_\odot$ assuming a linewidth of 100~km~s$^{-1}$.  The first
velocity measured using optical spectroscopy was
25662$\pm74$~km~s$^{-1}$ but with low reliability \citep{dcyhy96}. The
observed very extended $UV$ morphology (Fig.~\ref{fig-uvpec}) makes it
extremely unlikely that the galaxy is that distant.  More recently
\citet{gzsbb04} published an optical spectroscopic velocity of
$38$~km~s$^{-1}$ which we adopted for the H{\sc i}
observations. This redshift was recently confirmed (Crowl, private
communication) in a spectrum taken with LRIS at Keck. The velocity measured is
$\approx 0 km~s^{-1} \pm 50 km~s^{-1}$.
Despite our non detection it is still possible, that the
galaxy has a very faint and extended H{\sc i} disk.  If it were
smoothly distributed over an area of $3\times1$~arcmin$^2$ as it is in
$UV$, the noise goes up roughly by $\sqrt{3\times13.9}$ ($\approx6.5$)
with CS array (i.e. proportional to square root of number of
independent beams), which could have been missed in the VIVA study.
We note that it has also not been detected in ARECIBO-05 and with
ALFALFA, which covers the entire velocity range of the Virgo cluster.
It is quite possible that its H{\sc i} has been completely
stripped if the galaxy is a true member of Virgo and close to M87 (its
projected distance to M87 is only 0.28~Mpc). In that case
the $UV$ stream may have originated from recent star formation due to the
compression of the stripped H{\sc i} gas.

\noindent{\bf NGC~4501} 
This large Sc galaxy is mildly H{\sc i}-deficient.
To the SW the H{\sc i} disk is mildly truncated within the stellar disk
and has compressed contours, whereas to the NE the H{\sc i} is more extended and
diffuse.
These features have long been known from previous data
\citep{warmels88,cvgbk90},
and are suggestive of an ongoing ICM-ISM interaction \citep{cvgbk90}.
Our new data with better resolution and sensitivity
show additional key details, such as the
H{\sc i} arm in the outer NE region with disturbed kinematics.
Detailed comparisons with simulations
suggest that NGC~4501 is
in an early stage of ram pressure stripping \citep{vsc08}.
The compressed H{\sc i} contours in the SE are coincident with a
ridge of strongly enhanced radio polarization \citep{vsb07}
indicating the SE as the leading edge of the ICM-ISM interaction.
This is the side closest to M87, which suggests that
NGC~4501 is currently entering the high density region of the
cluster for the first time.

\noindent{\bf NGC~4522} ({\it Cashew})
The highly inclined Sc galaxy NGC~4522 is
one of the clearest cases of ongoing ram pressure stripping.
This galaxy was observed
by \citet{kvgv04} as a pilot study of the VIVA survey.
Its H{\sc i} has been stripped to well within the optical disk
(to 0.35R$_{25}$ along the major axis)  but
there is significant extraplanar H{\sc i} on one side of the disk.
The peaks in the extraplanar H{\sc i} are located just above the
gas truncation radius in the disk, a gas morphology indicative of ram pressure
stripping.
The old stellar disk ($R$-band image) is relatively undisturbed, implying that
ram pressure and not tidal interactions are responsible for the disturbed gas
distribution.
The extraplanar H{\sc i} in the west is kinematically distinct from the 
adjoining disk
gas,
with velocities offset toward the Virgo cluster mean velocity.
Enhanced radio polarization along the eastern edge, on the side opposite the
extraplanar H{\sc i}
\citep{vbkvg04}, and
a deficit of radio continuum emission relative to far-infrared emission beyond
the
polarized ridge \citep{mkhch09} both indicate strong active ram pressure.
In a comparison of simulations and data, \citet{vollmer06}
 find that the galaxy is in an active phase of
ram pressure stripping, and the best match to the H{\sc i} morphology and kinematics
is 50 Myr after peak pressure. Studies of the stellar population
by \citet[][2008]{ck06} show that star formation in the gas-free outer disk
stopped $\sim100~$Myr ago, consistent with the stripping time scale from
simulations. This galaxy is located
at 3.3 deg ($\sim$0.8 Mpc) from M87, where the estimated ICM pressure,
assuming a smooth and static ICM, is too low to remove the H{\sc i} from the
disk. \citet{kvgv04} have argued that motion or substructure
in the ICM, perhaps due to the merging of the M49 group with the main cluster,
could have increased the ram pressure on this galaxy.  In fact, NGC~4522 is
near a local peak in X-ray emission, a ridge where X-ray spectroscopy shows
high temperatures possibly indicating a shock front in the ICM \citep{smy01}.

\begin{figure}
\plotone{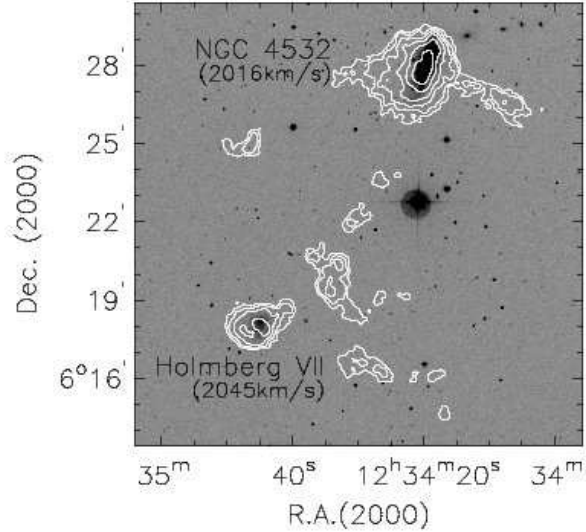}
\caption{The H{\sc i} distributions of NGC~4532 region is shown overlaid on the DSS image.
Only the C array data of \citet{hoff99} is shown while the total flux is well consistent with
theirs combined with the D array data. More sensitive H{\sc i} data taken with the
Arecibo data (ALFALFA survey; Koopmann et al. 2008) have revealed that NGC~4532 and
VCC~1581 (Holmberg~VII) are embedded in a common H{\sc i} envlope, which continues
to the south as a 500~kpc long H{\sc i} tail. \label{fig-4532}}
\end{figure}

\noindent{\bf NGC~4532} NGC~4532 is an H{\sc i}-rich, optically
peculiar Sm galaxy with strong H$\alpha$ emission indicating a high
star formation rate.  In Figure~\ref{fig-4532} as well as in the
atlas, we are presenting the same data as the C-array data of
\citet{hoff99}. The H{\sc i} distribution around this galaxy is patchy
with several clumps and a remarkably sharp east west extension.
Kinematically NGC~4532 appears to be in regular rotation with a warp
in the outer parts, the east west feature is kinematically decoupled
from the galaxy, and may be the tip of a huge tail seen by ALFALFA
\citep{kgh08}.  The total flux measured in this region
(55.9$\pm$1.2~Jy~km~s$^{-1}$) is consistent with what \citet{hoff99}
found in their C+D data (50.9~Jy~km~s$^{-1}$).
 The Arecibo Legacy Fast ALFA Survey (ALFALFA) showed recently that
 NGC~4532 and VCC~1581 (Holmberg~VII) are embedded within a large
 common envelope, which continues to the south as a 500 kpc long H{\sc
   i} tail \citep{kgh08}. The extended tail is far outside the primary
 beam of the VLA and could not have been detected in the VIVA survey.
 A comparison of the total flux seen by ALFALFA and VIVA shows that
 there is about as much H{\sc i} in the diffuse envelope around the
 pair as there is in the galaxies and small clouds seen by the VLA.
 It is particularly interesting to compare the velocity fields seen by
 the VLA and by ALFALFA. The giant tail connects to the galaxy pair in
 what is called the W cloud \citep{kgh08} at a velocity of about 1875
 km~s$^{-1}$, this is the same velocity seen by the VLA west of NGC~4532 in
 the east-west feature. We conclude then that the huge tail ends in
 the east west feature seen by the VLA and crosses NGC~4532.  NGC~4532
 is possibly influenced both by tidal effects and gas accretion.

\noindent{\bf VCC~1581, Holmberg VII} 
This is an optically faint low-mass galaxy
that lies within the huge H{\sc i} envelope (see Fig.~\ref{fig-4532}) 
that also covers NGC~4532 \citep{kgh08}.
Optically the galaxy has no clear nucleus or symmetry.
The  H{\sc i} distribution and kinematics are fairly regular and
symmetric, except for a cloud with distinct kinematics 1$'$ NE of the center.

\noindent{\bf NGC~4535} ({\it Snail}) 
The H{\sc i} content is relatively normal and
H{\sc i} extends somewhat beyond the stellar disk.
Sharp HI cutoffs to the N and NW, towards M87 while further extension
to the opposite side of the disk, are suggestive of weak
ongoing ram pressure.
\citet{vsb07} and \citet{vlcont07} detect polarized radio continuum 
in the center and the south west of the outer galactic disk. They argue
that this asymmetry is mostly due to an interaction with the ICM. 
The outermost H{\sc i} spiral arm to the W and SW is kinematically distinct,
with a sharp velocity gradient rather than the more
gradual and continuous curvature characteristic of spiral arm streaming motions.
There is also no other spiral arm in the galaxy that shows such kinematic
distinctness, supporting the picture that the
W/SW arm might be  due to ram pressure rather than spiral density
waves.
It appears similar to the kinematically distinct gas-stripped outer arm of
NGC~4569.
An H{\sc i} depression is found in the central 5 kpc
where a stellar bar is present. 
The radio continuum from the nuclear region is quite strong, probably due to an
AGN.

\begin{figure}
\plotone{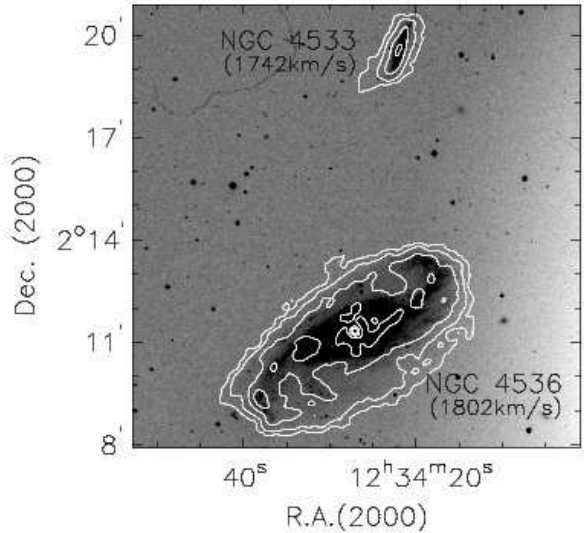}
\caption{The H{\sc i} distribution of NGC~4536 and NGC~4533 are
  overlaid on the DSS image. This pair of galaxies, about 39~kpc apart from each other in projection with similar $v_{\rm HI}$, show some morphological and kinematical peculiarities.\label{fig-4536}}
\end{figure}

\noindent{\bf NGC~4533} 
This galaxy has been detected at $\sim8'$
(39~kpc) distance to the north of NGC~4536 (Fig. \ref{fig-4536}). The H{\sc i} is symmetric
in the inner parts and has a diffuse asymmetric outer envelope
and a short H{\sc i} tail extends to the south east.
Optically it is a small Sd galaxy ($B_T=14.2$
and $D_{25}=2.1'$) and its stellar disk also looks asymmetric in the
sense that the northwest side is slightly warped and the southeast
disk is broader than the north west disk. There are two stellar tails
to the southeast, where the H{\sc i} tail is present.
Since this galaxy is close spatially and in velocity to NGC~4536 it
is plausible that the two galaxies are gravitationally interacting.

\noindent{\bf NGC~4536} The H{\sc i} disk is about the same size as
the stellar disk and high column density H{\sc i} ridges coincide
nicely with the optical spiral arms. The galaxy is located in the
southern outskirts of the cluster ($>2.8~$Mpc from M87) and shows no
clear signature of any kind of large disturbance. It is only 39~kpc
away in projection from NGC~4533 with $\Delta v_{\rm
  opt}<10~$km~s$^{-1}$ (or $\Delta v_{\rm HI}\approx50$~km~s$^{-1}$).
It is possible that this galaxy is responsible for NGC~4533's
peculiarities. During the close encounter with NGC~4533, this galaxy
may have been less affected because it is more massive (by a factor of
$\gtrsim4$) while NGC~4533 has been quite disturbed.  It does however
have a small bar in the central $\sim$1$'$, apparent in both the
optical morphology and the H{\sc i} PVDs, and the bar might have
originated from a tidal interaction.

\noindent{\bf NGC~4548} 
The H{\sc i} is mildly truncated within the optical disk and has a low surface
density
($D_{\rm HI}^{\rm iso}/D_{25}=0.86$ and $def_{\rm HI}=0.81\pm0.01$), resulting
in a rather large H{\sc i} deficiency.
An H{\sc i} hole is found in the central 5 kpc where a
strong stellar bar is present. This galaxy was also observed by 
\citet{vcbbd99} with a data quality very similar to ours. \citet{vcbbd99}
point out that the faint outer H{\sc i} arm in the north appears 
kinematically slightly distinct from the adjacent disk, and suggest
that this could have been caused by a past episode of ram pressure.
We find no evidence for ongoing ICM pressure or tidal stripping.
Kinematically distinct H{\sc i} is often found in the outer parts of 
spiral galaxies in the field and it needs not be related to the cluster 
environment. However, its H$\alpha$ morphology does suggest that NGC~4548
is anemic. The radio continuum emission is extremely weak, even for  
an anemic spiral \citep{vlcont07}.

\noindent{\bf NGC~4561}
The H{\sc i} in this small Sm or Sc galaxy shows the very high surface
density in the center while it is very extended, with two open
symmetric gas spiral arms reaching far beyond the stellar disk, 
to 2-3~R$_{25}$. The gas spiral arms appear to be superposed
on a very low surface brightness outer H{\sc i} disk.
There is no evidence of either young or old stars in these H{\sc i} arms.
The H{\sc i} velocity field shows strong non-circular motions with m=2 
symmetry, even within the optical disk.
There is a small stellar bar in the central 30$''$, but this small feature
cannot account for the widespread non-circular motions.
The H{\sc i} morphology could either be the result of a merger between two
small sytems, in which the outer   H{\sc i} is now falling back and
forming a new disk  \citep[e.g.][]{barnj02}, or alternatively it could
have been pulled out during an encounter with a nearby 
galaxy
IC~3605 which is $\approx150~$kpc away with $\Delta v=300~$km~s$^{-1}$).
The m=2 symmetry places strong constraints on any interaction scenario.
Lastly the outer H{\sc i} could be the remains of a giant low surface
brightness disk that is now being disturbed by the non axisymmetric 
stellar body in the center. 

\begin{figure}
\plotone{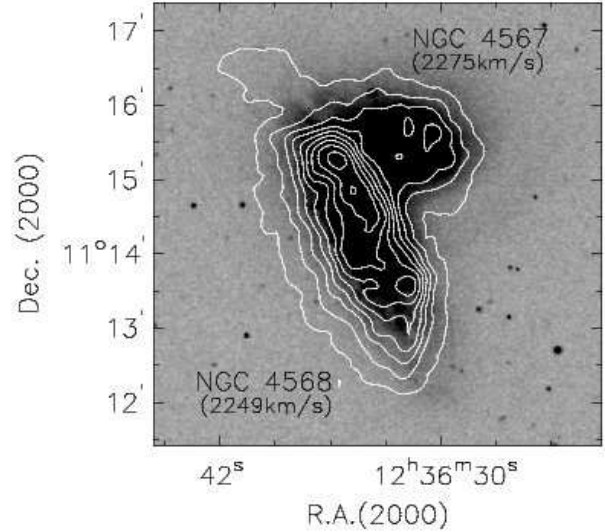}
\caption{The H{\sc i} distributions of NGC~4567 and NGC~4568 are shown in contours overlaid
on the DSS image. The two galaxies do not just overlap due to projection but are actually
interacting with each other as the channel maps or the PVDs show clear connections.
\label{fig-4567}}
\end{figure}

\noindent{\bf NGC~4567 and NGC~4568} 
This pair of Sc galaxies overlap on the sky (Fig.~\ref{fig-4567}), and
their line-of-sight velocities match where the galaxies overlap.
Optically neither shows significant disturbances in their
inner disks while both look mildly disturbed in their outer parts.
Likewise, the H{\sc i} morphology and kinematics look
fairly undisturbed within the brighter parts of the stellar disks.
In the outer parts of NGC~4568,
H{\sc i} contours  are more compressed in the east, toward NGC~4567,
and more extended in the west, where the kinematics suggest a warp or
other disturbance.
A prominent dust lane at the apparent intersection region between
the two galaxies suggests they are physically connected.
A little H{\sc i} tail extends northwest of the pair, and appears to be
an extension of the dust lane.
The H{\sc i} morphology suggests that the galaxies are
gravitationally interacting, but are in a phase before closest approach
so are not yet strongly disturbed.
The total H{\sc i} flux of the pair is consistent with single dish measurements.
Since the galaxies overlap both spatially and in velocity, it is difficult
(and perhaps not too meaningful) to ascribe an accurate  H{\sc i} flux for each
galaxy.
The fluxes presented in Table~\ref{tbl-hiprop} are based on assigning all the
flux in the overlap region to NGC~4568, since the  H{\sc i} contours suggest
that this may be appropriate.

\begin{figure}
\plotone{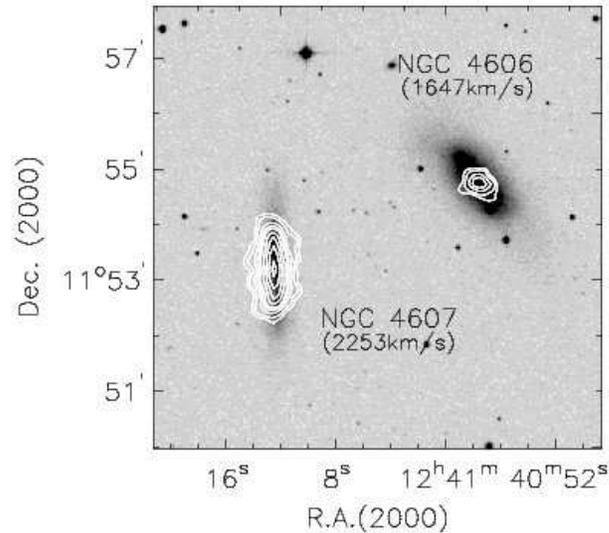}
\caption{The H{\sc i} distributions of NGC~4606 and NGC~4607 are shown in contours
overlaid on the DSS image. The projected distance between the two galaxies is $\lesssim$20~kpc while $\Delta v\approx600~$km~s$^{-1}$.
\label{fig-4606}}
\end{figure}

\noindent{\bf NGC~4569}
The H{\sc i} disk is only extended about one third of the stellar disk. 
Anomalous arms of H{\sc i} and H$\alpha$ emission to the west lie behind 
the stellar disk (Kenney et al. in prep). This extraplanar H{\sc i}, which is 
kinematically distinct from the disk gas,
is very likely gas stripped from the disk by ram pressure.
A stellar population study of the gas-free outer stellar disk \citep{ck08}
and a comparison of the H{\sc i} morphology and kinematics with simulations
\citet{vbcvdh04}
both indicate that the gas was stripped from the outer disk $\sim$300 Myr ago.
Radio continuum emission associated with star formation
is detected throughout the remaining gas disk \citep[see also][]{boselli06}.
At fainter levels there is radio continuum emission
near the minor axis on both sides extending 24 kpc from the center,
likely arising from a nuclear outflow \citep{csbvbbu06}.
Due to its negative velocity and large appearance
($\approx10'$), its cluster membership has been controversial for a long time
\citep[e.g.][]{rf70,sky86}.
Recent Tully-Fisher based distance estimates place NGC 4569 somewhat closer than
the mean Virgo distance \citep{sssgh02,cortes08}.
The H{\sc i} evidence for ICM-ISM stripping
strongly suggests that NGC~4569 is part of the cluster.

\noindent{\bf NGC~4579} 
This galaxy is moderately H{\sc i}-deficient, and
its H{\sc i} is mildly truncated within the optical disk.
But the H{\sc i} distribution and kinematics appear symmetric and regular,
with no indications of any ongoing interaction.
There is a deep depression in the central 4 kpc coincident with a stellar bar.
The star formation rate is relatively normal in the disk.
It has a well-known Seyfert 2 nucleus, with radio jets \citep{contini04}.
It shows a strong radio continuum emission, both from the nuclear source
and the extended star-forming disk.

\noindent{\bf NGC~4580} 
The H{\sc i} is very deficient and severely
truncated within the optical disk ($D_{\rm HI}^{\rm iso}/D_{25}<0.5$
and $def_{\rm HI}>1$). The H$\alpha$ is also truncated with a sharp
edge. The H{\sc i} peak is slightly offset from the optical center and
more emission is present to the southeast. The radio continuum appears
more extended than the H{\sc i} disk in the south.
The outer stellar disk appears undisturbed, although it still contains spiral arms.
The H{\sc i} truncation and undisturbed stellar disk
strongly suggest an ICM-ISM interaction. 
The stellar population study of \citet{ck08} indicates
that star formation stopped in the gas-free outer stellar disk
500 Myr ago. This galaxy is so far from the cluster core
that it could not have reached its current location by travelling from the core
in only 500 Myr, so it must have been stripped outside the core.

\noindent{\bf NGC~4606} 
The H{\sc i} extends only one tenth of the
stellar disk. The H{\sc i} peak is offset from the optical center with
slightly more emission to the west.  A faint low surface brightness
H{\sc i} feature extends to the east near the minor axis.  The
H$\alpha$ is strong in the central kpc but severely truncated within
the optical disk and asymmetric in the same way as the H{\sc i}. The
radio continuum is quite strong in the center with almost the same
extent as the H{\sc i} disk.  The stellar disk is disturbed
\citep{corphd05}, with non-elliptical isophotes and a
greater extension to the northeast.  Disturbed dust lanes are also
apparent.  This galaxy may have experienced some type of
gravitational interaction, although  it is not clear what the relative roles of ram
pressure and tidal interactions may have been in shaping the H{\sc i} properties.

\noindent{\bf NGC~4607}
The H{\sc i} in this  edge-on galaxy is truncated to within the stellar
disk and the galaxy is strongly deficient in H{\sc i} ($D_{\rm HI}^{\rm
 iso}/D_{25}\approx0.7$ and $def_{\rm HI}=0.82$).
As shown in Figure~\ref{fig-4606}, NGC~4607 is located near NGC~4606 with a
projected distance of only about 20~kpc. However, the two galaxies 
have very different velocities with $\Delta
v\approx600~$km~s$^{-1}$ making it unlikely that they are
gravitationally bound.
Although NGC~4606 looks optically somewhat disturbed, we find no evidence
for a tidal disturbance in the optical or H{\sc i} appearance of NGC~4607.
  
\begin{figure*}
\plotone{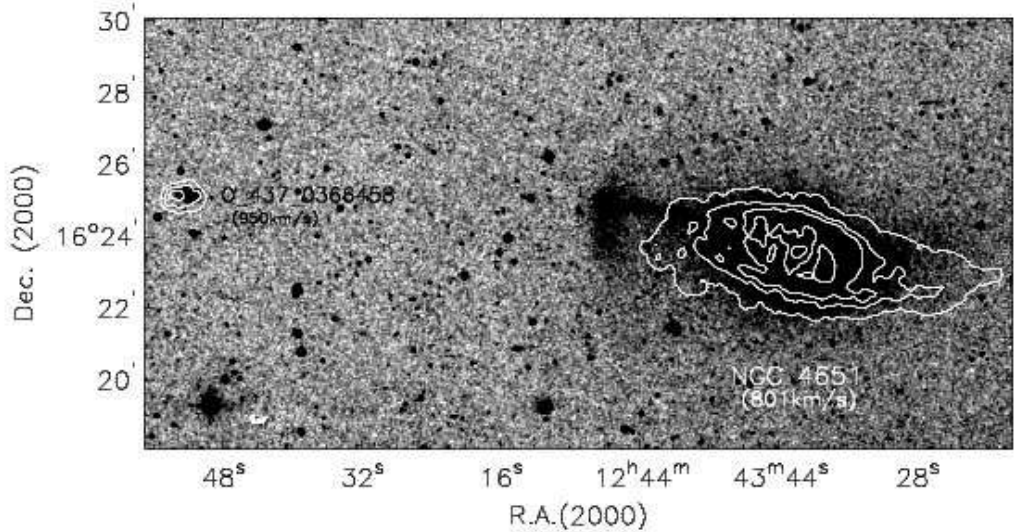}
\caption{The H{\sc i} contours of NGC~4651 are shown overlaid on a high contrast
DSS image. The H{\sc i} distribution is asymmetric in a sense that it is more extended
to the west. In the opposite side of the H{\sc i} tail, a stellar tail is found,
which ends with a shell. On the same side of the disk, we also have detected a small
H{\sc i} cloud of $\approx2.2\times10^7~M_\odot$ at $\approx80~$kpc distance from 
the center of NGC~4651. The cloud has an optical counterpart, MAPS-NGP~O\_437\_0366458
\citep{cabanela99}.\label{fig-4651}}
\end{figure*}

\noindent{\bf NGC~4651} 
While the bright central part of this Sc
galaxy (inside R$_{25}$) appears relatively symmetric and undisturbed
both in the optical and in H{\sc i}, the H{\sc i} beyond the stellar
disk looks more disturbed, possibly warped, and on the western side the
H{\sc i} appears to become more narrow and form a tail-like feature.
The H{\sc i} kinematics of
the outer galaxy has a different position angle from the inner galaxy,
confirming that the outer H{\sc i} may be warped.  While the inner
optical disk is quite symmetric, a high contrast optical image reveals
a remarkably straight optical tail to the east which ends at a low
surface brightness arc (Figure ~\ref{fig-4651}).  Although no H{\sc i}
is at the position of the optical tail and arc, Figure ~\ref{fig-4651}
shows that 50~kpc away from the stellar shell (and at about 80~kpc
distance from the center of NGC~4651) in the direction of the stellar
tail there is a small H{\sc i} cloud of $S_{\rm
HI}=0.36~$Jy~km~s$^{-1}$ ($M_{\rm HI}=2.2\times10^7~M_\odot$).  The
H{\sc i} cloud has an optical counterpart
\citep[MAPS-NGP\footnote{Minnesota Automated Plate Scanner - North
Galactic Pole}~O\_437\_0366458;][]{cabanela99}. This dwarf galaxy 
may well have a tidal origin.
The H{\sc i} surface density drops steeply in the central 2 kpc. 
Although there is quite  strong radio emission found in the center of the galaxy
(50 mJy), the low surface brightness makes it unlikely that the depression
in H{\sc i} is due to absorption. 


\begin{figure*}
\plotone{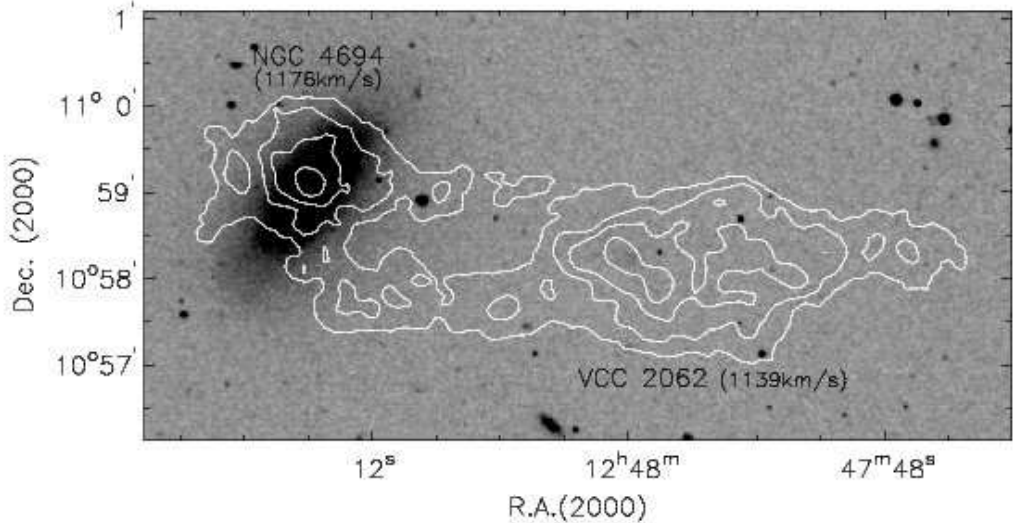}
\caption{The H{\sc i} distribution of NGC~4694 region is shown overlaid on the DSS image.
The H{\sc i} is stripped on both sides of the stellar disk while short and long extensions
are found to the northeast and southwest along the minor axis. The H{\sc i} feature in
the southwest also covers a low surface brightness galaxy, VCC~2062, and extends more
than 30~kpc from the center of NGC~4694. \label{fig-4694}}
\end{figure*}


\noindent{\bf NGC~4654} 
We present the same data published in
\citet{pm95}. The H{\sc i} map shows an extended tail
to the southeast and compressed contours to the
northwest. \citet{pm95} have attributed its H{\sc i} peculiarities to
the ram-pressure due to the ICM. The stellar disk extends beyond the H{\sc i} in the
northwest, suggesting an ICM-ISM interaction.
However, the inferred ICM pressure alone may be too low to strip the H{\sc i} at
this
location $\sim1~$Mpc from M87 \citep{cvgkv07}.
Alternatively, \citet{voll03} has shown that the  H{\sc i}
morphology and kinematics are best reproduced by a combination of both
ram-pressure and a gravitational interaction (e.g., with its neighbor
NGC~4639).  In
fact, the stellar disk also appears to be somewhat disturbed with more
diffuse emission to the southeast, which makes it more plausible that
a gravitational disturbance took place as well. See also
\citet{cvgkv07}.

\noindent{\bf NGC~4689} 
The H{\sc i} in this galaxy is mildly truncated
 within the stellar disk with a moderate H{\sc i} deficiency. The
 H{\sc i} morphology and kinematics are fairly regular and symmetric,
 showing no signatures of ongoing ram pressure or tidal interactions.
 Weak radio continuum emission is detected over about half the H{\sc
 i} disk.  The galaxy is also mildly truncated in H$\alpha$.

\noindent{\bf NGC~4694} 
The H{\sc i} in this galaxy is highly disturbed and irregular, and bears little
relation to the
stellar disk, as shown in Figure~\ref{fig-4694}. The H{\sc i} extent along the
major axis
is much smaller than the stellar disk, but along the minor axis
is relatively more extended.  The southwest extension is
connected to its optically faint neighbor, VCC~2062, and continues well beyond
that galaxy.
The H{\sc i} peak within the stellar disk almost coincides with
the optical center, however the morphology and kinematics are quite
asymmetric along both the major and the minor axes. Overall the H{\sc i} properties
are
consistent with an accretion event, and inconsistent with ram pressure
stripping.
Strong radio continuum emission is present at the
center.  Optically it has a large bulge with an extended but low surface
brightness disk.
The lack of H{\sc i} and star formation in the outer stellar disk, and the disturbed
appearance of
the central bulge-dominated region, in part due to irregular dust lanes,
contribute to its peculiar SB0 or Amorphous classification.

\noindent{\bf VCC~2062} 
 A huge H{\sc i} cloud is found to cover both
this low surface brightness dwarf galaxy and part of the nearby disturbed spiral
NGC~4694 (Fig.~\ref{fig-4694}). The H{\sc i} peak
almost coincides with the position of the highest surface brightness
in the optical. Neither the H{\sc i} nor the optical morphology is
well structured while a relatively smooth H{\sc i} velocity gradient
is present across the stellar disk. No radio continuum is found. It is
obvious that VCC~2062 is tidally interacting with NGC~4694, although the
origin of the stellar component of VCC~2062 is rather unclear, i.e. whether it
has tidally formed or it has
been destroyed due to the tidal interaction. Recently \citet{dblbb07}
have reported their {\sc CO} detection from this system.


\noindent{\bf NGC~4698} ({\it Fish tail}) 
The H{\sc i} disk is almost a factor 2 larger
than the stellar disk. Its H{\sc i} morphology and kinematics are fairly
undisturbed
but two short tails are found to the southeast, while to the northwest, which is
toward
the cluster center, the H{\sc i} appears to be fairly compressed. Thus the
galaxy may be
experiencing weak ram pressure as it approaches the cluster core.
Optically it has a low surface brightness
ring at the location where $\Sigma_{\rm HI}\approx1~M_\odot~$pc$^2$ as well as
an inner stellar ring at smaller radii. This galaxy is known for its
orthogonally
rotating bulge which was optically discovered \citep{bertola99}.
H$\alpha$ ionized gas kinematics show a clear decoupling of the gas and stellar
kinematics
in the central few kpc \citep{corphd05}, and the  H{\sc i} kinematics show hints of
this.
The galaxy likely experienced a merger over 1 Gyr ago, and is now experiencing
an unrelated episode of early stage ISM-ICM ram pressure stripping.

\noindent{\bf NGC~4713} ({\it Crab})
This galaxy has one of the most extended H{\sc i} disks in the 
cluster as compared to the optical disk and shares many properties
with NGC~4808, one of its nearest large neighbours in the southern
outskirts of the cluster.
The  H{\sc i} velocity field suggests a disturbed warp for the gas
beyond the stellar disk. There are different  H{\sc i} velocity
gradients and position angles on the two sides of the disk.
It is classified as an Sc or Sd galaxy, with
very strong star formation distributed somewhat irregularly throughout the
stellar disk. The H{\sc i} surface density distribution is
has  a depression in the center
coincident with a stellar bar.
A prominent optical spiral arm in the NW lies just inside an outer H{\sc i} arm.
Our H{\sc i} data for this galaxy are much older and noisier than than most of the
VIVA data. Yet our measured total H{\sc i} flux is only 10\% less than the flux measured in
the ALFALFA survey, despite our noisy data we recovered almost all of the H{\sc i}.
The radio continuum distribution is also irregular, with an extended
off-nuclear peak in the east, and a secondary peak in the west just inside
the prominent spiral arm.

\noindent{\bf NGC~4772} 
This galaxy was imaged previously by \citet{hjbbm00} in C array.
Although our data were taken with CS array and better sofware
was available, including the option to use robust weighting,
our results are similar to those of \citet{hjbbm00}.
 A prominent outer H{\sc i} ring is present.
Compared to the inner H{\sc i} disk,
the outer H{\sc i} ring is at a different position angle and has a lower peak LOS
velocity,
and may therefore lie in a different plane.
The inner H{\sc i} appears slightly warped at the end of the stellar disk,
and the warp feature may connect the inner disk and outer H{\sc i} ring.
Its optical morphology resembles that of NGC~4698 with a strong
bulge and a low surface brightness disk. As in NGC~4698 the H{\sc i}
extends well beyond the optical disk.
A comparison of stellar and H$\alpha$ gas velocities by  \citet{hjbbm00}
show central gas components with anomolous velocities.
Both the H{\sc i} properties and the central gas kinematics are
suggestive of a minor merger \citep{hjbbm00}, and the relatively
symmetric H{\sc i} and stellar morphologies suggest an older event.

\begin{figure}
\plotone{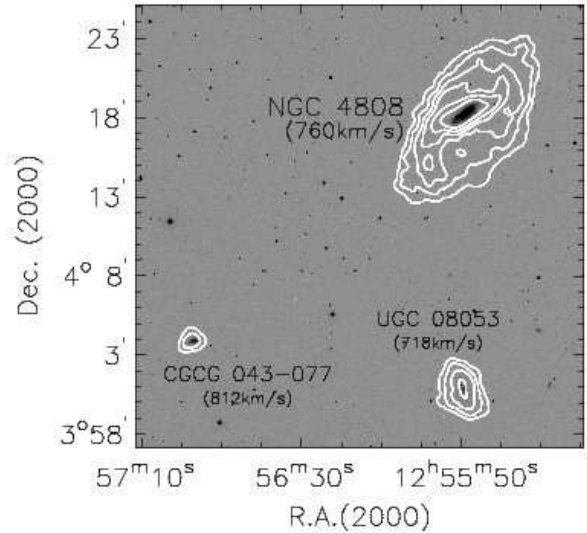}
\caption{The H{\sc i} distributions of NGC~4808 and nearby dwarfs are overlaid on
the DSS image. The combined data of the C and the D configurations are presented.
The dwarfs have been discovered in our recent D array follow-up observations.
\label{fig-4808}}
\end{figure}

\noindent{\bf NGC~4808}  This galaxy is very H{\sc i}-rich.
The H{\sc i} is much more extended than the stellar disk, and
both the H{\sc i} morphology and kinematics suggest a strong and asymmetric warp.
Optically the galaxy has been classified as Scd with a very weak bulge.
There is strong star formation throughout the stellar disk, with an
asymmetric and patchy distribution.
H{\sc i} has been detected from two nearby dwarf galaxies in the VLA D array
data (Fig.~\ref{fig-4808}).
Although no evidence for tidal interactions is found,  the three galaxies
are located within $\lesssim60~$kpc distances from each other with
similar velocities ($\Delta v<100~$km~s$^{-1}$) and it is very likely
that these galaxies are under the influence of one another.
Apart from the nearby dwarf galaxies,
NGC~4808 shares many properties with NGC~4713, one of its nearest large
neighbors
in the southern outskirts of the cluster.

\end{document}